\documentclass[11pt,a4paper]{article}
\pdfoutput=1
\usepackage{amssymb,amsmath,amsfonts, mathtools, mathrsfs}
\usepackage[utf8]{inputenc} 
\usepackage[dvipsnames]{xcolor}

\newif\ifnatbibsort\natbibsorttrue

\relax

\ifnatbibsort\RequirePackage[numbers,sort&compress]{natbib}\else\RequirePackage[numbers,compress]{natbib}\fi
\RequirePackage[colorlinks=true
,urlcolor=blue
,anchorcolor=blue
,citecolor=blue
,filecolor=blue
,linkcolor=blue
,menucolor=blue
,pagecolor=blue
,linktocpage=true
,pdfproducer=medialab
,pdfa=true
]{hyperref}

\usepackage{graphicx}
\usepackage{caption}
\usepackage{tensor}
\usepackage{subfigure}
\usepackage{enumerate}
\usepackage{dsfont}
\usepackage{verbatim}
\usepackage{amsfonts,euscript,amssymb,stmaryrd,braket}
\usepackage{tikz}
\usetikzlibrary{arrows,decorations.markings,patterns}
\usepackage{slashed}

\def\clock{{\count0=\time
		\divide\count0 60
		\ifnum\count0<10 0\fi\the\count0
		\multiply\count0 -60 \advance\count0 \time
		:\ifnum\count0<10 0\fi \the\count0
}}
\newcommand{\timestamp}{{\small\vbox{\hbox{\tt\jobname.tex}
			\hbox{\the\day/\the\month/\the\year, \clock}}}}

\newcommand{\bea}{\begin{eqnarray}}
\newcommand{\eea}{\end{eqnarray}}

\DeclareMathOperator{\diag}{diag}

\newcommand{\be}{\begin{equation}}
\newcommand{\ee}{\end{equation}}

\makeatletter
\let\old@startsection=\@startsection
\let\oldl@section=\l@section
\renewcommand{\@startsection}[6]{\old@startsection{#1}{#2}{#3}{#4}{#5}{#6\mathversion{bold}}}
\renewcommand{\l@section}[2]{\oldl@section{\mathversion{bold}#1}{#2}}
\makeatother

\numberwithin{equation}{section}

\usepackage{color}

\begin{document}

\renewcommand{\thefootnote}{\arabic{footnote}}

	\overfullrule=0pt
	\parskip=2pt
	\parindent=12pt
	\headheight=0in \headsep=0in \topmargin=0in \oddsidemargin=0in

	\vspace{ -3cm} \thispagestyle{empty} \vspace{-1cm}
	\begin{flushright} 
		\footnotesize
		\textcolor{red}{\phantom{print-report}}
	\end{flushright}

\begin{center}
	\vspace{.5cm}

%
	
	{\Large\bf \mathversion{bold}
	Subsystem complexity after a local quantum quench \\
	}

	\vspace{0.8cm} {
		Giuseppe Di Giulio
		and 
		Erik Tonni
	}
	\vskip  0.7cm
	
	\small
	{\em
		SISSA and INFN Sezione di Trieste, via Bonomea 265, 34136, Trieste, Italy 
	}
	\normalsize
	
\end{center}

\vspace{0.3cm}
\begin{abstract} 
We study the temporal evolution of the circuit complexity 
after the local quench where two harmonic chains are suddenly joined,
choosing the initial state as the reference state.
We discuss numerical results for the complexity for the entire chain 
and the subsystem complexity for a block of consecutive sites,
obtained by exploiting the Fisher information geometry of the covariance matrices. 
The qualitative behaviour of the temporal evolutions of the subsystem complexity 
depends on whether the joining point is inside the subsystem.
The revivals and a logarithmic growth observed during these temporal evolutions are discussed. 
When the joining point is outside the subsystem,
the temporal evolutions of the subsystem complexity 
and of the corresponding entanglement entropy
are qualitatively similar.
\end{abstract}

\newpage

\tableofcontents


\section{Introduction}
\label{sec:intro}

A quantum circuit constructs a target state from a given reference state
through a sequence of gates chosen within a set of allowed gates. 
The circuit complexity has been introduced in quantum information theory 
as the minimum number of allowed gates employed to construct the circuit 
\cite{Nielsen06,NielsenDowling06, DowlingNielsen08, Watrous2008quantum, Aaronson:2016vto}.
During the past few years the circuit complexity 
has been investigated in the context of quantum gravity 
through the gauge/gravity (holographic) correspondence 
\cite{Susskind:2014rva,Susskind:2014jwa, Roberts:2014isa, Stanford:2014jda, Susskind:2014moa,
Alishahiha:2015rta, Brown:2015bva,Brown:2015lvg,
Barbon:2015ria,Carmi:2016wjl}.
It is worth exploring the circuit complexity also 
in quantum field theory and in quantum many-body systems.

Within the class given by the quantum many-body systems, 
it is natural to start with quantum circuits made by Gaussian states in free systems
\cite{Weedbrook12b,Serafini17book}.
The complexity of these circuits 
when only pure states are allowed has been investigated,
obtaining explicit expressions 
both for bosonic and for fermionic lattices
\cite{Jefferson:2017sdb,Chapman:2018hou,Guo:2018kzl,Hackl:2018ptj,Khan:2018rzm,Braccia:2019xxi,Chapman:2019clq,Doroudiani:2019llj,guo2020circuit}.

It is important to quantify the complexity also for quantum circuits made by mixed states,
which are characterised by density matrices 
\cite{aharonov1998quantum,Agon:2018zso,Caceres:2019pgf,DiGiulio:2020hlz,Ruan:2020vze,Camargo:2020yfv}.
The reduced density matrices provide an important class of mixed states that are crucial to study 
the bipartite entanglement. 
Given the spatial bipartition $A\cup B$ 
of a quantum system in a state characterised by the density matrix $\rho$
whose Hilbert space can be factorised accordingly as $\mathcal{H} = \mathcal{H}_A \otimes \mathcal{H}_B$,
the reduced density matrix $\rho_A \equiv \textrm{Tr}_{\mathcal{H}_B} \rho$  of the spatial subsystem $A$
(normalised by $\textrm{Tr}_{\mathcal{H}_A} \, \rho_A =1$)
characterises a mixed state.
An important quantity to consider is the entanglement entropy $S_A = - \,\textrm{Tr} (\rho_A \log \rho_A)$.
When the entire system is in a pure state, $S_A = S_B$ 
(see \cite{Eisert:2008ur,Casini:2009sr,Calabrese:2009qy,Peschel_2009, 
Rangamani:2016dms,Headrick:2019eth,Tonni:2020bjq} for reviews).
The subsystem complexity $\mathcal{C}_A$ is defined as the complexity of a circuit 
where both the reference state and the target state are reduced density matrices
associated to the same spatial subsystem $A$.

It is insightful to compare $\mathcal{C}_A$ and $S_A$.
At equilibrium, this comparison has been discussed  
 in lattice models \cite{DiGiulio:2020hlz,Caceres:2019pgf} 
and in various gravitational backgrounds within the holographic correspondence  
\cite{Alishahiha:2015rta,Carmi:2016wjl,Abt:2018ywl,Agon:2018zso,Alishahiha:2018lfv,Auzzi:2019vyh},
finding that a major distinction occurs in the leading divergence:
while for $S_A$ it is determined by the volume of the boundary of $A$ 
(this law is violated in a 2D conformal field theory in its ground state,
e.g. when $A$ is an interval and therefore the boundary of $A$ is made by two points 
\cite{Callan:1994py,Holzhey:1994we,Calabrese:2004eu}),
for $\mathcal{C}_A$ it grows like the volume of $A$,
with a power that depends on the choice of the cost function \cite{Caceres:2019pgf,DiGiulio:2020hlz}.

Quantum quenches are interesting protocols to study
the dynamics of isolated quantum systems out of equilibrium 
(see \cite{Essler:2016ufo,Calabrese:2016xau} for reviews).
Given a system prepared in a state $| \psi_0 \rangle$,
consider a sudden change at $t=0$ that provides
the time-evolved state $| \psi(t) \rangle =  e^{-\textrm{i} H t}  | \psi_0 \rangle$ for $t>0$.
Since typically $| \psi_0 \rangle$ is not an eigenstate of the evolution Hamiltonian $H$,
this time-evolved pure state is highly non trivial.
The kind of sudden change leads to identify two main classes of quantum quenches.
Global quenches are characterised by sudden changes that involve the entire system 
(e.g. a modification of a parameter in the Hamiltonian) 
\cite{Calabrese_2005,Calabrese:2006rx,Calabrese:2007rg}.
Instead, in local quenches the sudden change occurs only at a point.
For instance, local quenches where 
either two systems are joined together \cite{Eisler_2007,Calabrese:2007mtj}
or a local operator is inserted at some point \cite{Nozaki:2014hna,Nozaki:2014uaa}
have been explored. 
The temporal evolutions of the entanglement entropy $S_A$
after various quantum quenches (either global or local) have been widely studied during the past few years
\cite{Sotiriadis:2010si,Cotler:2016acd,Eisler_2008local,Cardy:2011zz,Dubail_2011,Asplund:2013zba, Modak:2020faf}.
For systems in a finite volume, 
revivals occur in some temporal evolutions
\cite{Cardy:2011zz,Cardy:2014rqa,daSilva:2014zva}.

It is worth investigating the temporal evolutions 
of the circuit complexity for the entire system and of the subsystem complexity
after different quantum quenches. 
For some global quenches,
holographic prescriptions have been employed to determine numerically
the temporal evolutions of the complexity for the entire system \cite{Moosa:2017yvt,Chapman:2018dem,Chapman:2018lsv} 
and of the subsystem complexity \cite{Chen:2018mcc,Auzzi:2019mah,Ling:2019ien,Zhou:2019xzc}.
In free lattice models,
the temporal evolutions of the complexity after some global quenches have been studied,
both for the entire system 
\cite{Alves:2018qfv,Camargo:2018eof,Ali:2018fcz,Jiang:2018gft,Ali:2018aon,Bhattacharyya:2020iic} 
and for subsystems
\cite{DiGiulio:2021oal,Camargo:2018eof, Bhattacharyya:2020iic}.
The temporal evolution of the holographic entanglement entropy after a local quench 
has been explored in 
\cite{Nozaki:2013wia,Ugajin:2013xxa,Astaneh:2014fga,Asplund:2014coa,Jahn:2017xsg,Shimaji:2018czt,Caputa:2019avh}.
For the local quench corresponding to an operator insertion
the temporal evolution of the holographic subsystem complexity has been considered \cite{ Ageev:2018nye,Ageev:2019fxn},
while for the local quench where two systems are joined together
only the holographic entanglement entropy has been studied \cite{Ugajin:2013xxa,Shimaji:2018czt}.

\begin{figure}[t!]
{
\subfigure
{\hspace{-.27cm}
\includegraphics[width=1.02\textwidth]{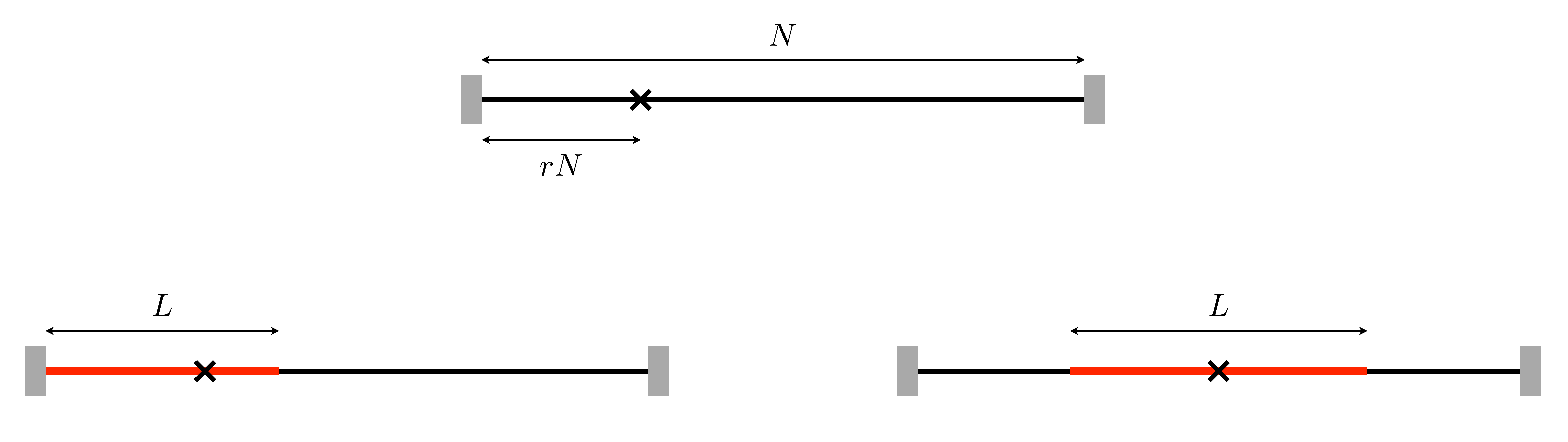}}
}
\caption{The local quench considered in this manuscript:
two harmonic chains containing $N_\textrm{\tiny l} = rN$ and $N_\textrm{\tiny r} =(1-r)N$ sites
are joined together at $t=0$  into a single chain made by $N \equiv N_\textrm{\tiny l} + N_\textrm{\tiny r}$ sites
(top panel).
The subsystem $A$ is a block of $L$ consecutive sites (red segment),
either adjacent to the left boundary (bottom left panel)
or centered at the midpoint of the chain, when $r=1/2$ (bottom right panel).
}
\vspace{0.5cm}
\label{fig:intro-configs}
\end{figure}

In this manuscript, we are interested in the temporal evolution of the circuit complexity after a local quench.
We consider the local quench described by Eisler and Zimbor\'as in \cite{Eisler_2014},
where two harmonic chains containing $r N$ and $(1-r)N$ sites 
are joined at $t=0$ (here $r$ is a rational number $0 \leqslant r \leqslant 1$),
as shown in the top panel in Fig.\,\ref{fig:intro-configs}.
We focus on circuits made only by Gaussian states. 
First we study the temporal evolution of the complexity for the entire chain;
then we investigate the temporal evolution of the subsystem complexity $\mathcal{C}_A$
for the subsystem $A$ given by a block of $L$ consecutive sites
(the spatial bipartitions considered in the manuscript are shown in the bottom panels of Fig.\,\ref{fig:intro-configs}),
by employing the complexity for mixed states based on the Fisher information geometry
\cite{DiGiulio:2020hlz}.

The outline of this manuscript is as follows. 
In Sec.\,\ref{subsec:setup}  we describe the local quench protocol,
introducing the covariance matrices characterising the states involved in the construction of the optimal circuit.
In Sec.\,\ref{sec:PureStateComp} we evaluate numerically the temporal evolution of the circuit complexity for the entire chain,
choosing the (pure) initial state at $t=0$ as the reference state 
and the (pure) state at generic time $t>0$ after the local quench as the target state. 
In Sec.\,\ref{sec-subsystem-comp} we discuss our numerical analysis of 
the  temporal evolutions of the subsystem complexity for a block of consecutive sites.
Some conclusions and open questions are drawn in Sec.\,\ref{sec:conclusions}.
The Appendices\,\ref{app:Newglobalquench}, \ref{app:EulerDecomposition} and \ref{app:Initialgrowth}
contain further technical details and supplementary results.


\section{Covariance matrix after the quench}
\label{subsec:setup}

The Hamiltonian of the harmonic chain made by $N$ sites (we set  $\hbar=1$) reads
\be
\label{HC ham-1d}
\widehat{H} 
\,=\, 
\sum_{i=0}^{N} \left(\,
\frac{1}{2m}\,\hat{p}_i^2+\frac{m\omega^2}{2}\,\hat{q}_i^2 
+ \frac{\kappa}{2}(\hat{q}_{i+1} -\hat{q}_{i})^2
\right)
\ee
where the position and the momentum operators  $\hat{q}_i$ and $\hat{p}_i$
are hermitean operators satisfying the canonical commutation relations $[\hat{q}_i , \hat{q}_j]=[\hat{p}_i , \hat{p}_j] = 0$ 
and $[\hat{q}_i , \hat{p}_j]= \textrm{i} \delta_{i,j}$.
The Dirichlet boundary conditions (DBC) 
$\hat{q}_0=\hat{q}_{N+1}=0$ and $\hat{p}_0=0$
are imposed at the endpoints.

The initial state is given by  the following pure state
\be
\label{InitialState general}
|\Psi_0^{\textrm{\tiny (l,r)}}\rangle
\equiv
|\psi_\textrm{\tiny l}\rangle\otimes|\psi_\textrm{\tiny r}\rangle
\ee
where $|\psi_\textrm{\tiny l}\rangle$  
is the ground state of the Hamiltonian $\widehat{H}_{\textrm{\tiny l}}$,
defined by (\ref{HC ham-1d}) for the sites labelled by $0\leqslant i \leqslant N_\textrm{\tiny l}$,
with the physical parameters $m_\textrm{\tiny l}$, $\omega_\textrm{\tiny l}$ and $\kappa_\textrm{\tiny l}$ 
and DBC imposed at $i=0$ and $i=N_\textrm{\tiny l}+1$.
Similarly, $|\psi_\textrm{\tiny r}\rangle$  
is the ground state of the Hamiltonian $\widehat{H}_{\textrm{\tiny r}}$
in (\ref{HC ham-1d}) for the chain made by the sites labelled by 
$N_\textrm{\tiny l} \leqslant i \leqslant N_\textrm{\tiny l}+ N_\textrm{\tiny r} +1$,
with parameters $m_\textrm{\tiny r}$, $\omega_\textrm{\tiny r}$ and $\kappa_\textrm{\tiny r}$ 
and DBC imposed at $i=N_\textrm{\tiny l}$ and $i=N_\textrm{\tiny l}+N_\textrm{\tiny r}+1$.
Thus, the initial state (\ref{InitialState general}) depends on 
$N_\textrm{\tiny l}$, $m_\textrm{\tiny l}$, $\omega_\textrm{\tiny l}$, $\kappa_\textrm{\tiny l}$, 
$N_\textrm{\tiny r}$, $m_\textrm{\tiny r}$, $\omega_\textrm{\tiny r}$ and $\kappa_\textrm{\tiny r}$.
The total number of sites is $N\equiv N_\textrm{\tiny l}+N_\textrm{\tiny r}$.
Equivalently, we can describe the initial state in terms of $N$ 
and of the position parameter $0\leqslant r \leqslant 1$ (in unit of $N$), 
that determines the separation between the left and the right chain;
indeed $N_\textrm{\tiny l}=N r$ and $N_\textrm{\tiny r}=N (1-r)$.

Given the state (\ref{InitialState general}) at $t=0$,
the time evolved state at $t>0$ through (\ref{HC ham-1d}) is
\be
\label{EvolvedState general}
|\Psi^{\textrm{\tiny (l,r)}}(t)\rangle=
e^{-\textrm{i}\widehat{H} t}|\Psi^{\textrm{\tiny (l,r)}}_0\rangle\,.
\ee

This setup describes different quantum quenches. 
A global quench can be obtained by setting $N_\textrm{\tiny l}=N$ and $N_\textrm{\tiny r}=0$ (or viceversa, equivalently), 
$\kappa_\textrm{\tiny l}=\kappa_\textrm{\tiny r}\equiv\kappa_0$, $m_\textrm{\tiny l}=m_\textrm{\tiny r}\equiv m_0$ and $\omega_\textrm{\tiny l}=\omega_\textrm{\tiny r}\equiv\omega_0$. 
In this case the initial state is the ground state of a single chain made by $N$ sites.  
If $\kappa_0\neq \kappa$, $m_0\neq m$ and $\omega_0\neq \omega$, 
the global quench involves all the parameters occurring in the Hamiltonian (\ref{HC ham-1d}).
An important special case is  the global quench of the frequency parameter \cite{Calabrese:2006rx}
discussed in Appendix\,\ref{subapp:massquench}.
In Appendix\,\ref{subapp:kappaquench} we consider
the global quench of the spring constant and of the frequency,
which corresponds to  $m_0= m$, $\kappa_0\neq \kappa$ and $\omega_0\neq \omega$.

In this manuscript we consider the local quench described in \cite{Eisler_2014}, 
where two disconnected harmonic chains, 
containing  $N_\textrm{\tiny l}= r N $ and $N_\textrm{\tiny r}= (1-r)N$ sites,
are joined at $t=0$,
as represented pictorially in the top panel of Fig.\,\ref{fig:intro-configs}.
This quench protocol corresponds to 
$m_\textrm{\tiny r}=m_\textrm{\tiny l}=m$, 
$\kappa\textrm{\tiny r}=\kappa\textrm{\tiny l}=\kappa$
and $\omega\textrm{\tiny r}=\omega_\textrm{\tiny l}=\omega$.

The bosonic Gaussian states in harmonic lattices are fully characterised by their covariance matrix
\cite{Weedbrook12b,Serafini17book}.
In the quench protocol that we are considering \cite{Eisler_2014},
the initial state (\ref{InitialState general}) is Gaussian 
and the time evolution generated by (\ref{HC ham-1d}) preserves its Gaussian nature;
hence (\ref{EvolvedState general}) is Gaussian too, for any $t>0$.
The time evolved state (\ref{EvolvedState general})
is completely characterised by the $2N\times 2N$ covariance matrix $\gamma^{\textrm{\tiny (l,r)}}(t)$, 
whose generic element is defined as 
\be
\label{CM-def}
\gamma_{i,j}^{\textrm{\tiny (l,r)}}(t) 
\,=\,\textrm{Re} \big[ \langle \hat{r}_i(t)\, \hat{r}_j(t)\rangle \big]
\;\;\qquad\;\;
\hat{\boldsymbol{r}}(t) \equiv \big(\hat{q}_1(t) , \dots , \hat{q}_N(t), \hat{p}_1(t), \dots, \hat{p}_N(t) \big)^{\textrm{t}}\,.
\ee

The covariance matrix of the initial state (\ref{InitialState general}) at $t=0$ reads \cite{Eisler_2014}
\be
\label{initial CM general}
\gamma_0^{\textrm{\tiny (l,r)}}
\,=\,
V_0^{\textrm{t}} \,\Gamma_0 \,V_0 
\ee
where the superscript indicates that this covariance matrix corresponds to an initial configuration 
made by two disjoint chains, containing $N_\textrm{\tiny l}$ and $N_\textrm{\tiny r}$ sites  respectively,
and 
\be
\label{Q0-P0 mat diag}
\Gamma_0 \,=\, \mathcal{Q}_0 \oplus \mathcal{P}_0
\hspace{2.5cm}
\mathcal{Q}_0 \,=\, \frac{1}{2} \, \mathcal{T}_0 ^{-1}
\;\;\qquad\;\;
\mathcal{P}_0 \,=\, \frac{1}{2} \, \mathcal{T}_0 
\ee
with $\mathcal{T}_0 $ being the following diagonal matrix
\be
\label{T0 mat diag}
\mathcal{T}_0 \,\equiv\,
\textrm{diag} \Big(
m_\textrm{\tiny l} \,\Omega^{\textrm{\tiny (l)}}_1,
\dots,
m_\textrm{\tiny l} \,\Omega^{\textrm{\tiny (l)}}_{N_\textrm{\tiny l}},
m_\textrm{\tiny r} \,\Omega^{\textrm{\tiny (r)}}_1,
\dots,
m_\textrm{\tiny r} \,\Omega^{\textrm{\tiny (r)}}_{N_\textrm{\tiny r}}
\Big)
\ee
written in terms of the dispersion relations
\be
\label{dispersion relation general}
\Omega^{\textrm{\tiny (s)}}_k=\sqrt{\omega_\textrm{\tiny s}^2+\frac{4\kappa_\textrm{\tiny s}}{m_\textrm{\tiny s}}
\bigg[ \sin\bigg(\frac{\pi k}{2(N_\textrm{\tiny s}+1)}\bigg)\bigg]^2}
\;\;\qquad\;\;
1 \leqslant k \leqslant N_\textrm{\tiny s}
\;\;\qquad\;\;
\textrm{s} \in \big\{ \textrm{l} , \textrm{r} \big\}\,.
\ee

The matrix $V_0$ in (\ref{initial CM general}) is block diagonal too and it can be written as 
\be
\label{Vtilde0 mat}
V_0 \,\equiv\,\widetilde{V}_0\oplus\widetilde{V}_0
\hspace{2cm}
\widetilde{V}_0=\widetilde{V}_{N_\textrm{\tiny l}}\oplus \widetilde{V}_{N_\textrm{\tiny r}}
\ee
where the elements of the $\widetilde{V}_{N_\textrm{\tiny l}}$ 
and $\widetilde{V}_{N_\textrm{\tiny r}}$ are given by 
\be
\label{VtildeNover2 mat}
\big(\widetilde{V}_{N_\textrm{\tiny s}}\big)_{j,k}
\,=\,
\sqrt{\frac{2}{N_\textrm{\tiny s}+1}}\;
\sin\! \bigg(\frac{\pi j k}{N_\textrm{\tiny s}+1}\bigg)
\;\;\qquad\;\;
1 \leqslant j,k \leqslant N_\textrm{\tiny s}
\;\;\qquad\;\;
\textrm{s} \in \big\{ \textrm{l} , \textrm{r} \big\}\,.
\ee

Notice that, since the matrices $\widetilde{V}_{N_\textrm{\tiny s}}$ defined by (\ref{VtildeNover2 mat}) are orthogonal, 
$\widetilde{V}_0$ is symplectic and orthogonal. 
Thus, also $V_0$  in (\ref{Vtilde0 mat}) is symplectic and orthogonal.

We find it worth remarking that,
in the expression (\ref{initial CM general}) for $\gamma^{\textrm{\tiny (l,r)}}_0$,
the parameters $m_\textrm{\tiny l}$, $\omega_\textrm{\tiny l}$, $\kappa_\textrm{\tiny l}$, 
$m_\textrm{\tiny r}$, $\omega_\textrm{\tiny r}$ and $\kappa_\textrm{\tiny r}$ 
occur in  $\mathcal{Q}_0$ and $\mathcal{P}_0$, 
while $V_0$ depends only on $N_\textrm{\tiny l}$ and $N_\textrm{\tiny r}$.

The covariance matrix (\ref{CM-def}) of the pure state (\ref{EvolvedState general}) 
at any $t>0$ after the quench 
is written in terms of the covariance matrix of the initial state (\ref{initial CM general})
as \cite{Eisler_2014}
\be
\label{CM-local-evolved general}
\gamma^{\textrm{\tiny (l,r)}}(t) 
\,=\,
 E(t) \,\gamma_0^{\textrm{\tiny (l,r)}} \, E(t)^{\textrm t}
\ee
where the time dependence occurs only through the matrix $E(t)$.

A convenient block decomposition of (\ref{CM-local-evolved general}),
which is employed in Sec.\,\ref{subsec-subcomp-gen}, reads
\be
\label{CM block decomposed}
\gamma^{\textrm{\tiny (l,r)}}(t) 
=\,
\bigg( 
\begin{array}{cc}
Q(t)  \,& M(t) \\
M(t)^{\textrm{t}}  \,& P(t)  \\
\end{array}   \bigg)
\ee
where $Q(t)$, $P(t)$ and $M(t)$ are the $N \times N$ correlation matrices 
whose elements are 
$Q(t)_{i,j}=\langle \psi_0 |\, \hat{q}_i(t)  \,\hat{q}_j(t) \, | \psi_0 \rangle $,
$P(t)_{i,j}=\langle \psi_0 |\, \hat{p}_i(t)  \,\hat{p}_j(t) \, | \psi_0 \rangle $
and 
$M(t)_{i,j}=\textrm{Re}\big[\langle \psi_0 |\, \hat{q}_i(t)  \,\hat{p}_j(t) \, | \psi_0 \rangle \big]$
respectively.
All these three matrices provide a non trivial temporal dependence.

An insightful decomposition for the matrix $E(t)$ in (\ref{CM-local-evolved general}) is \cite{Eisler_2014}
\be
\label{mat E def}
E(t)\,=\,V^{\textrm{t}} \,\mathcal{E}(t) \, V 
\;\;\qquad\;\;
V=\widetilde{V}_{N}\oplus\widetilde{V}_{N}
\ee
in terms of the matrix $\widetilde{V}_{N}$,
whose generic element  is given by (\ref{VtildeNover2 mat}) with $N_\textrm{\tiny s}$ replaced by $N$,
and of the matrix $\mathcal{E}(t)$,
whose block decomposition reads
\be
\label{diagmat E}
\mathcal{E}(t)
\equiv
\bigg( 
\begin{array}{cc}
\mathcal{D}(t)  \,& \mathcal{A}(t)
\\
 \mathcal{B}(t)  \,&  \mathcal{D}(t)
\end{array}\bigg)
\ee
with $\mathcal{D}$, $\mathcal{A}$ and $\mathcal{B}$ being diagonal matrices whose elements are respectively
\be
\label{blocks diagonal matrices}
\mathcal{D}_k(t) \,\equiv\, \cos\!\big(\Omega_k t\big)
\;\qquad\;
\mathcal{A}_k(t) \,\equiv\, \frac{\sin\!\big(\Omega_k t\big)}{m\,\Omega_k}
\;\qquad\;
\mathcal{B}_k(t) \,\equiv\, -\, m\,\Omega_k \sin\!\big(\Omega_k t\big)
\ee
where $\Omega_k$  is the dispersion relation given by
\be 
\label{dispersion relation evolution}
\Omega_k
=
\sqrt{\omega^2+\frac{4\kappa}{m}
\bigg[ \sin\bigg(\frac{\pi k}{2(N+1)}\bigg)\bigg]^2}
\;\;\qquad\;\;
1 \leqslant k \leqslant N\,.
\ee
Since $\widetilde{V}_{N}$ is orthogonal, $V$ is symplectic and orthogonal.
This observation and the fact that $\mathcal{E}(t=0)=\boldsymbol{1}$ lead to $E(t=0)=\boldsymbol{1}$;
hence, from (\ref{CM-local-evolved general}), we have that
$\gamma^{\textrm{\tiny (l,r)}}(t=0)  = \gamma_0^{\textrm{\tiny (l,r)}}$, 
as expected.
Notice that,
by using (\ref{blocks diagonal matrices}), one finds that $\mathcal{E}(t)$ in (\ref{diagmat E}) is symplectic;
hence, since $V$ is symplectic too, we conclude that $E(t)$ in (\ref{mat E def}) is symplectic.
Thus, $E(t)$ implements on the initial covariance matrix the unitary transformation on the initial state 
given in (\ref{EvolvedState general}).

In order to investigate the circuit complexity, 
we find it worth employing also the Williamson's decompositions \cite{Williamson36,Weedbrook12b}
of the covariance matrices
of the reference and of the target states. 

The Williamson's decomposition of the initial covariance matrix $\gamma_0^{\textrm{\tiny (l,r)}}$ in (\ref{initial CM general}) reads
\be
\label{Williamson initial}
\gamma_0^{\textrm{\tiny (l,r)}}
=
\frac{1}{2}\, W^{\textrm t}_0 \,W_0
\;\;\qquad\;\;
W_0\equiv
\mathcal{X}_0\,V_0
\;\;\qquad\;\;
\mathcal{X}_0 \,\equiv\,
\mathcal{T}_0 ^{\,-1/2} \oplus  \mathcal{T}_0 ^{1/2} 
\ee
where the symplectic matrix $V_0$ has been defined in (\ref{Vtilde0 mat}).
Since the initial state (\ref{InitialState general}) is pure,
all the symplectic eigenvalues of its covariance matrix $\gamma_0^{\textrm{\tiny (l,r)}}$ 
are identical and equal to $1/2$.
Notice that $\mathcal{X}_0^2=\Gamma_0$, where $\Gamma_0$ has been introduced in (\ref{initial CM general}).

Plugging the Williamson's decomposition (\ref{Williamson initial}) into (\ref{CM-local-evolved general}),
it is straightforward to obtain the Williamson's decomposition of the covariance matrix 
$\gamma^{\textrm{\tiny (l,r)}}(t)$ at any $t>0$,
which characterises the pure state (\ref{EvolvedState general}). 
It reads
\be
\label{Williamson gamma t}
\gamma^{\textrm{\tiny (l,r)}}(t)
\,=\,
\frac{1}{2}\, W(t)^{\textrm t} \,W(t)
\,\,\qquad\,\,
W(t) \equiv W_0\,E(t)^{\textrm{t}}
\ee
where the matrices $E(t)$ and $W_0$ have been defined in (\ref{mat E def}) and (\ref{Williamson initial}) respectively. 
Since both $W_0$ and $E(t)$ are symplectic matrices, the matrix $W(t)$ is symplectic too.

\section{Complexity for the harmonic chain}
\label{sec:PureStateComp}

In this section we discuss the temporal evolution of the complexity 
for the entire chain after the local quench defined in Sec.\,\ref{subsec:setup};
hence both the reference and the target states are pure.
We focus on the simplified setup where the quantum circuits 
are made only by bosonic Gaussian states with vanishing first moments.

\subsection{Optimal circuit and complexity}
\label{subsec:PureStateCompTheory}

The reference and the target states are fully characterised by their covariance matrices, 
which are $\gamma_{\textrm{\tiny R}}$ and $\gamma_{\textrm{\tiny T}}$ respectively. 
The circuit complexity obtained from the Fisher-Rao distance between 
$\gamma_{\textrm{\tiny R}}$ and $\gamma_{\textrm{\tiny T}}$ reads
\cite{Atkinson81,Bhatia07book}
\be
\label{c2 complexity}
\mathcal{C}
\,\equiv\,
\frac{1}{2\sqrt{2}}\;
\sqrt{\,
\textrm{Tr}\, \Big\{ \big[ \log \big( \gamma_{\textrm{\tiny T}} \,\gamma_{\textrm{\tiny R}}^{-1}  \big) \big]^2 \Big\}
}\,.
\ee
This complexity, which corresponds to the $F_2$ cost function,
has been studied for both pure states \cite{Jefferson:2017sdb,Chapman:2018hou}
and mixed states \cite{DiGiulio:2020hlz}.

The optimal circuit that allows to construct $\gamma_{\textrm{\tiny T}}$ from $\gamma_{\textrm{\tiny R}}$ 
is made by the following sequence of covariance matrices
\cite{Bhatia07book}
\be
\label{optimal circuit}
G_s(\gamma_{\textrm{\tiny R}} \, , \gamma_{\textrm{\tiny T}})
\,\equiv \,
\gamma_{\textrm{\tiny R}}^{1/2} 
\Big(  \gamma_{\textrm{\tiny R}}^{- 1/2}  \,\gamma_{\textrm{\tiny T}} \,\gamma_{\textrm{\tiny R}}^{-1/2}  \Big)^s
\gamma_{\textrm{\tiny R}}^{1/2} 
\;\; \qquad \;\;
0 \leqslant s \leqslant 1
\ee
which satisfies $G_0(\gamma_{\textrm{\tiny R}} \, , \gamma_{\textrm{\tiny T}}) = \gamma_{\textrm{\tiny R}}$ 
and $G_1(\gamma_{\textrm{\tiny R}} \, , \gamma_{\textrm{\tiny T}}) = \gamma_{\textrm{\tiny T}}$ .

Denoting by $t_{\textrm{\tiny R}}$ and $t_{\textrm{\tiny T}}$ the values of time $t$
corresponding to the reference and to the target states respectively, 
for their covariance matrices we have 
\be
\label{gamma-RT-time-def}
\gamma_{\textrm{\tiny R}} = \gamma^{\textrm{\tiny (l,r)}}(t_{\textrm{\tiny R}})
\;\;\qquad\;\;
\gamma_{\textrm{\tiny T}} = \gamma^{\textrm{\tiny (l,r)}}(t_{\textrm{\tiny T}})\,.
\ee
In the most general setup, these matrices depend on the sets of parameters given by 
$\mathcal{Y}_{\textrm{\tiny S}}\equiv \{m_{\textrm{\tiny l,S}}, \kappa_{\textrm{\tiny l,S}}, \omega_{\textrm{\tiny l,S}},m_{\textrm{\tiny r,S}}, \kappa_{\textrm{\tiny r,S}}, \omega_{\textrm{\tiny r,S}},m_{\textrm{\tiny S}}, \kappa_{\textrm{\tiny S}}, \omega_{\textrm{\tiny S}}\}$,
with $\textrm{S} = \textrm{R}$ and $\textrm{S} = \textrm{T}$ 
for the reference and the target state respectively. 
The corresponding states can be interpreted as the states obtained through 
the time evolutions at $t=t_\textrm{\tiny R} \geqslant 0$ and $t=t_\textrm{\tiny T} \geqslant t_\textrm{\tiny R}$
respectively, through two different quenches determined by the parameters
$\mathcal{Y}_{\textrm{\tiny R}}$ and $\mathcal{Y}_{\textrm{\tiny T}}$ respectively, 
 as described in Sec.\,\ref{subsec:setup}.

The circuit complexity (\ref{c2 complexity}) can be evaluated by finding  
the eigenvalues of $\gamma_{\textrm{\tiny T}} \,\gamma_{\textrm{\tiny R}}^{-1}$.

From (\ref{initial CM general}), (\ref{CM-local-evolved general}) and (\ref{mat E def}) 
for the reference and the target states (where $V$ is orthogonal)
with $\mathcal{E}_\textrm{\tiny R} \equiv \mathcal{E}(t_\textrm{\tiny R})$
and $\mathcal{E}_\textrm{\tiny T} \equiv \mathcal{E}(t_\textrm{\tiny T})$,
we find
\bea
\label{Delta TR general}
\gamma_{\textrm{\tiny T}} \,\gamma_{\textrm{\tiny R}}^{-1}
&=&
\big( V^{\textrm{t}} \, \mathcal{E}_\textrm{\tiny T} \,V \big) 
\big(V_0^{\textrm{t}} \, \Gamma_{0,\textrm{\tiny T}} \, V_0\big)
\big(V^{\textrm{t}} \, \mathcal{E}^{\textrm{t}}_\textrm{\tiny T} \,V\big)
\big(V^{\textrm{t}} \, \mathcal{E}^{-\textrm{t}}_\textrm{\tiny R} \,V\big)
\big(V_0^{\textrm{t}} \, \Gamma^{-1}_{0,\textrm{\tiny R}} \, V_0\big)
\big(V^{\textrm{t}} \, \mathcal{E}^{-1}_\textrm{\tiny R} \, V\big)
\nonumber
\\
&=&
V^{\textrm{t}} \, \mathcal{E}_\textrm{\tiny T} \,V \,
V_0^{\textrm{t}} \, \Gamma_{0,\textrm{\tiny T}} \, V_0 \,
V^{\textrm{t}} \, \mathcal{E}^{\textrm{t}}_\textrm{\tiny T} \,
\mathcal{E}^{-\textrm{t}}_\textrm{\tiny R} \, V \,
V_0^{\textrm{t}} \, \Gamma^{-1}_{0,\textrm{\tiny R}} \, V_0 \,
V^{\textrm{t}} \, \mathcal{E}^{-1}_\textrm{\tiny R} \, V\,.
\eea
This expression is difficult to deal with mainly because of  $V \,V_0^{\textrm{t}}$,
which encodes the spatial geometries before and after the local quench.

Notice that, when $t_\textrm{\tiny R}=t_\textrm{\tiny T}=0$, 
we have $\mathcal{E}_\textrm{\tiny R}=\mathcal{E}_\textrm{\tiny T}=\boldsymbol{1}$;
hence (\ref{Delta TR general}) simplifies to
\be
\label{Delta TR general tr tt zero}
\gamma_{\textrm{\tiny T}} \,\gamma_{\textrm{\tiny R}}^{-1}
=
V_0^{\textrm{t}}\, \Gamma_{0,\textrm{\tiny T}}\, \Gamma^{-1}_{0,\textrm{\tiny R}}\, V_0\,.
\ee
Since $V_0$ defined in (\ref{Vtilde0 mat}) is orthogonal 
and the matrices $\Gamma_{0,\textrm{\tiny T}}$ and $\Gamma_{0,\textrm{\tiny R}}$ in (\ref{Q0-P0 mat diag}) 
are diagonal, 
the eigenvalues of $\gamma_{\textrm{\tiny T}} \,\gamma_{\textrm{\tiny R}}^{-1}$ 
in (\ref{Delta TR general tr tt zero}) 
are the ratios of the entries of $\Gamma_{0,\textrm{\tiny T}}$ and $\Gamma_{0,\textrm{\tiny R}}$.
By employing this observation and (\ref{c2 complexity}), we obtain the following expression for the circuit complexity
\be
\label{comp Delta TR general}
\mathcal{C}
\,=\,
\frac{1}{2} \,\sqrt{
\,\sum_{k=1}^{N_\textrm{\tiny l}}
\Bigg\{ \!
\log\! \Bigg[
\frac{m_{\textrm{\tiny l,T}}\,\Omega^{\textrm{\tiny (l)}}_{\textrm{\tiny T},k}}{
m_{\textrm{\tiny l,R}}\,\Omega^{\textrm{\tiny (l)}}_{\textrm{\tiny R},k}}
\Bigg]
\Bigg\}^2
\! +\,
\sum_{k=1}^{N_\textrm{\tiny r}}
\Bigg\{ \!
\log\! \Bigg[
\frac{m_{\textrm{\tiny r,T}}\,\Omega^{\textrm{\tiny (r)}}_{\textrm{\tiny T},k}}{
m_{\textrm{\tiny r,R}}\,\Omega^{\textrm{\tiny (r)}}_{\textrm{\tiny R},k}}
\Bigg]
\Bigg\}^2\,
}
\ee
in terms of the dispersion relations (\ref{dispersion relation general}).
In the special case where $m_{\textrm{\tiny l,R}}=m_{\textrm{\tiny l,T}}$, $m_{\textrm{\tiny r,R}}=m_{\textrm{\tiny r,T}}$
and either $N_\textrm{\tiny l} =0$ or $N_\textrm{\tiny r}=0$,
the expression (\ref{comp Delta TR general}) becomes the result obtained in \cite{Jefferson:2017sdb}.

The above results can be employed to study the temporal evolution of the complexity after a global quench,
as already mentioned in Sec.\,\ref{subsec:setup}.
In Appendix\,\ref{subapp:massquench} we discuss the case of the quench of the mass parameter,
showing that the analysis of \cite{Eisler_2014}
allows to recover the correlators obtained in  \cite{Calabrese:2007rg},
which have been employed to evaluate the temporal evolutions
both of the complexity of the entire chain \cite{Ali:2018fcz,DiGiulio:2021oal}
and of the subsystem complexity \cite{DiGiulio:2021oal} after this kind of global quench.
Instead, in Appendix\,\ref{subapp:kappaquench} 
the temporal evolution of the complexity after the global quench of the spring constant is mainly considered,
with the initial state (which is also the reference state) 
given by the unentangled product state
(i.e. the ground state of the hamiltonian (\ref{HC ham-1d}) with a certain frequency and vanishing spring constant),
that has been adopted as the reference state in various studies about complexity 
\cite{Jefferson:2017sdb,Chapman:2018hou,Guo:2018kzl,Alves:2018qfv,Caceres:2019pgf}.

In this manuscript we are interested in 
circuits whose reference and the target states 
are pure states along the time evolution of a given local quench
at different times $t_\textrm{\tiny R}$ and $t_\textrm{\tiny T}$.
This can be done by choosing the parameters introduced in Sec.\,\ref{subsec:setup} as follows
\be
\label{parameters local quench}
\begin{array}{l}
m \equiv
m_{\textrm{\tiny l,R}}=m_{\textrm{\tiny l,T}}=m_{\textrm{\tiny r,R}}
=m_{\textrm{\tiny r,T}}=m_{\textrm{\tiny R}}=m_{\textrm{\tiny T}}
\\
\kappa
\equiv
\kappa_{\textrm{\tiny l,R}}=\kappa_{\textrm{\tiny l,T}}
=\kappa_{\textrm{\tiny r,R}}=\kappa_{\textrm{\tiny r,T}}=\kappa_{\textrm{\tiny R}}=\kappa_{\textrm{\tiny T}}
\\
\omega
\equiv
\omega_{\textrm{\tiny l,R}}=\omega_{\textrm{\tiny l,T}}
=\omega_{\textrm{\tiny r,R}}=\omega_{\textrm{\tiny r,T}}=\omega_{\textrm{\tiny R}}=\omega_{\textrm{\tiny T}}\,.
\end{array}
\ee
From (\ref{Q0-P0 mat diag}), we have that 
 $\Gamma_{0,\textrm{\tiny T}}$ and $\Gamma_{0,\textrm{\tiny R}}$ do not depend on time
and the setting given by (\ref{parameters local quench}) leads to 
$\Gamma_{0,\textrm{\tiny T}}=\Gamma_{0,\textrm{\tiny R}}\equiv \Gamma_{0}$.
Thus, (\ref{Delta TR general}) simplifies to
\be
\label{Delta TR same quench}
\gamma_{\textrm{\tiny T}} \,\gamma_{\textrm{\tiny R}}^{-1}
\,=\,
V^{\textrm{t}} \, \mathcal{E}_\textrm{\tiny T} \,V \,
V_0^{\textrm{t}} \, \Gamma_{0} \, V_0 \,
V^{\textrm{t}} \, \mathcal{E}^{\textrm{t}}_\textrm{\tiny T} \,
\mathcal{E}^{-\textrm{t}}_\textrm{\tiny R} \, V \,
V_0^{\textrm{t}} \, \Gamma^{-1}_{0} \, V_0 \,
V^{\textrm{t}} \, \mathcal{E}^{-1}_\textrm{\tiny R} \, V\,.
\ee

In our analysis we mainly consider 
the initial state (\ref{InitialState general}) as the reference state. 
This choice corresponds to set $t_{\textrm{\tiny R}}=0$ in (\ref{gamma-RT-time-def}) 
and $\gamma_\textrm{\tiny R}=\gamma_0^{\textrm{\tiny (l,r)}}$ 
given by (\ref{initial CM general}).
In this case, 
from (\ref{diagmat E}) and (\ref{blocks diagonal matrices}),
one finds that $\mathcal{E}_\textrm{\tiny R}=\mathcal{E}(t=0)=\boldsymbol{1}$
and that (\ref{Delta TR same quench}) simplifies to
\be
\label{Delta TR initial state ref}
\gamma_{\textrm{\tiny T}} \,\gamma_{\textrm{\tiny R}}^{-1}
=
V^{\textrm{t}} \, \mathcal{E} \,V \,
V_0^{\textrm{t}} \,\Gamma_{0} \,V_0 \,
V^{\textrm{t}} \, \mathcal{E}^{\textrm{t}} \,V \,
V_0^{\textrm{t}} \, \Gamma^{-1}_{0} \,V_0
\ee
where the notation $\mathcal{E}_\textrm{\tiny T}=\mathcal{E} $ has been introduced.
Finding the eigenvalues of (\ref{Delta TR initial state ref}) analytically is complicated;
hence we study them numerically.

When both the reference and the target states are pure, 
the Williamson's decompositions of their covariance matrices (\ref{gamma-RT-time-def})
read respectively 
\be
\gamma_\textrm{\tiny R}=\frac{1}{2}\, W_\textrm{\tiny R}^{\textrm t} \,W_\textrm{\tiny R}
\;\;\qquad\;\;
\gamma_\textrm{\tiny T}=\frac{1}{2}\, W_\textrm{\tiny T}^{\textrm t} \,W_\textrm{\tiny T}
\ee
where $W_\textrm{\tiny R}$ and $W_\textrm{\tiny T}$ are symplectic matrices. 
By introducing the following symplectic matrix
\be
\label{W_TR-def}
W_\textrm{\tiny TR} \equiv W_\textrm{\tiny T}W_\textrm{\tiny R}^{-1}
\ee
it is straightforward to realise that 
the complexity (\ref{c2 complexity}) becomes \cite{Chapman:2018hou}
\be
\label{c2 complexity WTR}
\mathcal{C}
\,=\,
\frac{1}{2\sqrt{2}}\;
\sqrt{\,
\textrm{Tr}\, \Big\{ \big[ \log \big( W_\textrm{\tiny TR}^{\textrm{t}}W_\textrm{\tiny TR} \big) \big]^2 \Big\}
}\;.
\ee

In the case where the reference and the target states are pure states along the time evolution of a given local quench 
and, furthermore, the initial state is chosen as the reference state, 
we have that $W_\textrm{\tiny R}=W_0$ and $W_\textrm{\tiny T}=W(t)$, 
where $W_0$ and $W(t)$ are defined in (\ref{Williamson initial}) and (\ref{Williamson gamma t}) respectively.
From the expressions of $W_0$ and $W(t)$
and the fact that $V_0$ in (\ref{initial CM general}) is orthogonal, for (\ref{W_TR-def}) one obtains 
\be
\label{WTR}
W_\textrm{\tiny TR}
=
W_0 \,E(t)^{\textrm{t}} \, W_0^{-1}
=
\mathcal{X}_0\,V_0 \,V^{\textrm{t}} \, \mathcal{E}(t)^{\textrm{t}}
\, V \,V_0^{\textrm{t}}\,\mathcal{X}_0^{-1}\,.
\ee
By using that 
$\mathcal{X}_0^2=\Gamma_0$, where $\Gamma_0$ is given in (\ref{initial CM general}),
we find that
\be
\label{WTRtrans WTR}
W_\textrm{\tiny TR}^{\textrm{t}}W_\textrm{\tiny TR}
\,=\,
\mathcal{X}_0^{-1}\,V_0 \,V^{\textrm{t}}\,\mathcal{E}(t)\,
V \,V_0^{\textrm{t}}\,\Gamma_0\,V_0 V^{\textrm{t}}\,
\mathcal{E}(t)^{\textrm{t}} \, V \,V_0^{\textrm{t}}\,\mathcal{X}_0^{-1}\,.
\ee
The diagonalisation of this matrix is as difficult as the one of (\ref{Delta TR initial state ref}).
However, this form could be helpful in future attempts to obtain analytic results for the complexity (\ref{c2 complexity WTR}).

The  Euler decomposition (also known as Bloch-Messiah decomposition) of a symplectic matrix $S$ reads
\cite{Arvind:1995ab} 
\be
\label{EulerDec}
S=L \, \mathcal{X}  \, R 
\;\;\qquad\;\; 
\mathcal{X}=e^\Lambda \oplus e^{-\Lambda}
 \;\;\qquad\;\; 
 L,R \in K(N)\equiv \textrm{Sp}(2N, \mathbb{R}) \cap O(2N)
\ee
where the diagonal matrix $ \Lambda=\textrm{diag}(\Lambda_1,\dots,\Lambda_N)$
contains the squeezing parameters $\Lambda_j \geqslant 0$.

From (\ref{c2 complexity WTR}), it is straightforward to realise that
the complexity of circuits made by pure states 
can be written in terms of the squeezing parameters $(\Lambda_{\textrm{\tiny TR}})_j$ 
corresponding to the symplectic matrix $W_{\textrm{\tiny TR}}$ as follows \cite{Chapman:2018hou}
\be
\label{complexity-squeezing}
\mathcal{C}=\sqrt{\sum_{j=1}^{N}\big(\Lambda_{\textrm{\tiny TR}}\big)^2_j}\,.
\ee

The symplectic matrix $\mathcal{E}(t)^{\textrm{t}}$ 
can be decomposed into four $N\times N$ blocks which are diagonal matrices 
(see (\ref{diagmat E})); hence we can find its Euler decomposition
$\mathcal{E}(t)^{\textrm{t}}=L_\mathcal{E} \, \mathcal{X}_\mathcal{E}  \, R_\mathcal{E}$
(where all the three matrices can depend on $t$)
by following the procedure discussed in Appendix \ref{app:EulerDecomposition}.
Plugging this decomposition into (\ref{WTR}), one obtains
\be
\label{WTR v2}
W_\textrm{\tiny TR}
\,=\,
\mathcal{X}_0\,V_0 V^{\textrm{t}}L_\mathcal{E} \, \mathcal{X}_\mathcal{E}  \, 
R_\mathcal{E}\,V \,V_0^{\textrm{t}}\,\mathcal{X}_0^{-1}\,.
\ee
This expression does not provide  the Euler decomposition of $W_\textrm{\tiny TR}$ 
because of the occurrence of the diagonal matrix $\mathcal{X}_0$, which is not orthogonal.
By contradiction, if $\mathcal{X}_0$ were orthogonal,
(\ref{WTR v2}) would be the Euler decomposition of $W_\textrm{\tiny TR}$
with the squeezing parameters given by $\mathcal{X}_\mathcal{E}$
because $L_\mathcal{E}$, $R_\mathcal{E}$, $V$ and $V_0$ are symplectic and orthogonal matrices.
This would lead to a complexity (\ref{complexity-squeezing}) 
independent of the position of the joining point
because $\mathcal{X}_\mathcal{E}$ depends only on 
$\mathcal{E}$ in (\ref{diagmat E}), which is 
determined by the parameters characterising the evolution Hamiltonian. 
The numerical analysis performed in Sec.\,\ref{subsec:PureStateNumerics}
shows that this is not the case.

\subsection{Initial growth}
\label{subsec:Initialgrowth}

It is worth exploring the leading term of the initial growth 
of the temporal evolution of the complexity (\ref{c2 complexity}) for the entire chain
when the reference state is the initial state 
(i.e. $\gamma_{\textrm{\tiny R}}=\gamma_0^{\textrm{\tiny (l,r)}}$ in (\ref{initial CM general}))
and the target state is the state at time $t$ after the local quench that we are exploring
(i.e. $\gamma_{\textrm{\tiny T}}=\gamma^{\textrm{\tiny (l,r)}}(t)$ in (\ref{CM-local-evolved general})).

By expanding $\mathcal{E}(t)$ in (\ref{diagmat E}) as $t\to 0$, one finds
\be
\label{Ediag smalltimes}
\mathcal{E}(t)=
\bigg( 
\begin{array}{cc}
\boldsymbol{1}  \,& \tfrac{t}{m}\boldsymbol{1}
\\
t\,\mathcal{N}  \,&  \boldsymbol{1}
\end{array}\bigg)+O\big(t^2\big)
\;\;\qquad\;\;
\mathcal{N}
\equiv 
-\,m\; \textrm{diag}\Big(\Omega_1^2,\dots,\Omega_N^2\Big)
\ee
where $\Omega_k$ is given in (\ref{dispersion relation evolution}).
By employing (\ref{Ediag smalltimes}) in (\ref{mat E def})
and the fact that $\widetilde{V}_N$ is orthogonal, we obtain
\be
\label{Emat smalltimes}
E(t)=\boldsymbol{1}+t\,E_{(1)}
+O\big(t^2\big)
\;\;\qquad\;\;
E_{(1)}=
\bigg( 
\begin{array}{cc}
\boldsymbol{0}  \,& \tfrac{1}{m}\boldsymbol{1}
\\
\widetilde{V}_N^{\textrm{t}}\,\mathcal{N}\,\widetilde{V}_N  \,&  \boldsymbol{0}
\end{array}\bigg)
\ee
where the $N\times N$ matrix 
$\widetilde{V}_N^{\textrm{t}}\,\mathcal{N}\,\widetilde{V}_N$
is not diagonal. 
By using the expansion (\ref{Emat smalltimes}), 
for the covariance matrix (\ref{CM-local-evolved general}) we find
\be
\gamma^{\textrm{\tiny (l,r)}}(t)
\,=\,
\gamma_0^{\textrm{\tiny (l,r)}}
+
\Big(
(E_{(1)}\gamma_0^{\textrm{\tiny (l,r)}}
+\gamma_0^{\textrm{\tiny (l,r)}}E_{(1)}^{\,\textrm{t}}
\Big)
\,t
+O\big(t^2\big)
\ee
where  the $O(t)$ term is symmetric, as expected. 
This straightforwardly leads to 
\be
\label{DeltaTR smallt}
\gamma_{\textrm{\tiny T}}\,\gamma_{\textrm{\tiny R}}^{-1}=\boldsymbol{1}
+
\Big[\,
E_{(1)}+ \gamma_0^{\textrm{\tiny (l,r)}}E_{(1)}^{\,\textrm{t}}\big(\gamma_0^{\textrm{\tiny (l,r)}}\big)^{-1} \,\Big]
\; t
+O\big(t^2\big)\,.
\ee

This expansion provides the following linear growth for the complexity (\ref{c2 complexity}) 
\be
\label{initial growth}
\mathcal{C}
\,=\,
c_1\, t
+O\big(t^2\big)
\ee
where for the coefficient $c_1$ we find
(see the Appendix\,\ref{app:Initialgrowth} for its derivation)
\bea
\label{slope smallt final}
c_1^2
&=&
\frac{1}{4}\; \Bigg\{\,
\sum_{k=1}^{N_{\textrm{\tiny l}}} \big[ \Omega^{\textrm{\tiny (l)}}_k\big]^2
+
\sum_{k=1}^{N_{\textrm{\tiny r}}}\big[\Omega^{\textrm{\tiny (r)}}_k\big]^2
-
2\sum_{k=1}^{N}\Omega_k^2
\nonumber
\\
& & \hspace{1cm}
+ \,
\textrm{Tr}\Big[
\widetilde{V}_N^{\textrm{t}}\,\mathcal{N}\,\widetilde{V}_N\,
\widetilde{V}_0^{\textrm{t}}\,\mathcal{Q}_0\,\widetilde{V}_0\,
\widetilde{V}_N^{\textrm{t}}\,\mathcal{N}\,\widetilde{V}_N\,
\widetilde{V}_0^{\textrm{t}}\,\mathcal{P}_0^{-1}\,\widetilde{V}_0
\Big]
\Bigg\}\,.
\eea
Simplifying further the last term in this expression is complicated, hence we evaluate it numerically,
as done in the bottom panel of Fig.\,\ref{fig:PureStateMassiveLargetimes} to determine the dashed straight line. 

As a consistency check for (\ref{slope smallt final}),
let us consider the trivial case where the quench does not occur, 
which corresponds to set $N_{\textrm{\tiny l}}=N$ and $N_{\textrm{\tiny r}}=0$ (or viceversa),
implying that $\widetilde{V}_0=\widetilde{V}_N$.
By using that $\widetilde{V}_N$ is orthogonal, (\ref{Q0-P0 mat diag}) and (\ref{T0 mat diag}),
one finds that the last term in (\ref{slope smallt final}) 
simplifies to $\sum_{k=1}^{N}\Omega_k^2$.
Then, since $N_{\textrm{\tiny r}}=0$, the second sum in (\ref{slope smallt final}) does not occur
and therefore $c_1 = 0$,
as expected, consistently with the fact that the initial state does not evolve.

\subsection{Numerical results}
\label{subsec:PureStateNumerics}

\begin{figure}[t!]
\vspace{-.5cm}
\subfigure
{\hspace{-1.25cm}
\includegraphics[width=.57\textwidth]{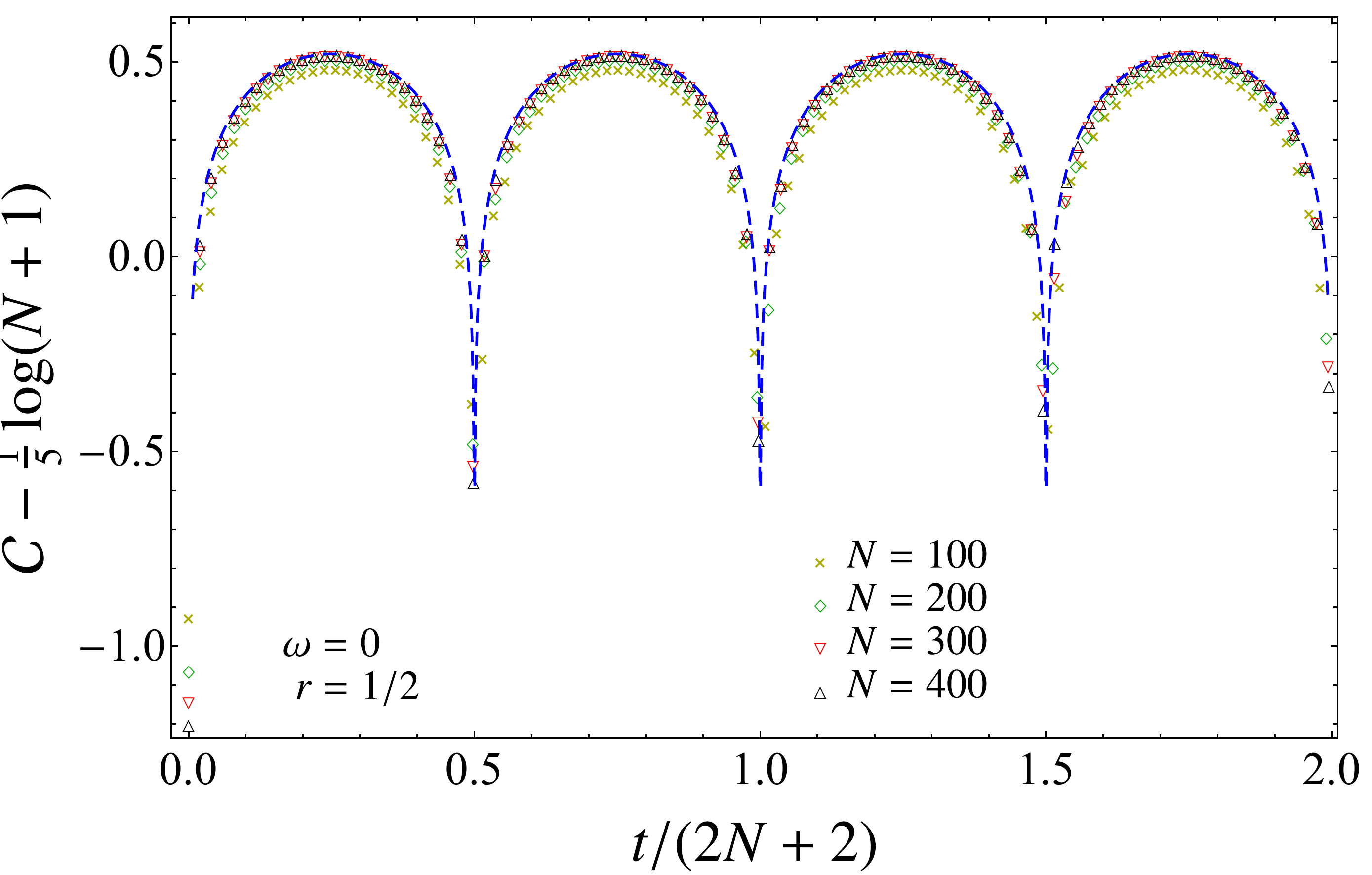}}
\subfigure
{
\hspace{-0.05cm}\includegraphics[width=.57\textwidth]{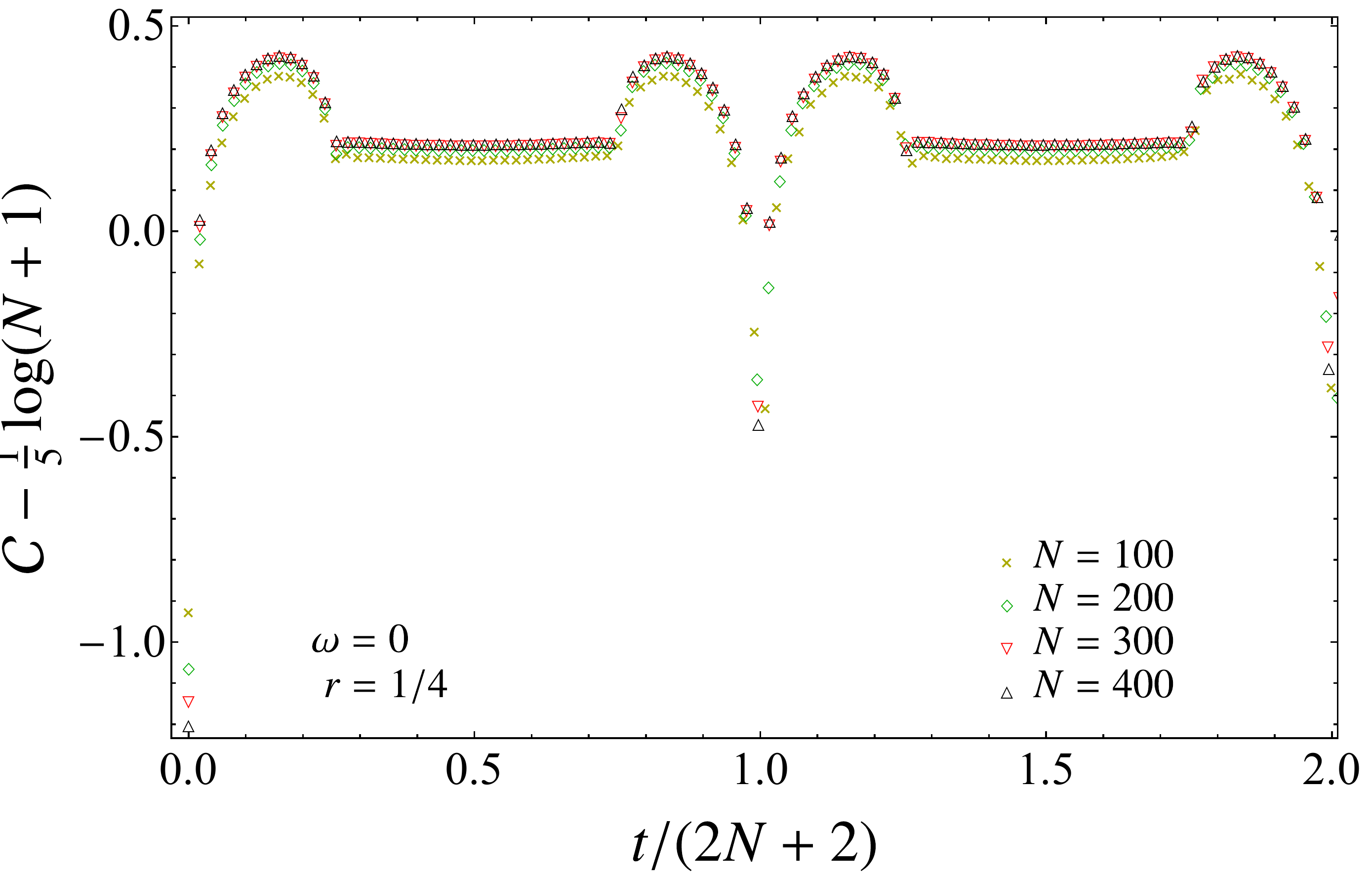}}
\subfigure
{
\hspace{-1.25cm}\includegraphics[width=.57\textwidth]{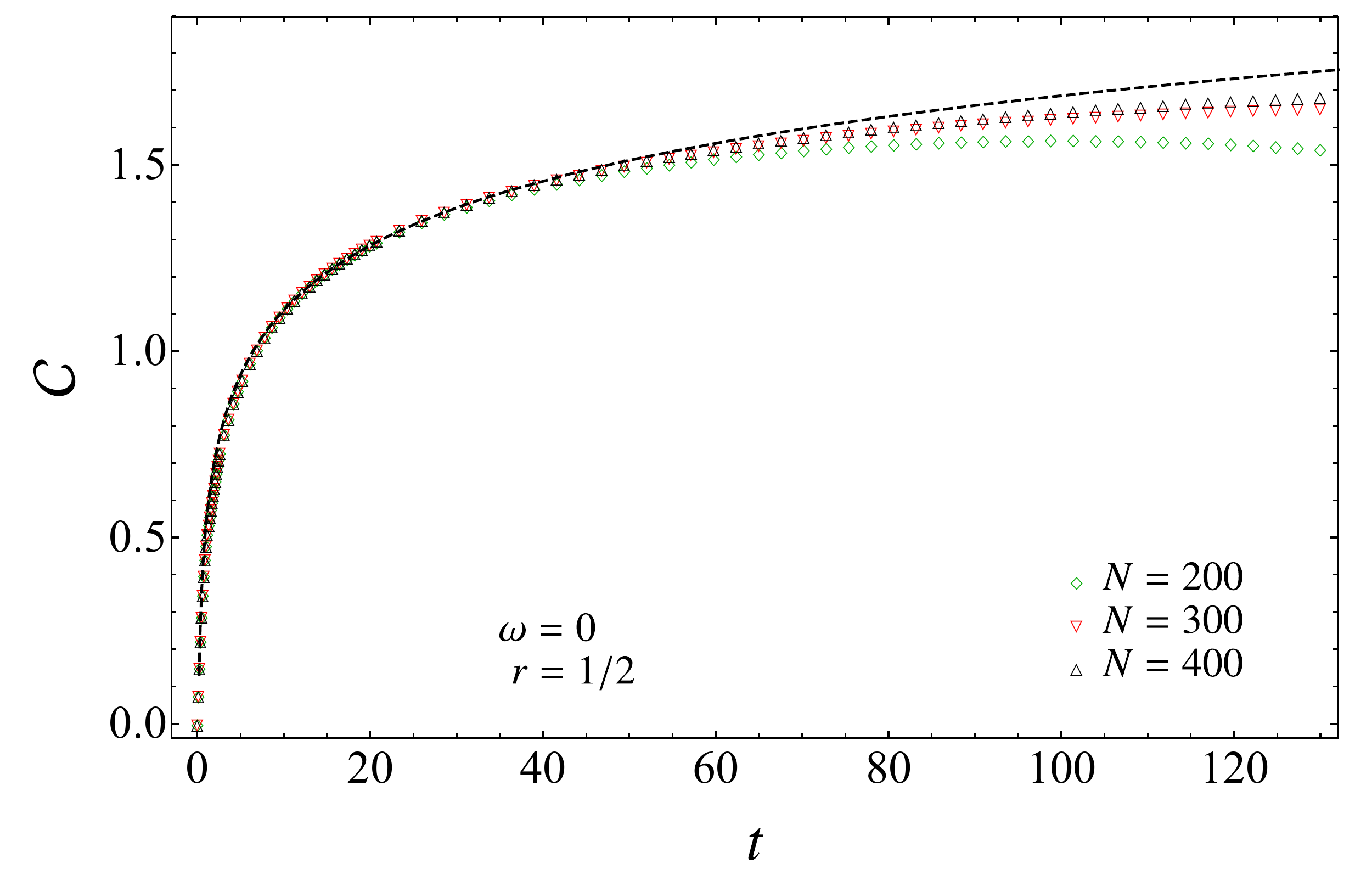}}
\subfigure
{\hspace{-0.05cm}
\includegraphics[width=.57\textwidth]{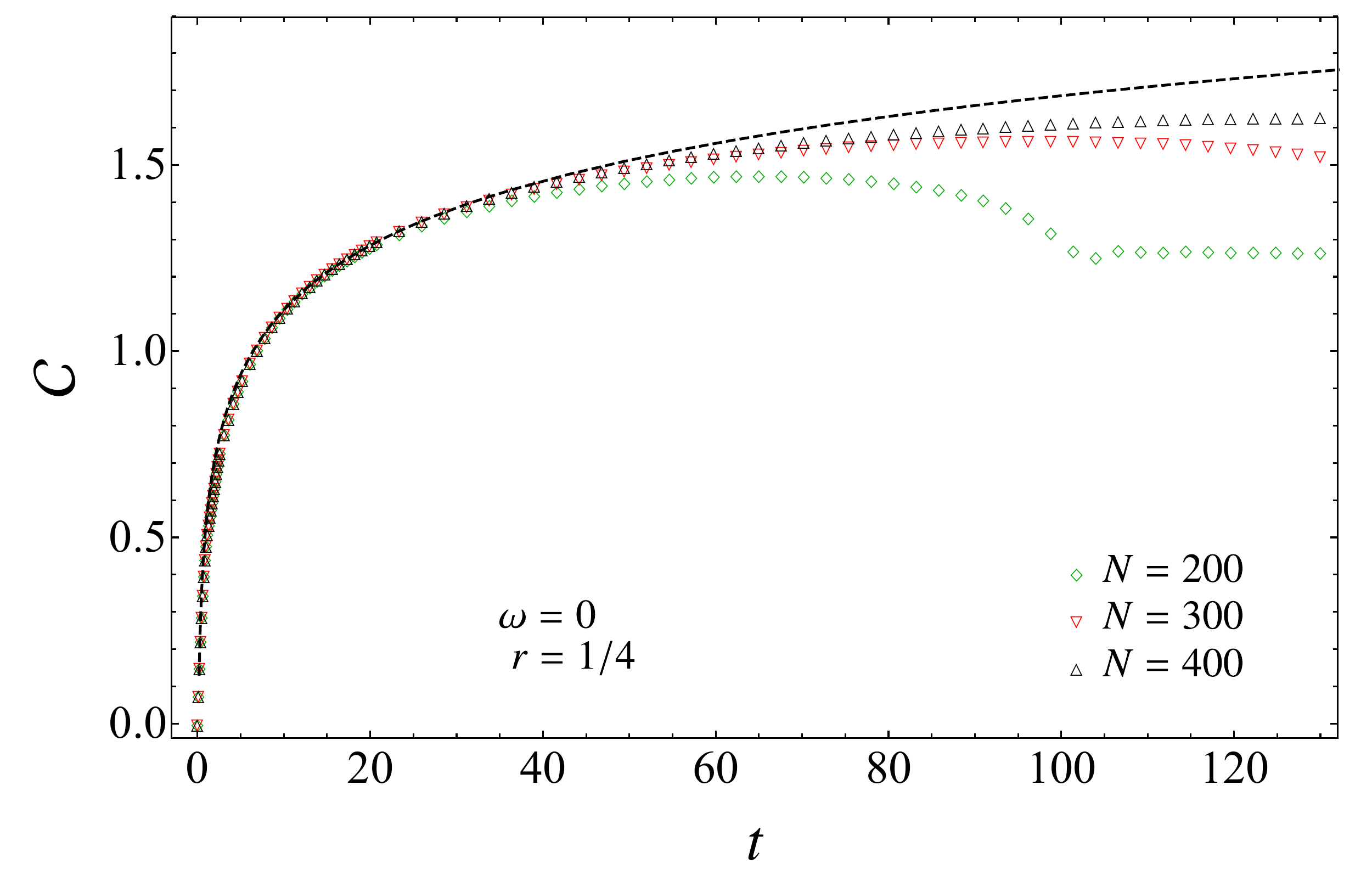}}
\caption{Temporal evolution of the complexity $\mathcal{C}$ 
in (\ref{c2 complexity})
for the entire chain (made by $N$ sites)
after a local quench with $\omega  = 0$ w.r.t. the initial state at $t=0$.
Here either $r=1/2$ (left panels) or $r=1/4$ (right panels). 
The dashed curve in the top left panel corresponds to (\ref{guess complexity pure r1over2}), 
while the ones in the bottom panels are obtained from (\ref{guess complexity initial}).
}
\vspace{0.4cm}
\label{fig:PureStateCritical}
\end{figure}

In this section we discuss some temporal evolutions of the complexity  
for the entire chain after a local quench where two chains are joined (see Sec.\,\ref{subsec:setup}),
evaluated numerically through (\ref{c2 complexity}).
The reference and the target states are respectively 
the initial state ($t_\textrm{\tiny R}=0$) 
and the pure state corresponding to a generic value of $t_\textrm{\tiny T}\equiv t\geqslant 0$ 
along the evolution after the quench. 
The parameters of this quench protocol are set as in (\ref{parameters local quench}).
The data points reported in all the figures shown in the main text
have been obtained for $m=1$ and $\kappa=1$.

\begin{figure}[t!]
\vspace{-.5cm}
{
\subfigure
{\hspace{-1.05cm}
\includegraphics[width=1.03\textwidth]{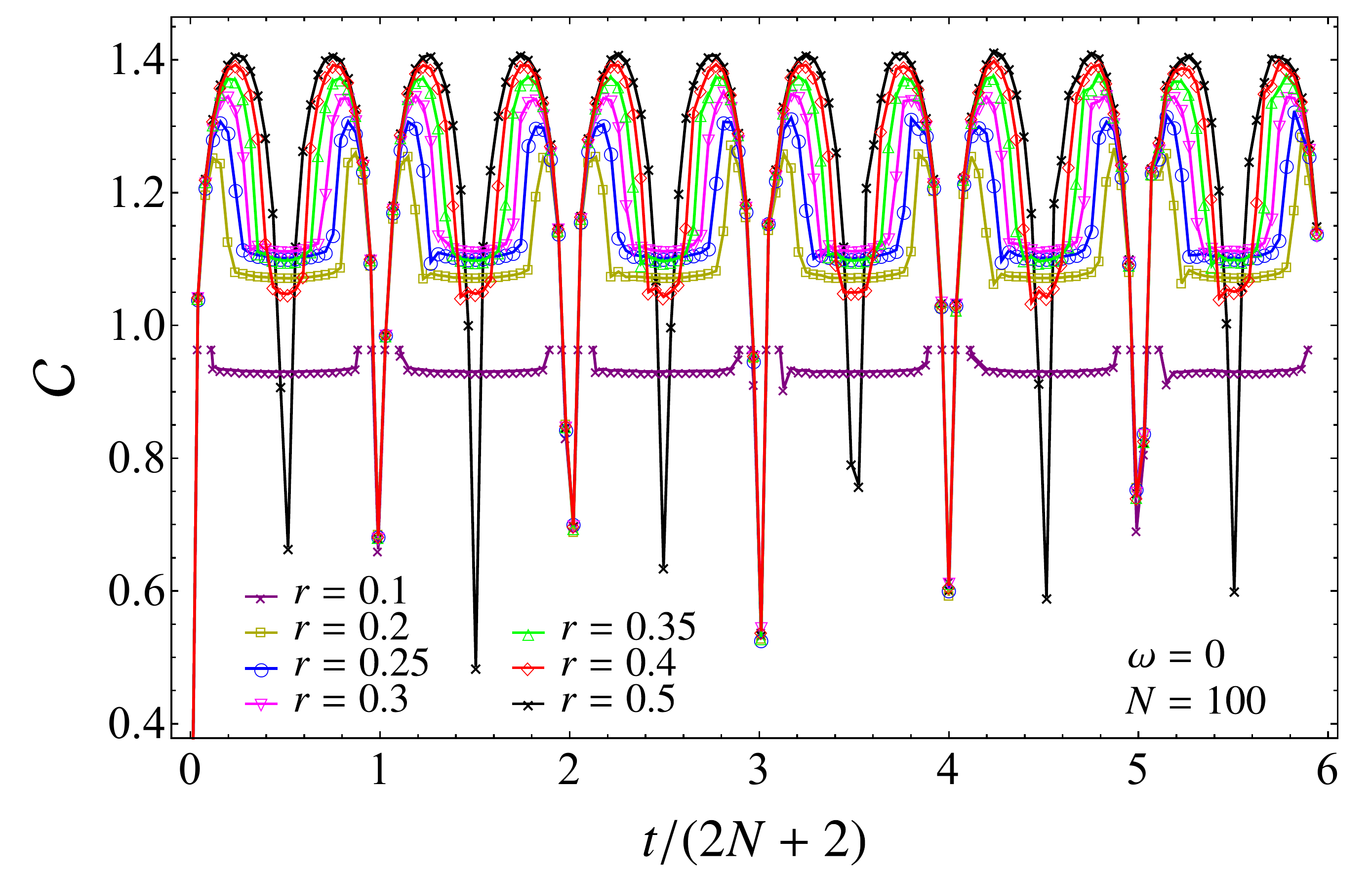}}
}
\caption{
Temporal evolution of the complexity $\mathcal{C}$ in (\ref{c2 complexity})
for the chain with $N=100$
after a local quench with $\omega  = 0$ w.r.t. the initial state,
for various positions of the joining point
(see the top panel of Fig.\,\ref{fig:intro-configs}).
}
\vspace{0.4cm}
\label{fig:PureStateCriticalLargeTimes}
\end{figure}

In Fig.\,\ref{fig:PureStateCritical} and Fig.\,\ref{fig:PureStateCriticalLargeTimes}, 
the temporal evolutions corresponding to critical Hamiltonians are considered; i.e. $\omega=0$.
Since the volume is kept finite, revivals are observed,
as already discussed for the temporal evolutions of other quantities \cite{Cardy:2014rqa}.
The different cycles correspond to $p<t/(2N+2)<p+1$, with $p$ being a non-negative integer.
This approximate periodic behaviour is observed also in the correlators providing the covariance matrix. 
For instance,  
in Fig.\,\ref{fig:PureStateCritical} the cycles corresponding to $p=0$ and $p=1$ are displayed.
\\
Within each cycle we can identify three temporal regimes:
(I) $p<t/(2N+2)<p+r$,
characterised by an initial growth, a local maximum and a subsequent decrease;
(II) $p+r<t/(2N+2)<p+1-r$,
where the evolution is almost stationary 
(a slight convexity of the curves is observed by zooming in);
(III) $p+1-r<t/(2N+2)<p+1$,
characterised by a growth until a local maximum is reached and a subsequent decrease.
The last regime is very similar the first one, after a time reversal;
indeed, the curve of $\mathcal{C}(t)$ within each cycle remains roughly invariant after
a reflection with respect to the value of $t$ corresponding to the center of the cycle.

In the special case of  $r=1/2$ 
(see the top left panel of Fig.\,\ref{fig:PureStateCritical} and the black symbols in Fig.\,\ref{fig:PureStateCriticalLargeTimes}), 
the second regime does not occur;
hence the cycles correspond to $p<t/(N+1)<p+1$, 
with $p$ being a non-negative integer.

In the top panels of Fig.\,\ref{fig:PureStateCritical},
the temporal evolutions of  $\mathcal{C} - \tfrac{1}{5}\log(N+1)$ 
are displayed for $r=1/2$ (left panel) and $r=1/4$ (right panel).
When $N$ is large enough, the data for different values of $N$ nicely collapse,
except for the beginning and the end of each cycle, as discussed below.
In the top left panel of Fig.\,\ref{fig:PureStateCritical}, where $r=1/2$, 
also the following curve is shown
\be
\label{guess complexity pure r1over2}
\mathcal{C}(t)
\,=\,
\frac{1}{5}\,\log\bigg\{\bigg(\frac{N+1}{\pi}\bigg) \, \bigg|\sin\!\bigg(\frac{\pi \,t}{N+1}\bigg)\bigg|\,\bigg\}
\,+ \,\textrm{const}
\ee
which  nicely agrees with the data points in the middle of each cycle $p<t/(N+1)<p+1$,
when $N$ is large enough.

In the bottom panels of Fig.\,\ref{fig:PureStateCritical}, 
we consider the temporal regime of initial growth for $\mathcal{C}$
subsequent to the early linear growth (\ref{initial growth}).
We find  that the data corresponding to different values of $N$ nicely collapse
on the curve given by 
\be
\label{guess complexity initial}
\mathcal{C} \,=\,\frac{1}{4}\log(t)  + \textrm{const}
\ee
with $\textrm{const} \simeq 0.5346 $ within a temporal regime whose width increases with $N$.
Notice that (\ref{guess complexity initial}) does not correspond to the leading term of (\ref{guess complexity pure r1over2})
when $t/(N+1) \to 0$ because the coefficients multiplying the logarithms are different. 
This is consistent with the fact that the data in the top panels of Fig.\,\ref{fig:PureStateCritical} do not collapse
at the beginning and at the end of each cycle. 
Taking $t/(2N+2)$ instead of $t$ as the independent variable 
in the bottom panels of Fig.\,\ref{fig:PureStateCritical},
the data collapse is observed for $\mathcal{C} -  \tfrac{1}{4} \log(N+1)$
and not for $\mathcal{C} -  \tfrac{1}{5} \log(N+1)$, which is plotted 
in the top panels of the same figure.
This is consistent with the data corresponding to the black symbols
in the right panels Figs.\,\ref{fig:CompMixedMasslessr1over2} and \ref{fig:CompMixedMasslessr1over4} 
and in the left panels of Fig.\,\ref{fig:Massless2endpoints},
which describe the complexity of the entire chain. 
Let us anticipate that also for the subsystem complexity $\mathcal{C}_A$
different temporal regimes occur
where the data points for $\mathcal{C}_A- \alpha \log(N+1)$
corresponding to increasing values of $N$ collapse,
with different values of $\alpha$ in the different regimes
(see Sec.\,\ref{sec-subsystem-comp}).
By comparing the two bottom panels in Fig.\,\ref{fig:PureStateCritical},
we observe that the initial growth of the complexity is independent of the value of $r$.

In Fig.\,\ref{fig:PureStateCriticalLargeTimes} we consider a longer range of $t$, 
in order to include more cycles and to highlight the fact that the approximate periodicity persists,
for various values of $r$.
Notice that the values of the local maximum within each cycle increases with $r$ until $r=1/2$.
Furthermore, the height of the plateaux characterising the second temporal regime within each cycle
grows with $r$ until certain value $r_\ast$ 
(from Fig.\,\ref{fig:PureStateCriticalLargeTimes}, we have $0.25<r_*<0.35$),
then it decreases. 
Instead, the duration of this plateaux is always decreasing for $0 < r \leqslant 1/2$ 
and vanishes at $r=1/2$.
The symmetry of the problem straightforwardly leads to realise that 
the temporal evolution of the complexity for a given $r$ is equal to the one corresponding to $1-r$,
for the same choice of all the other parameters.
We have obtained numerical data for the temporal evolutions of the complexity displayed in 
Fig.\,\ref{fig:PureStateCriticalLargeTimes} also for $N=200$,
finding that the data points of $\mathcal{C} - \tfrac{1}{5} \log(N+1)$ for $N=100$ and $N=200$
approximatively collapse (see also the top panels of Fig.\,\ref{fig:PureStateCritical}).
In Fig.\,\ref{fig:PureStateCriticalLargeTimes} we have reported only the 
numerical curves for $N=100$ in order to display in a clear way
the qualitative changes in the temporal evolutions corresponding to different $r$.

\begin{figure}[t!]
\vspace{-.5cm}
\subfigure
{
\hspace{-1.25cm}\includegraphics[width=.57\textwidth]{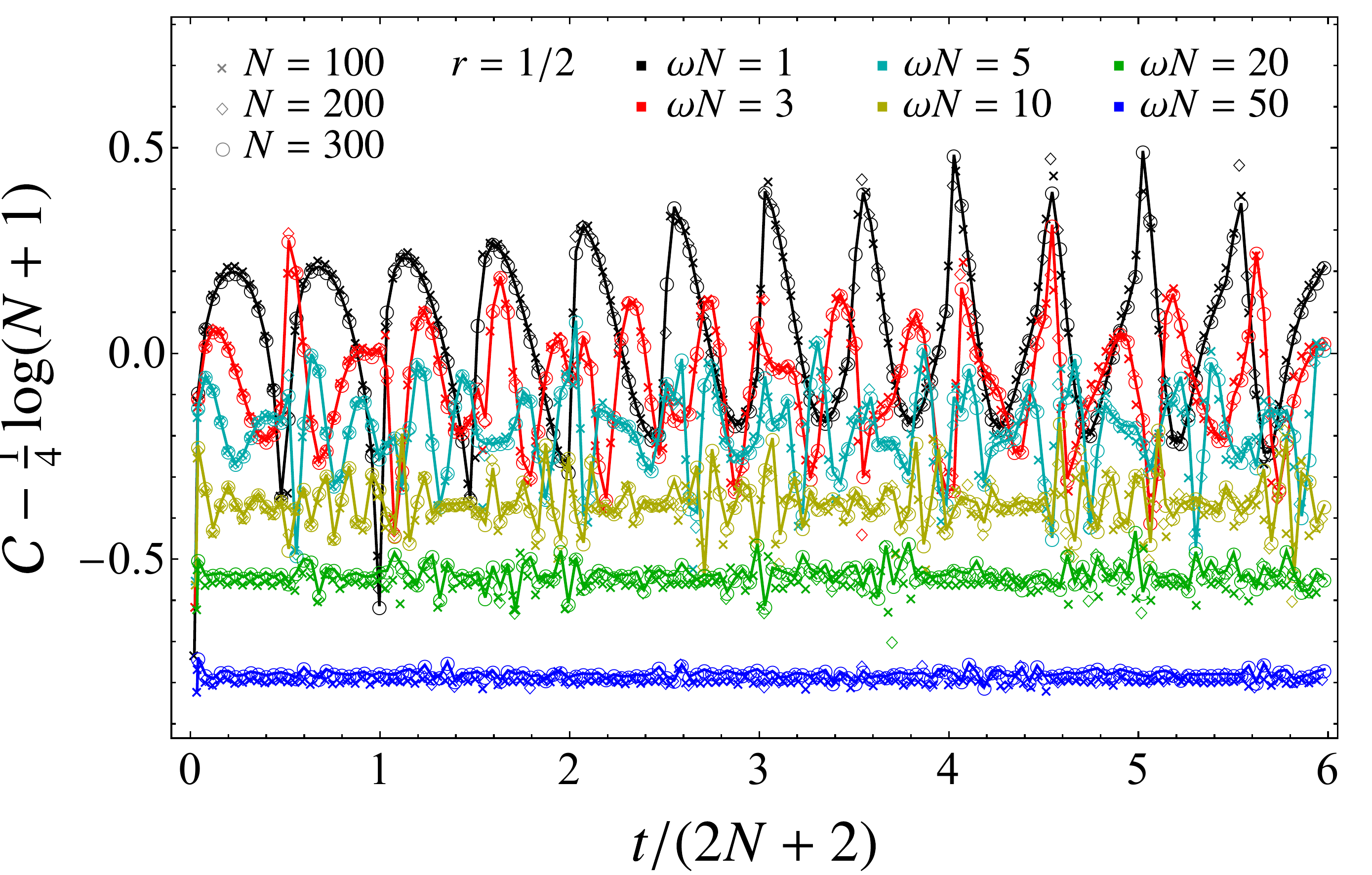}}
\subfigure
{
\hspace{-0.05cm}\includegraphics[width=.57\textwidth]{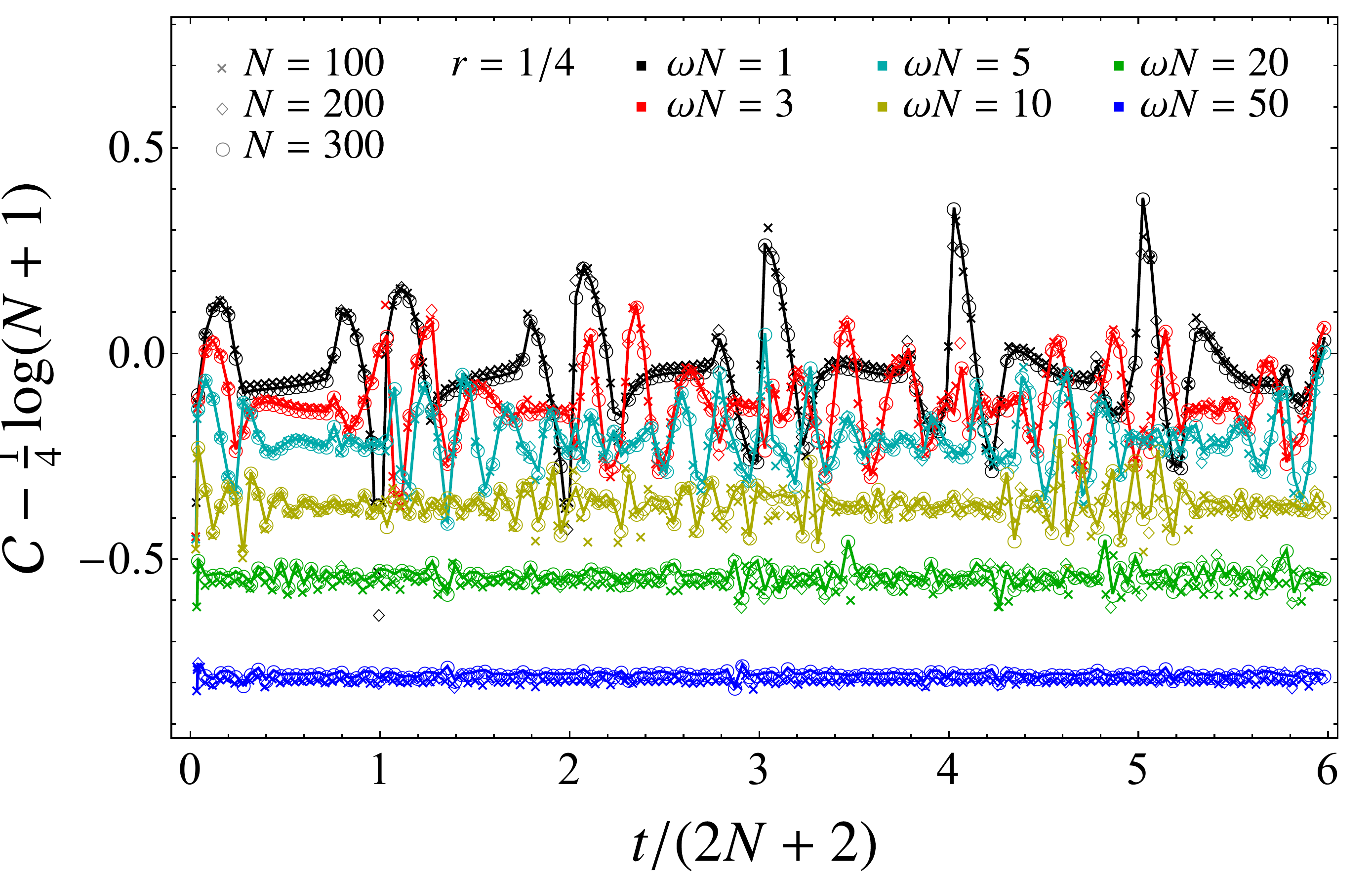}}
\subfigure
{
\hspace{2.9cm}\includegraphics[width=.57\textwidth]{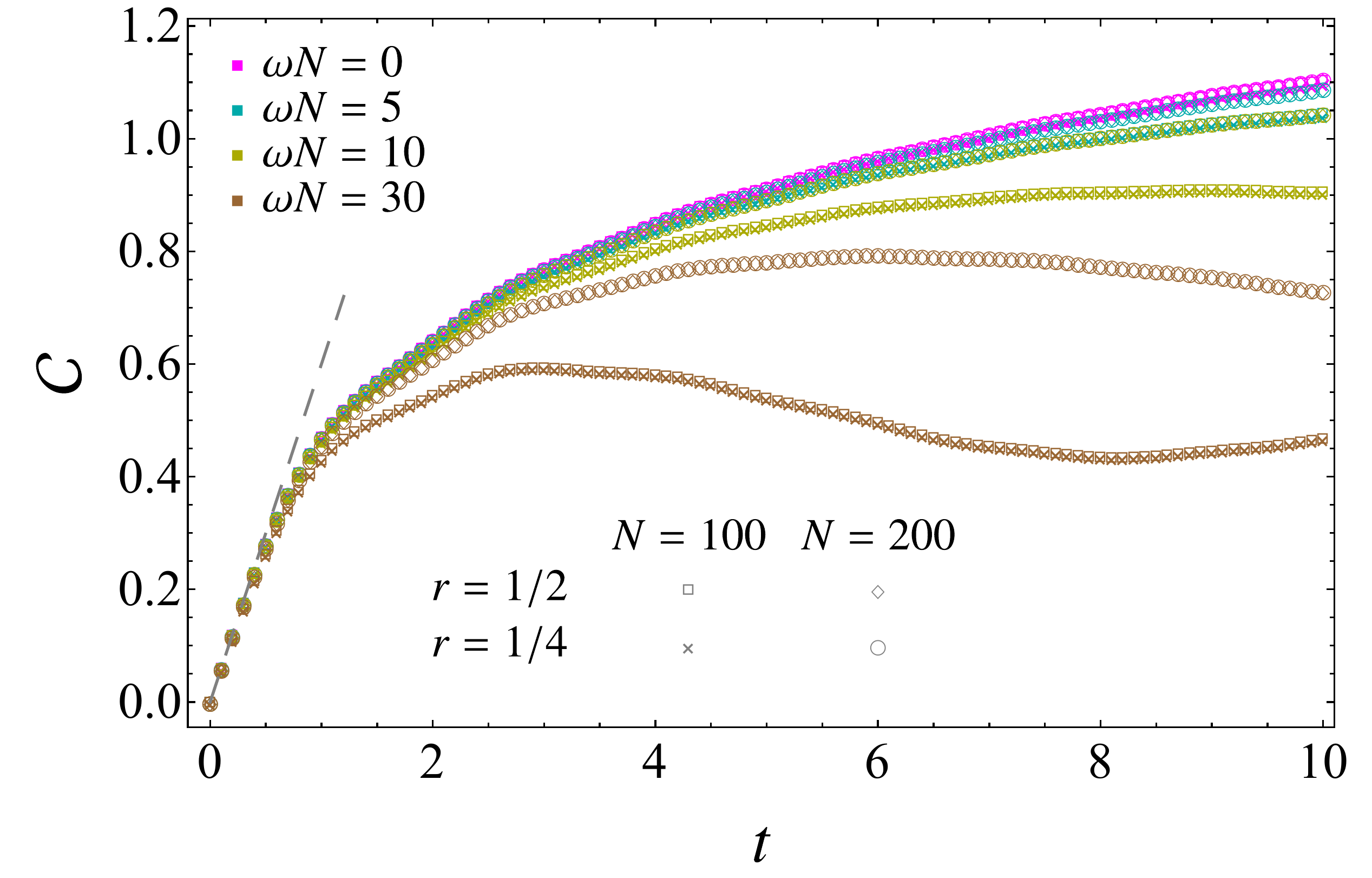}}
\caption{Temporal evolution of the complexity $\mathcal{C}$ in (\ref{c2 complexity})
for the entire chain 
after local quenches characterised by various $\omega > 0$ w.r.t. the initial state
(the data corresponding to $N=300$ have been joined through a piecewise line),
with either $r=1/2$ (top left panel) or $r=1/4$ (top right panel).
The initial growth is shown in the bottom panel, where 
the dashed grey line corresponds to (\ref{initial growth}) and (\ref{slope smallt final}) with $\omega=0$, $r=1/2$ and $N=200$. 
}
\vspace{0.4cm}
\label{fig:PureStateMassiveLargetimes}
\end{figure} 

Some temporal evolutions of the complexity determined by gapped Hamiltonians after the local quench 
are shown  in Fig.\,\ref{fig:PureStateMassiveLargetimes}, 
where the different coloured curves correspond to different values of $\omega N \leqslant 50$.

In the top panels of Fig.\,\ref{fig:PureStateMassiveLargetimes},
we show that the curves for $\mathcal{C} - \tfrac{1}{4}\log(N+1)$ corresponding to different values of $N$ collapse.
We remind that this collapse has been observed for $\mathcal{C} - \tfrac{1}{5}\log(N+1)$ when $\omega = 0$
(see the top panels of Fig.\,\ref{fig:PureStateCritical}).
It would be interesting to understand this numerical observation. 
 Furthermore, by comparing the temporal evolutions in the top panels of Fig.\,\ref{fig:PureStateMassiveLargetimes}
with the ones in Fig.\,\ref{fig:PureStateCriticalLargeTimes},
we notice that the periodicity highlighted for $\omega = 0$ does not occur when $\omega N >0$ in general.

When $\omega N\lesssim 1$, the initial part of the temporal evolution is similar 
to the one observed in the case of $\omega =0$ (see Fig.\,\ref{fig:PureStateCritical}),
as one realises from the curves corresponding to $\omega N =1$ in the top panels of Fig.\,\ref{fig:PureStateMassiveLargetimes}.
For large values of $\omega N \gtrsim 10$, 
the temporal evolution of  $\mathcal{C} - \tfrac{1}{4}\log(N+1)$ is roughly described by a complicated oscillation 
about a constant value.
This constant value decreases with $\omega N$ and, when $\omega N $ is large enough,
is independent of $r$.
Also the amplitude of the oscillations about this constant value decreases as $\omega N $ increases.

The bottom panel of Fig.\,\ref{fig:PureStateMassiveLargetimes} focuses on the initial growth of $\mathcal{C}$;
hence it is instructive to compare it against  the bottom panels of Fig.\,\ref{fig:PureStateCritical} where $\omega =0$.
In the temporal regime considered in the bottom panel of Fig.\,\ref{fig:PureStateMassiveLargetimes},
the curves corresponding to different values of $N$ nicely collapse. 
Furthermore, for small values of $\omega N$ a collapse is observed for  different values of $r$
(see also the bottom panels of Fig.\,\ref{fig:PureStateCritical}),
while they are clearly different for $\omega N =30$, after a certain time. 
For $t \lesssim 1/2$, all the numerical curves displayed in the panel collapse on the same 
approximate  dashed gray line, which has been obtained by setting 
$\omega=0$, $r=1/2$ and $N=200$ in (\ref{initial growth}) and (\ref{slope smallt final}).
Although the lines corresponding to the other values of $\omega$, $r$ and $N$ are different, 
they roughly overlap with the only one that we have displayed.

We find it worth mentioning some results about the temporal evolution of the complexity evaluated within the 
gauge/gravity correspondence. 

The temporal evolution of the holographic complexity in the Vaidya gravitational spacetimes,
which model the formation of a black hole through the collapse of a shell 
and have been exploited to study the gravitational duals of global quenches \cite{Hubeny:2007xt,AbajoArrastia:2010yt},
has been studied in 
\cite{Stanford:2014jda,Susskind:2014jwa,Brown:2015bva,Brown:2015lvg,Roberts:2014isa,Moosa:2017yvt,
Chapman:2018dem,Chapman:2018lsv}.
Qualitative comparisons between these results 
and the temporal evolution of the complexity in harmonic chains 
have been discussed in \cite{DiGiulio:2021oal,Camargo:2018eof,Alves:2018qfv}.

A gravitational background dual to the local quench obtained through the insertion of a local operator
\cite{Nozaki:2014hna,Nozaki:2014uaa} has been proposed in \cite{Nozaki:2013wia}.
The temporal evolution of the holographic complexity in this spacetime has been studied in 
\cite{Ageev:2018nye,Ageev:2019fxn}.
However, this local quench is very different from the one considered in this manuscript,
where two systems initially disconnected are glued together at some point. 
A gravitational dual for this local quench has been studied e.g. in \cite{Ugajin:2013xxa,Astaneh:2014fga,Shimaji:2018czt}
by employing the AdS/BCFT setup discussed in \cite{Takayanagi:2011zk,Fujita:2011fp}. 
It would be interesting to investigate the temporal evolution of the holographic complexity for the entire system 
in this spacetime.

\subsection{Evolution Hamiltonians made by two sites}
\label{subsec:twositespure}

In the simplest case, two separate systems containing only one site  are joined at $t=0$;
hence $N_\textrm{\tiny l} = N_\textrm{\tiny r} =1$ and $N=2$, i.e. the evolution Hamiltonian describes two sites.

In order to specialise (\ref{Delta TR initial state ref}) to this case, 
one  first observes that (\ref{VtildeNover2 mat}) gives
\be
\label{Vtilde few sites}
\widetilde{V}_{1}  \,=\, 1
\;\;\qquad\;\;
\widetilde{V}_2 \, =\, \frac{1}{\sqrt{2}}\,
\bigg(
\begin{array}{cc}
1 \; & 1
\\
1 \; &  -1
\end{array} 
\bigg)
\ee
which (from (\ref{Vtilde0 mat}) and (\ref{mat E def}))
provide respectively 
$V_0=\boldsymbol{1}$ and 
\be
V \,=\,
\frac{1}{\sqrt{2}} \,
\bigg[\,
\bigg(\begin{array}{cc}
1  \; & 1 \\ 1  \; &  -1
\end{array} \bigg)
\oplus
\bigg(\begin{array}{cc}
1 \; & 1
\\
1  \; &  -1
\end{array}\!\bigg)\,\bigg]
\ee
that is symmetric and orthogonal.
Since $V_0=\boldsymbol{1}$, in this case (\ref{Delta TR initial state ref}) simplifies to
\be
\label{Delta TR initial state ref 2 sites}
\gamma_{\textrm{\tiny T}} \,\gamma_{\textrm{\tiny R}}^{-1}
\,=\,
V \, \mathcal{E} \,V \, \Gamma_{0} \, V \, \mathcal{E}^{\textrm{t}} \, V \, \Gamma^{-1}_{0}
\ee
where
\be
\label{gamma0 2 sites}
\Gamma_0\,=\,
\frac{1}{2} \; \textrm{diag} \bigg( 
\big[ m\Omega^{(1)}_1\big]^{-1} , \, \big[ m\Omega^{(1)}_1\big]^{-1} ,
\,m\Omega^{(1)}_1 \,, \, m\Omega^{(1)}_1
\bigg)
\;\;\qquad\;\;
\Omega^{(1)}_1=\sqrt{\omega^2+2\kappa/m}
\ee
and, by using (\ref{diagmat E}), we have 
\be
\label{VEV block dec}
V \, \mathcal{E} \,V
\,=\,
\bigg(\begin{array}{cc}
\widetilde{V}_2\,\mathcal{D}\,\widetilde{V}_2 \;\; & \widetilde{V}_2\, \mathcal{A}\, \widetilde{V}_2
\\
\widetilde{V}_2\, \mathcal{B}\, \widetilde{V}_2 \;\; &  \widetilde{V}_2\, \mathcal{D}\, \widetilde{V}_2
\end{array}\bigg)
\;\;\;\qquad\;\;\;
V\, \mathcal{E}^{\textrm{t}} \, V
\,=\,
\bigg( \begin{array}{cc}
\widetilde{V}_2\, \mathcal{D}\, \widetilde{V}_2 \;\;  & \widetilde{V}_2\, \mathcal{B}\, \widetilde{V}_2
\\
\widetilde{V}_2\,\mathcal{A}\,\widetilde{V}_2 \;\; &  \widetilde{V}_2\, \mathcal{D}\,\widetilde{V}_2
\end{array} \bigg)
\ee
with $\mathcal{A}=\textrm{diag}(\mathcal{A}_1,\mathcal{A}_2)$, 
$\mathcal{B}=\textrm{diag}(\mathcal{B}_1,\mathcal{B}_2)$ 
and $\mathcal{D}=\textrm{diag}(\mathcal{D}_1,\mathcal{D}_2)$ 
being the diagonal matrices whose elements are given by (\ref{blocks diagonal matrices}) with $N=2$.
From (\ref{Delta TR initial state ref 2 sites}) and (\ref{VEV block dec})
one observes that 
the structure of $\gamma_{\textrm{\tiny T}} \,\gamma_{\textrm{\tiny R}}^{-1}$ is not very easy
already in this simple case of $N=2$.

From (\ref{gamma0 2 sites}) and (\ref{VEV block dec}), 
one obtains an explicit expression for $\gamma_{\textrm{\tiny T}} \,\gamma_{\textrm{\tiny R}}^{-1} $ 
in (\ref{Delta TR initial state ref 2 sites}),
which is not reported here because we find it not very insightful.
Its eigenvalues are
$g_{\textrm{\tiny TR},1} $, $g_{\textrm{\tiny TR},1}^{-1} $, 
$g_{\textrm{\tiny TR},2}  $ and $g_{\textrm{\tiny TR},2}^{-1}$,
in terms of
\bea
\label{eigenvalues deltaTR 2 sites}
g_{\textrm{\tiny TR},k}
&\equiv&
\frac{1}
{4 \big( (2k-1) \tfrac{\kappa}{m} +\omega^2\big)\, \big(2  \tfrac{\kappa}{m}+\omega^2 \big)} \;
\Bigg\{
\bigg((2k+1)\frac{\kappa}{m}+2\omega^2\bigg)^2
-\frac{\kappa^2}{m^2}\,\cos\!\big(2 \widetilde{\Omega}_k t \big)
\\
\rule{0pt}{1cm}
&&
-\; \sqrt{2}\; \frac{\kappa}{m}\, 
\sin\!\big(\widetilde{\Omega}_k t \big)\;
\sqrt{
\bigg( (2k+1)\frac{\kappa}{m}+2\omega^2\bigg)^2
+
4 \bigg( 2\,\frac{\kappa}{m}+\omega^2\bigg) \widetilde{\Omega}_k^2 
- \frac{\kappa^2}{m^2}\,
\cos\!\big(2 \widetilde{\Omega}_k t \big)
}
\, \Bigg\}
\nonumber
\eea
where  
$\widetilde{\Omega}_k \equiv \sqrt{\omega^2 + (2k-1) \tfrac{\kappa}{m}}$, 
with $k=1,2$.
Notice  that (\ref{eigenvalues deltaTR 2 sites}) 
depends only on the two dimensionless parameters $\widetilde{\omega}\equiv \omega / \sqrt{\kappa/m}$ and $\omega t $.
For any given $k$, the oscillatory behaviour is governed by the frequency  $\pi/ \widetilde{\Omega}_k$.
This result is qualitatively similar from the one obtained for the 
temporal evolution of the complexity after the global quench of the mass;
indeed, also in that case the temporal evolution associated to 
each mode is determined by a single frequency (see Eq.\,(3.7) of \cite{DiGiulio:2021oal}).
When the evolution is critical, 
(\ref{eigenvalues deltaTR 2 sites}) 
is written through $\widetilde{\Omega}_k=\widetilde{\Omega}_k|_{\omega =0} = \sqrt{(2k-1) \tfrac{\kappa}{m}}$, with $k=1,2$
as follows
\be
g_{\textrm{\tiny TR},k}\big|_{\omega =0}
\,=\,
\frac{1}{8(2k-1)}\;
\bigg\{(2k+1)^2
-
\cos \!\big(2 \widetilde{\Omega}_k t \big)
-
\sqrt{2}\,
\sin\!\big(\widetilde{\Omega}_k t \big)
\sqrt{
4k(k+5) - 7
-
\cos \!\big(2 \widetilde{\Omega}_k t \big)
}\;
\bigg\} .
\ee

By employing the above expressions for the eigenvalues of $\gamma_{\textrm{\tiny T}} \,\gamma_{\textrm{\tiny R}}^{-1} $,
we find that the complexity (\ref{c2 complexity}) in this case can be written as follows
\be
\label{ComplexityPure 2 sites}
\mathcal{C}
=
\frac{1}{2} \;
\sqrt{\big[\log g_{\textrm{\tiny TR},1}\big]^2+\big[\log g_{\textrm{\tiny TR},2}\big]^2}
\ee
in terms of the expressions in (\ref{eigenvalues deltaTR 2 sites}).

We find it worth investigating 
the asymptotic regime given by $\tilde{\omega} \equiv \omega/\sqrt{\kappa/m}\to\infty$, 
while $\omega t$ is kept fixed and finite; 
hence the condition $\omega t\ll \tilde{\omega}$ is imposed. 
In this regime, the expansion of (\ref{eigenvalues deltaTR 2 sites}) reads
\be
\label{eigenvalues deltaTR 2 sites largemass}
g_{\textrm{\tiny TR},k}
\,=\,
1 + \frac{b_1(t)}{\tilde{\omega}^2}
+
\frac{b_2(t)}{\tilde{\omega}^4}
+
O\big( 1/\tilde{\omega}^{6}\big)
\;\;\qquad\;\;
\tilde{\omega} \equiv \frac{\omega}{\sqrt{\kappa/m}}
\ee
where
\be
b_1(t)  \equiv  - \sin(\omega t)
\;\qquad\;
b_2(t) \equiv
\frac{1}{4}
\Big((2k-3)^2-2(2k-1)\,\omega t \, \cos(\omega t)-\cos(2\omega t)+2(2k+1)\sin(\omega t)\Big)
\ee
and we find it worth remarking that $b_1(t) $ is independent of $k$.
By employing (\ref{eigenvalues deltaTR 2 sites largemass}), it is straightforward to obtain 
the first terms in the expansion of the complexity (\ref{ComplexityPure 2 sites}) in this asymptotic regime.
The leading term reads
\be
\label{ComplexityPure 2 sites largemass}
\mathcal{C}
\,=\, \frac{|\sin(\omega t)|}{\sqrt{2}\,\tilde{\omega}^2}
+
O\big( 1/\tilde{\omega}^{4}\big)\,.
\ee

\begin{figure}[t!]
\vspace{-.5cm}
\subfigure
{\hspace{-1.25cm}
\includegraphics[width=.57\textwidth]{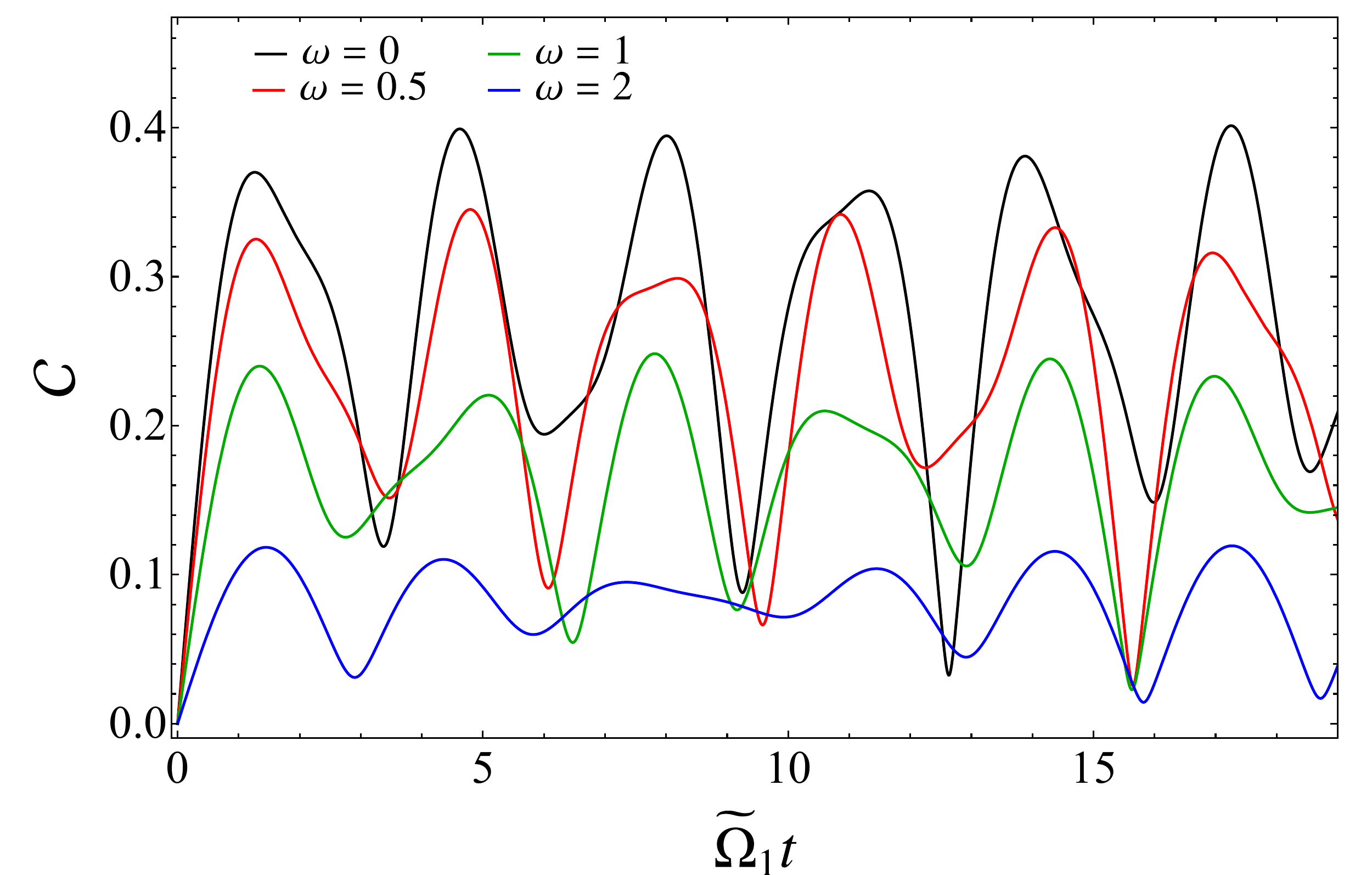}}
\subfigure
{
\hspace{-0.05cm}\includegraphics[width=.57\textwidth]{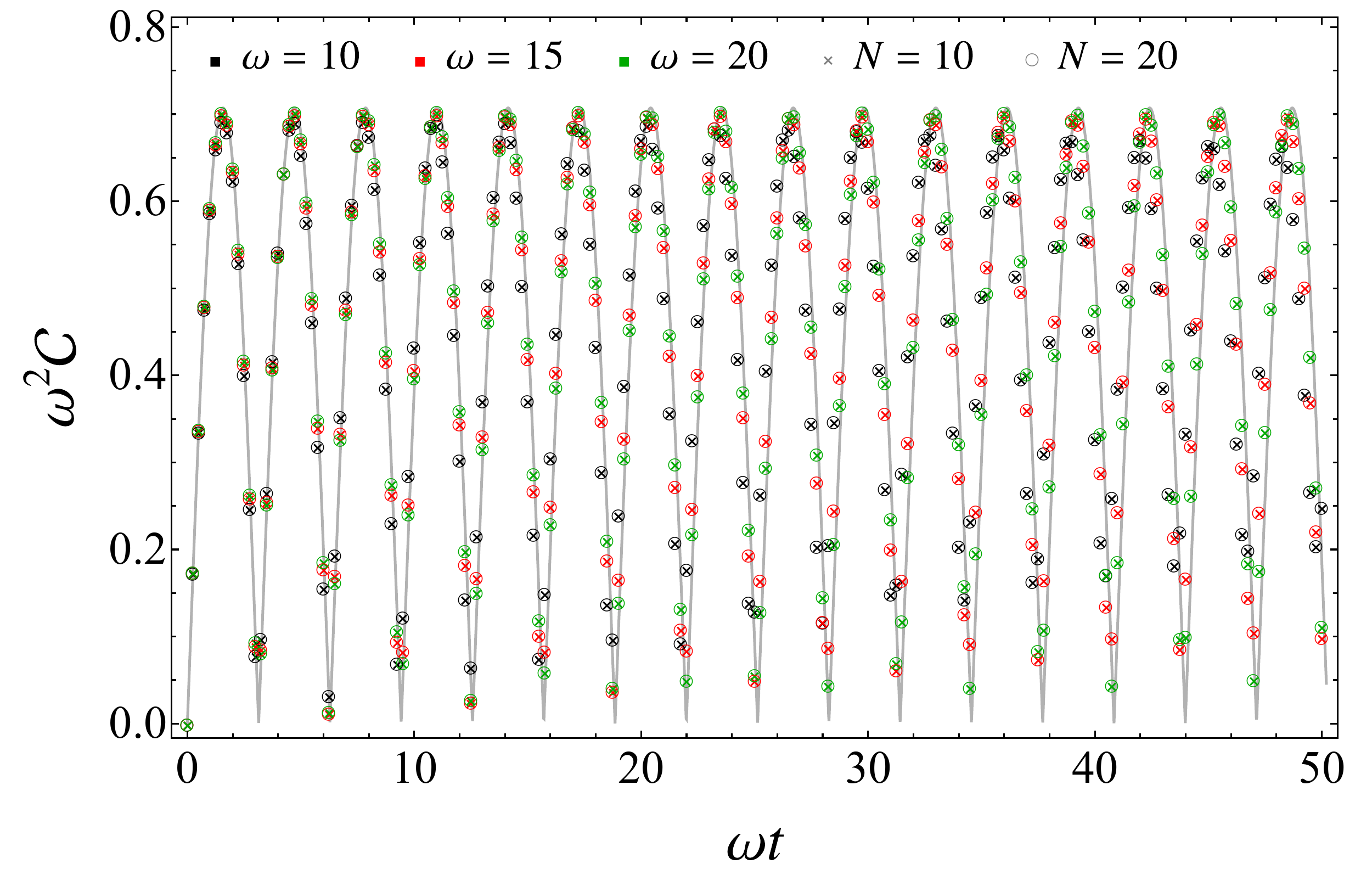}}
\caption{Temporal evolution of the complexity $\mathcal{C}$ for the chain made by two sites 
after a local quench with various $\omega$ w.r.t. the initial state.
Setting $\kappa=1$ and $m=1$,
in the left panel (\ref{ComplexityPure 2 sites}) is shown 
as function of $\widetilde{\Omega}_1t$ (see below (\ref{eigenvalues deltaTR 2 sites})),
while in the right panel (\ref{ComplexityPure 2 sites largemass}) 
is compared against data points obtained numerically for $N>2$ from (\ref{c2 complexity}).
}
\vspace{0.4cm}
\label{fig:PureState2sites}
\end{figure}

In Fig.\,\ref{fig:PureState2sites} we show some temporal evolutions for the complexity (\ref{ComplexityPure 2 sites}),
where the evolution Hamiltonian is made by two sites.

The expression (\ref{eigenvalues deltaTR 2 sites}) tells us that $g_{\textrm{\tiny TR},k}$ with $k=1,2$, 
which provide the complexity (\ref{ComplexityPure 2 sites}), 
are oscillating functions whose periods are $\pi/\widetilde{\Omega}_k$.
A straightforward  numerical inspection shows that  
$g_{\textrm{\tiny TR},1}>g_{\textrm{\tiny TR},2}$ in the whole ranges of $\tilde{\omega}$ and of $\omega t$.
This leads us to plot the complexity in terms of $\widetilde{\Omega}_1 t$ in the left panel of Fig.\,\ref{fig:PureState2sites},
where $\tilde{\omega}$ is not too large
(in this case this argument does not apply because of (\ref{eigenvalues deltaTR 2 sites largemass})).
Interestingly, we observe that the local extrema of the curves having different $\omega$ occur approximatively 
at the same values of $\widetilde{\Omega}_1 t$.

Notice that the temporal evolution corresponding to $\omega=0$ in the left panel of Fig.\,\ref{fig:PureState2sites}
is very different from the one displayed in  the top left panel of Fig.\,\ref{fig:PureStateCritical},
obtained for $N \gg 1$.

In the right panel of Fig.\,\ref{fig:PureState2sites},
the data points, 
obtained numerically for $N=10$ and $N=20$ and three values of $\omega$,
are compared against the leading term given in (\ref{ComplexityPure 2 sites largemass}).
We find it worth remarking that (\ref{ComplexityPure 2 sites largemass}) nicely agrees with 
the temporal evolution of the complexity for small values of $t$, even when $N>2$.
The agreement between the analytic curve and the data improves as $\omega$ grows, as expected.

\section{Subsystem complexity}
\label{sec-subsystem-comp}

In this section we investigate the temporal evolution of the subsystem complexity 
$\mathcal{C}_A$ after the local quench introduced in Sec.\,\ref{subsec:setup},
when the reference and the target states are the reduced density matrices of the block $A$ 
in the configurations shown in the bottom panels of Fig.\,\ref{fig:intro-configs}.

\subsection{Optimal circuit and subsystem complexity}
\label{subsec-subcomp-gen}

In the harmonic lattices in the pure states that we are considering, 
the reduced density matrix associated to a spatial subsystem $A$ 
characterises a mixed Gaussian state
which can be fully described through 
its reduced covariance matrix $\gamma_A$ \cite{Peschel03,Eisert:2008ur,Weedbrook12b},
defined as the $2L \times 2L$ real, symmetric and positive definite matrix 
($L$ denotes the number of sites in $A$) 
\be
\label{reduced CM}
\gamma_A(t)
=\,
\bigg( 
\begin{array}{cc}
Q_A(t)  \,& M_A(t) \\
M_A(t)^{\textrm{t}}  \,& P_A(t)  \\
\end{array}   \bigg)
\ee
where $Q_A(t)$, $P_A(t)$ and $M_A(t)$
are the reduced correlation matrices,
obtained by selecting the rows and the columns corresponding to $A$ in (\ref{CM block decomposed}),
namely $Q(t)_{i,j}$, $P(t)_{i,j}$ and $M(t)_{i,j}$, with $i,j\in A$.
The reduced correlation matrices usually
depend on the time $t$ after the quench.

In this section we study the circuit complexity when both the reference and the target states
are mixed states corresponding to a subsystem $A$.
In particular, we apply to the local quench that we are investigating 
the results for the circuit complexity 
of mixed states based on the Fisher information geometry \cite{DiGiulio:2020hlz},
as done in \cite{DiGiulio:2021oal} for a global quench.

We consider the reference state given by the reduced density matrix for the subsystem $A$
at time $t_\textrm{\tiny R} \geqslant 0$ 
obtained through the local quench protocol characterised by
$\{m_{\textrm{\tiny R}}, \kappa_{\textrm{\tiny R}}, \omega_{\textrm{\tiny R}}\}$
and the target state given by  the reduced density matrix of the same subsystem
at time $t_\textrm{\tiny T} \geqslant  t_\textrm{\tiny R}$,
constructed through the quench protocol described by
$\{m_{\textrm{\tiny T}}, \kappa_{\textrm{\tiny T}}, \omega_{\textrm{\tiny T}}\}$
(see Sec.\,\ref{subsec:PureStateCompTheory}).
The corresponding reduced covariance matrices, 
denoted by $ \gamma_{\textrm{\tiny R},A} (t_\textrm{\tiny R}) $
and $ \gamma_{\textrm{\tiny T},A} (t_\textrm{\tiny T}) $ respectively,
can be decomposed as done in (\ref{reduced CM}).

The approach to the circuit complexity of mixed states based on the Fisher information geometry \cite{DiGiulio:2020hlz} 
provides also the optimal circuit connecting 
$ \gamma_{\textrm{\tiny R},A} (t_\textrm{\tiny R}) $ to $ \gamma_{\textrm{\tiny T},A} (t_\textrm{\tiny T}) $
\cite{Bhatia07book}
\be
\label{optimal circuit rdm}
G_{s}( \gamma_{\textrm{\tiny R},A}(t_\textrm{\tiny R}) \, , \gamma_{\textrm{\tiny T},A}(t_\textrm{\tiny T}))
\,\equiv \,
\gamma_{\textrm{\tiny R},A}(t_\textrm{\tiny R})^{1/2} 
\Big(  \gamma_{\textrm{\tiny R},A}(t_\textrm{\tiny R})^{- 1/2} \,
\gamma_{\textrm{\tiny T},A}(t_\textrm{\tiny T}) \,
\gamma_{\textrm{\tiny R},A}(t_\textrm{\tiny R})^{-1/2} \Big)^s
\gamma_{\textrm{\tiny R},A}(t_\textrm{\tiny R})^{1/2}
\ee
where $0 \leqslant s \leqslant 1$, 
which is a covariance matrix for any  $s$
\cite{Bathia15}.
The length of the optimal circuit (\ref{optimal circuit rdm}) is proportional to its complexity 
\be
\label{c2-complexity-rdm}
\mathcal{C}_A
\,=\,
\frac{1}{2\sqrt{2}}\;
\sqrt{\,
\textrm{Tr} \,\Big\{ \big[
\log \!\big( \gamma_{\textrm{\tiny T},A}(t_\textrm{\tiny T}) \, \gamma_{\textrm{\tiny R},A}(t_\textrm{\tiny R})^{-1} \big)
\big]^2  \Big\}}\,.
\ee

Both the reduced covariance matrices in (\ref{c2-complexity-rdm}) 
have the form (\ref{reduced CM}),
obtained by restricting to $A$ the covariance matrix 
$\gamma(t)$ in (\ref{CM-local-evolved general}), 
as discussed above.

In our analysis we consider the simplest setup
where the reference state is the initial state (i.e. $t_\textrm{\tiny R}=0$)
and the target state corresponds to a generic value of $t_\textrm{\tiny T}=t  \geqslant 0$ after the local quench. 
The remaining parameters are fixed to
 $\omega_{\textrm{\tiny R}}=\omega_{\textrm{\tiny T}}\equiv \omega$, 
 $\kappa_{\textrm{\tiny R}}=\kappa_{\textrm{\tiny T}}\equiv \kappa$ and $m_{\textrm{\tiny R}}=m_{\textrm{\tiny T}} \equiv m$.
In this case the subsystem complexity (\ref{c2-complexity-rdm}) reads
\be
\label{c2-complexity-rdm-our-case}
\mathcal{C}_A
\,=\,
\frac{1}{2\sqrt{2}}\;
\sqrt{\,
\textrm{Tr} \,\Big\{ \big[
\log \!\big( \gamma_{A}(t) \, \gamma_{A}(0)^{-1} \big)
\big]^2  \Big\}}\,.
\ee

It is instructive to compare the temporal evolution of $\mathcal{C}_A$
against the temporal evolution of the entanglement entropy $S_A$ after the same local quench, 
which can be evaluated from the symplectic spectrum of 
$\gamma_{A}(t)$ in the standard way
\cite{Peschel_2009,Audenaert:2002xfl,Cramer:2005mx,Casini:2009sr}.
The considerations above can be easily adapted to harmonic lattices in 
any number of spatial dimensions.

 \subsection{Numerical results}

\begin{figure}[t!]
\vspace{-.5cm}
\subfigure
{\hspace{-1.25cm}
\includegraphics[width=.57\textwidth]{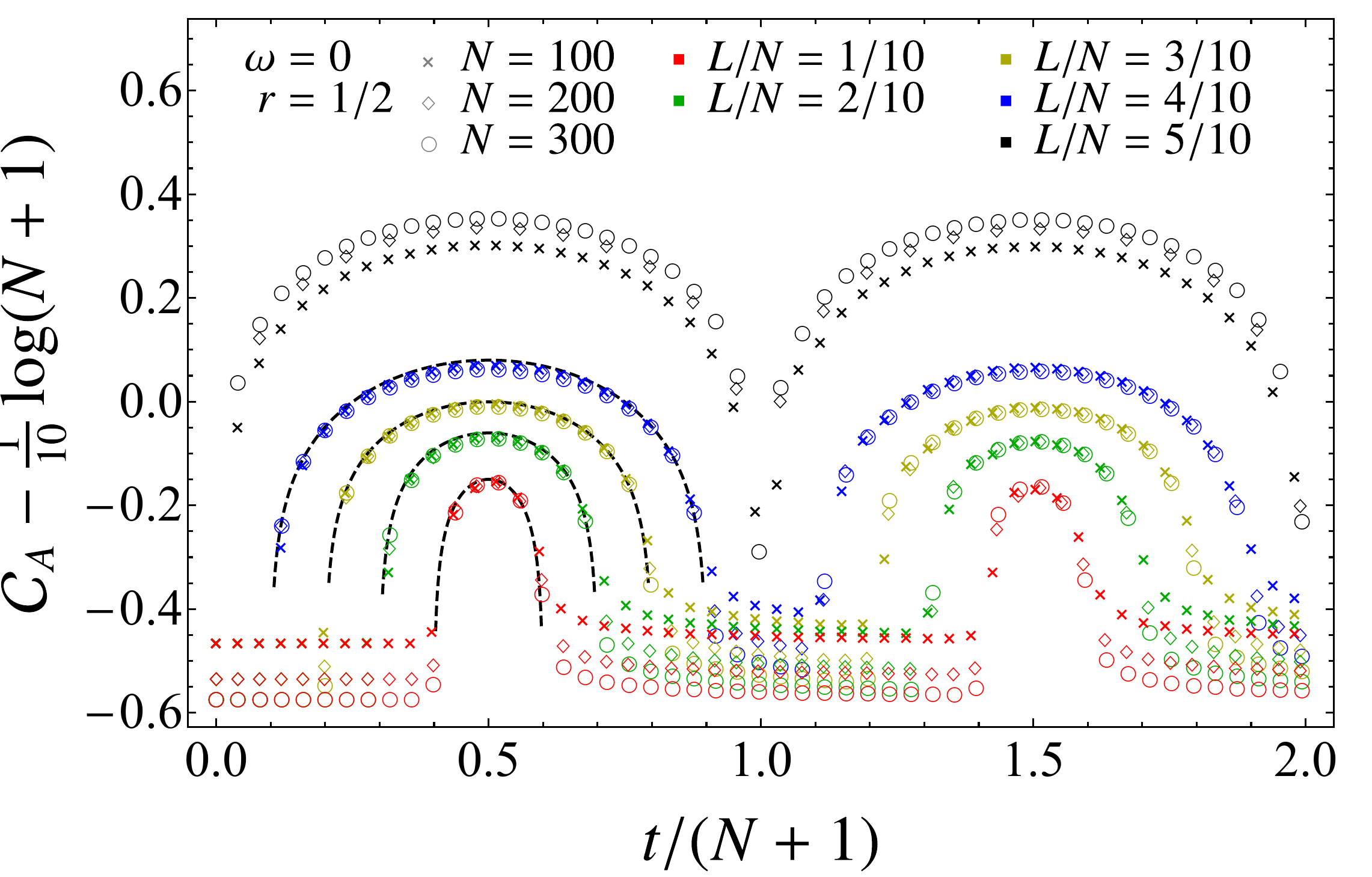}}
\subfigure
{
\hspace{-.0cm}\includegraphics[width=.57\textwidth]{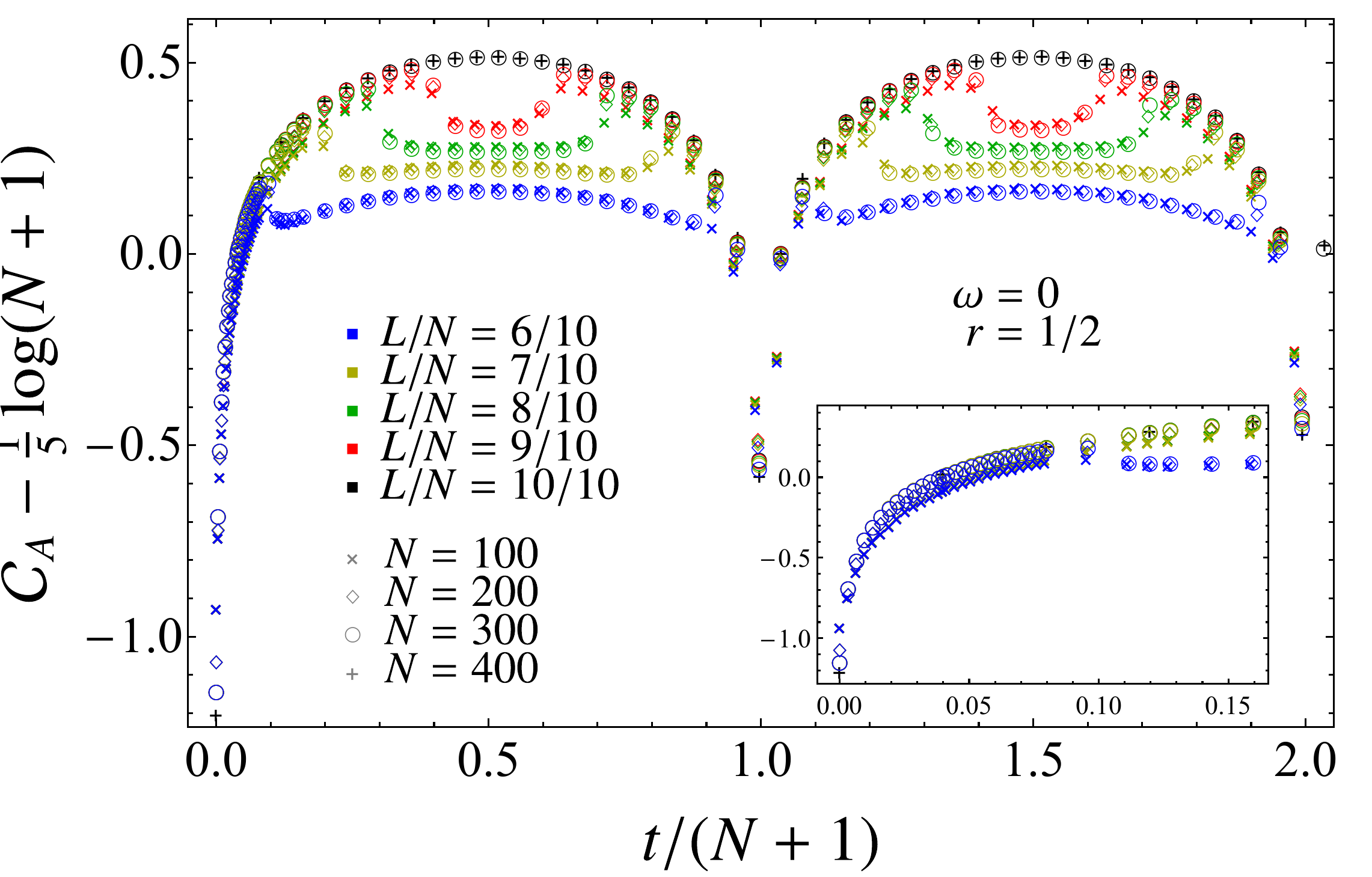}}
\subfigure
{\hspace{-1.25cm}
\includegraphics[width=.57\textwidth]{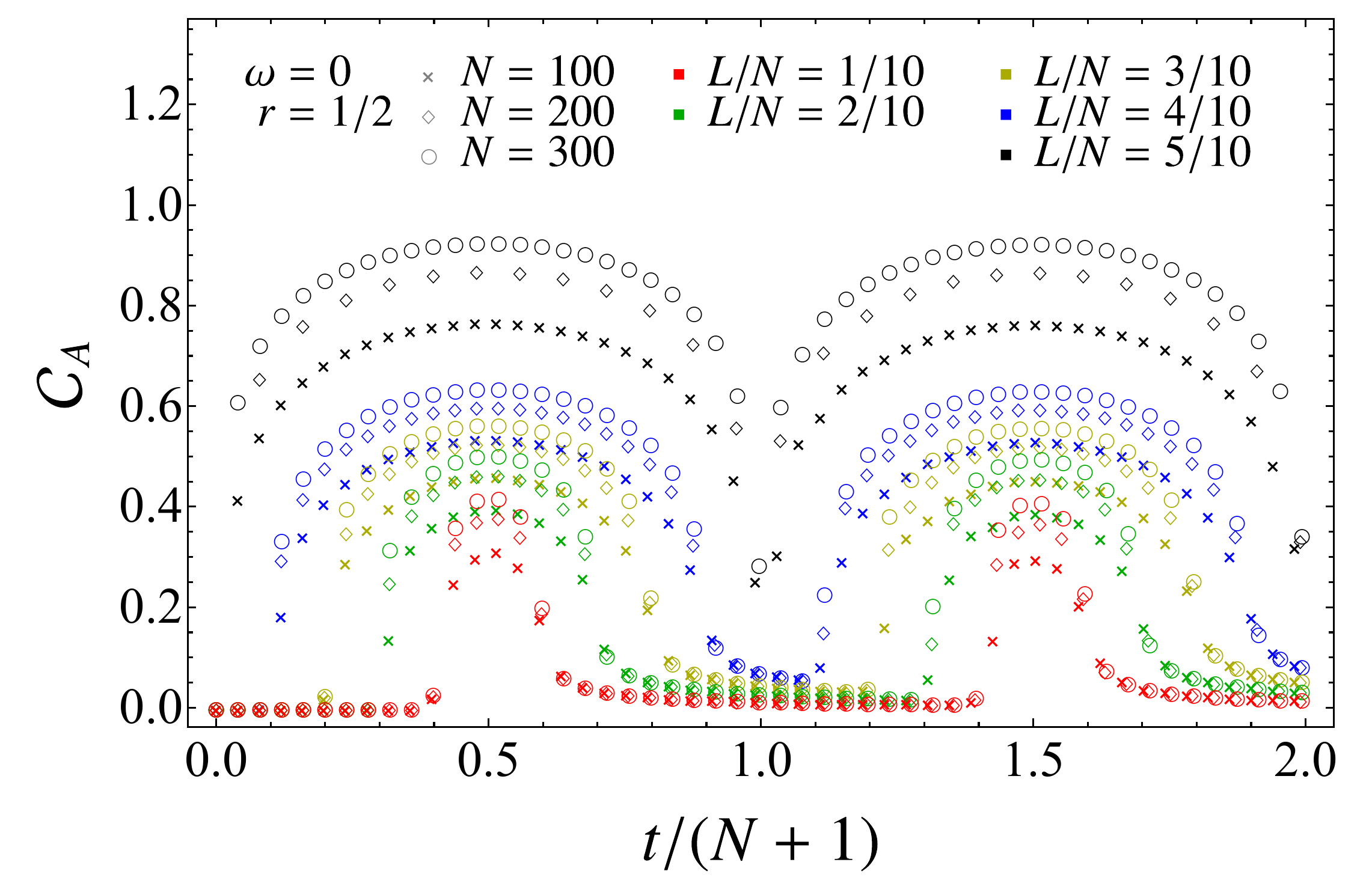}}
\subfigure
{
\hspace{-.0cm}\includegraphics[width=.57\textwidth]{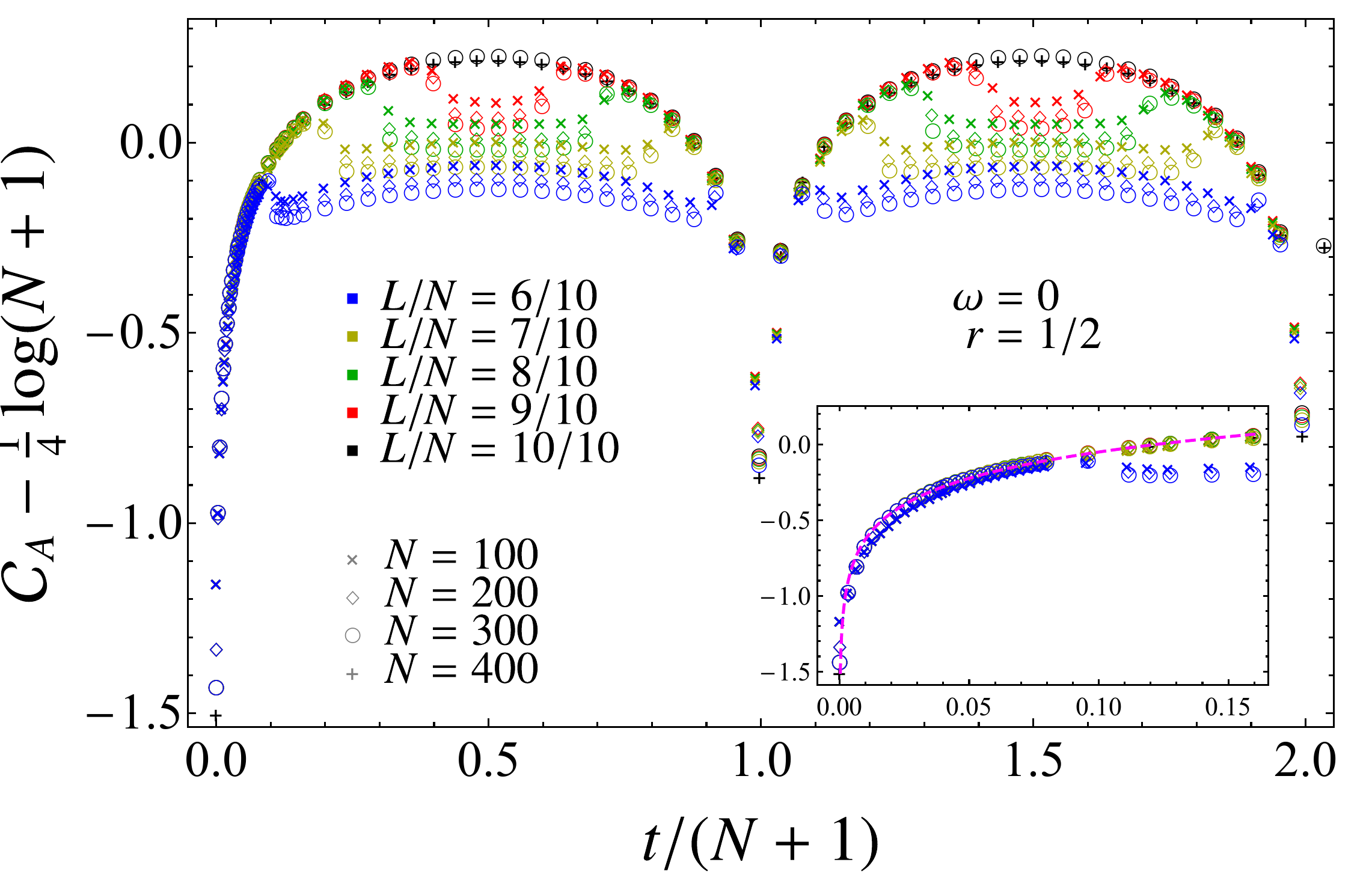}}
\caption{Temporal evolution of the subsystem complexity $\mathcal{C}_A$ 
in (\ref{c2-complexity-rdm-our-case})
for a block $A$ made by $L$ consecutive sites adjacent to the left boundary
of harmonic chains made by $N$ sites 
(see Fig.\,\ref{fig:intro-configs}, bottom left panel)
after a local quench with $\omega  = 0$ and $r=1/2$.
The size of the blocks is $L \leqslant N r$ in the left panels and $L>N r$ in the right panels. 
The black dashed curves in the top left panel correspond to (\ref{complexity guess}).
The insets zoom in on the initial growth 
(the dashed curve in the bottom right panel corresponds to (\ref{guess complexity initial})).
The data points corresponding to $L/N=1$ in the right panels
are also reported in the top left panel of Fig.\,\ref{fig:PureStateCritical}.
}
\vspace{0.4cm}
\label{fig:CompMixedMasslessr1over2}
\end{figure}

\begin{figure}[t!]
\vspace{-.5cm}
\subfigure
{\hspace{-1.25cm}
\includegraphics[width=.57\textwidth]{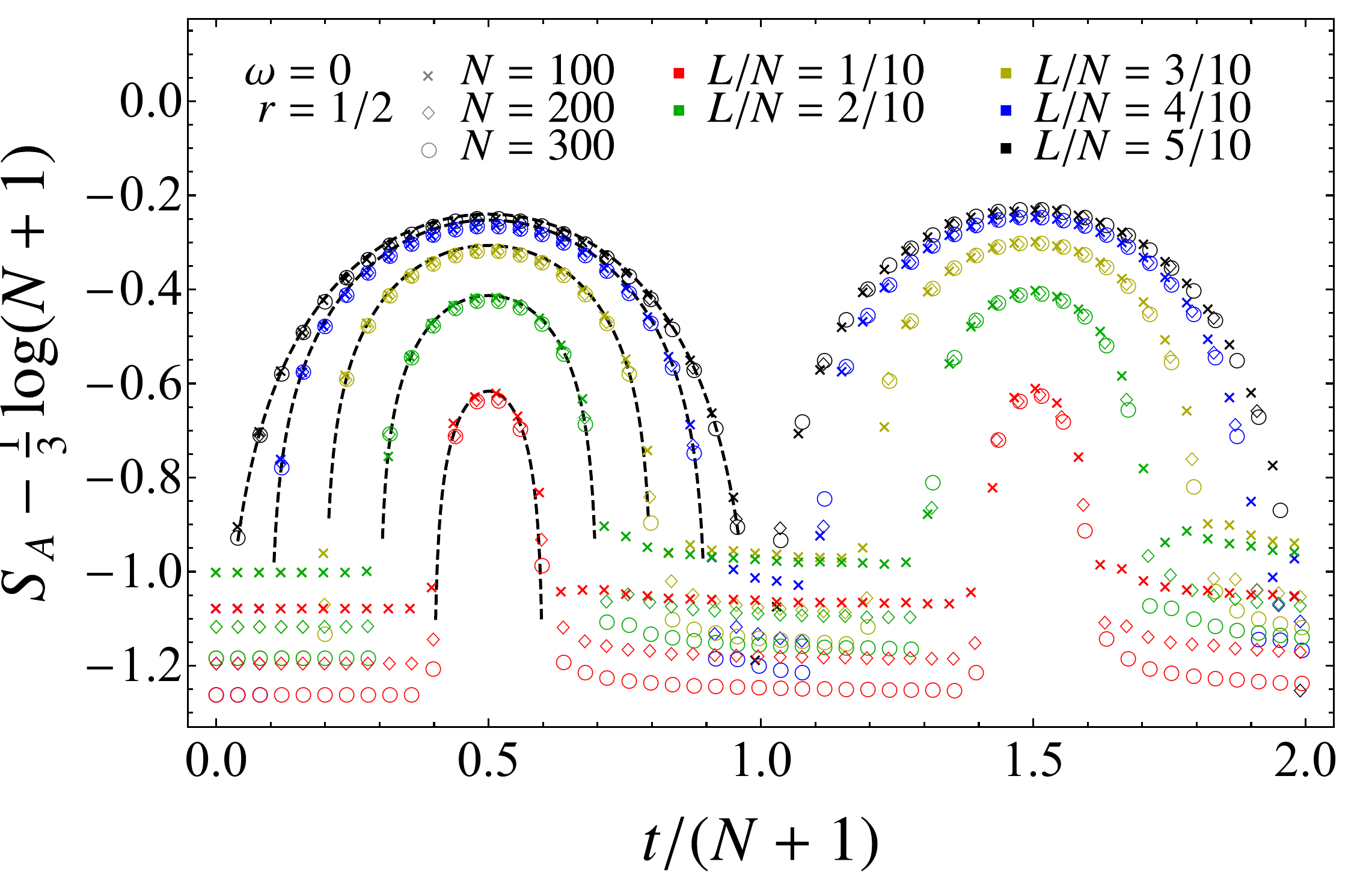}}
\subfigure
{
\hspace{-.0cm}\includegraphics[width=.57\textwidth]{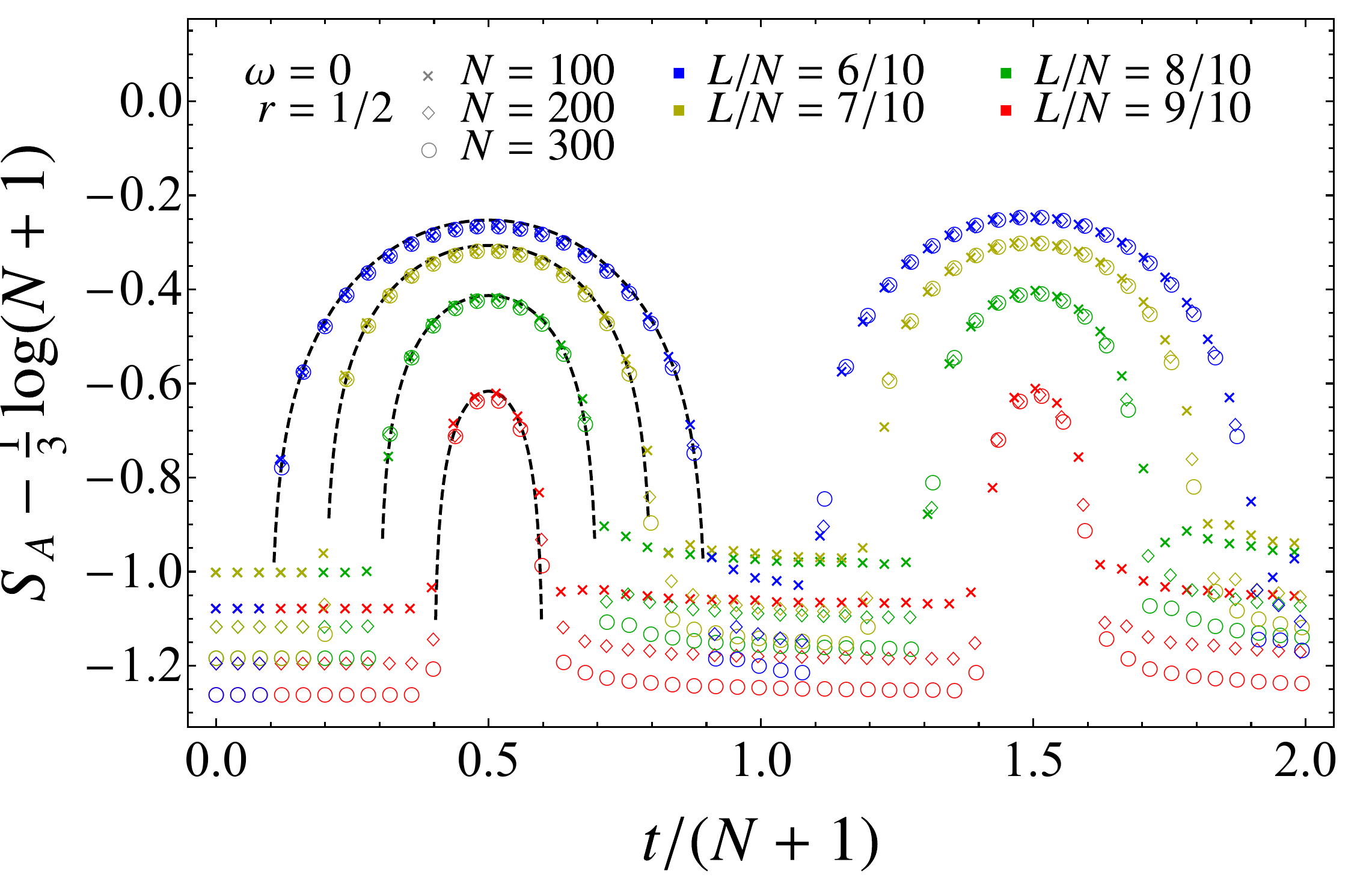}}
\subfigure
{\hspace{-1.25cm}
\includegraphics[width=.57\textwidth]{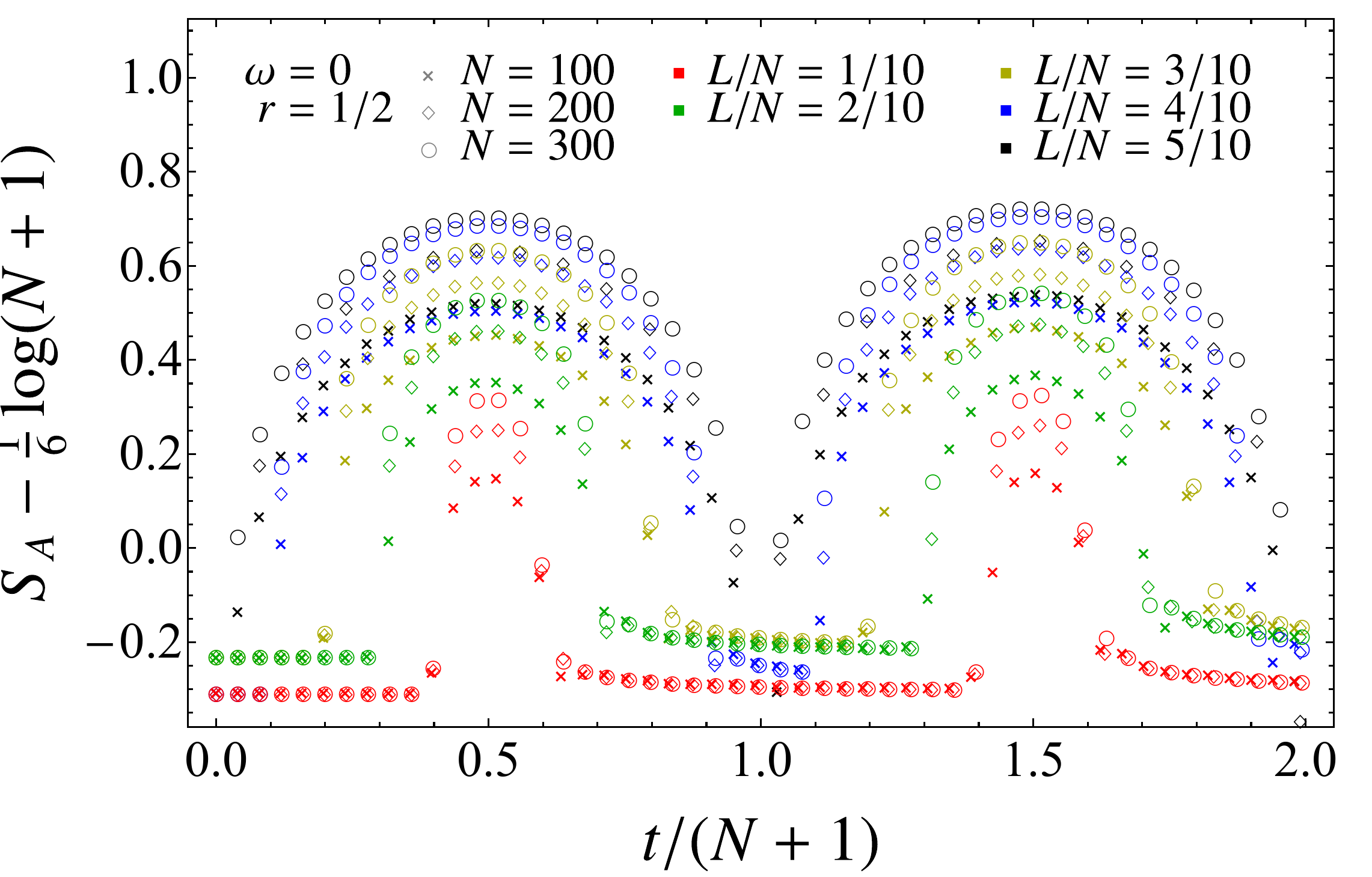}}
\subfigure
{
\hspace{-.0cm}\includegraphics[width=.57\textwidth]{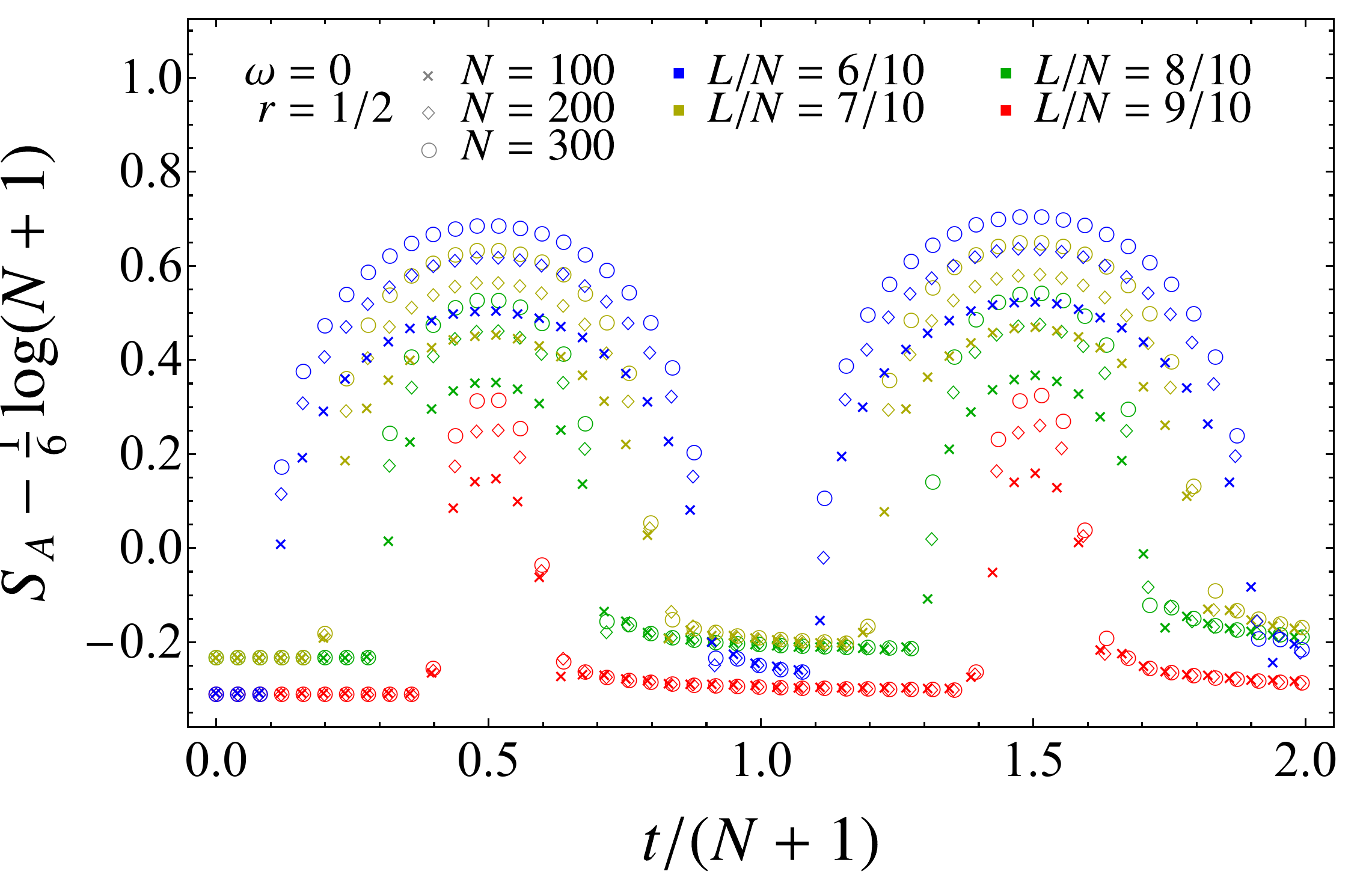}}
\caption{
Temporal evolution of the entanglement entropy $S_A$ 
in the same setup described for Fig.\,\ref{fig:CompMixedMasslessr1over2}.
The size of the blocks is $L \leqslant N r$ in the left panels and $L>N r$ in the right panels. 
The black dashed curves in the top panels correspond to (\ref{entropy CFT}).
}
\vspace{0.4cm}
\label{fig:EntMasslessr1over2}
\end{figure}

In the following we discuss some numerical
results for the temporal evolution after a local quench of the subsystem complexity (\ref{c2-complexity-rdm-our-case}) 
in the case where the subsystem $A$ is a block made by $L$
consecutive sites in harmonic chains made by $N$ sites.
Let us remind that 
the reference state is the initial state ($t_\textrm{\tiny R}=0$) 
and the target state corresponds to the state at
the generic value $t_\textrm{\tiny T}\equiv t\geqslant 0$ after the local quench,
whose protocol is specified by the values of the parameters in (\ref{parameters local quench})
with $m=1$ and $\kappa=1$. 
For a given local quench, 
we display both the temporal evolution of the subsystem complexity $\mathcal{C}_A$ 
and of the entanglement entropy $S_A$.

\begin{figure}[t!]
\vspace{-.5cm}
\subfigure
{\hspace{-1.25cm}
\includegraphics[width=.57\textwidth]{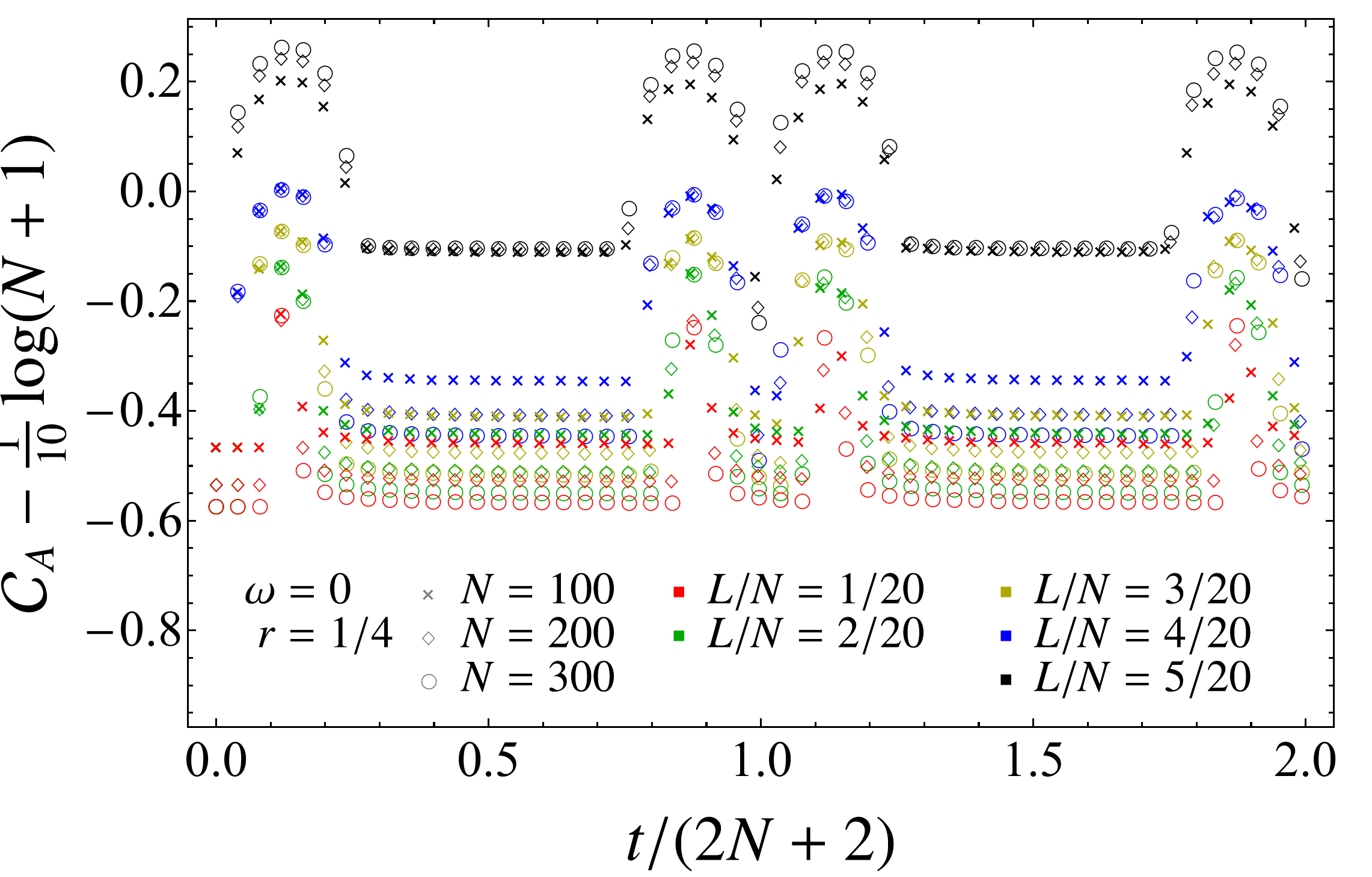}}
\subfigure
{
\hspace{-.0cm}\includegraphics[width=.57\textwidth]{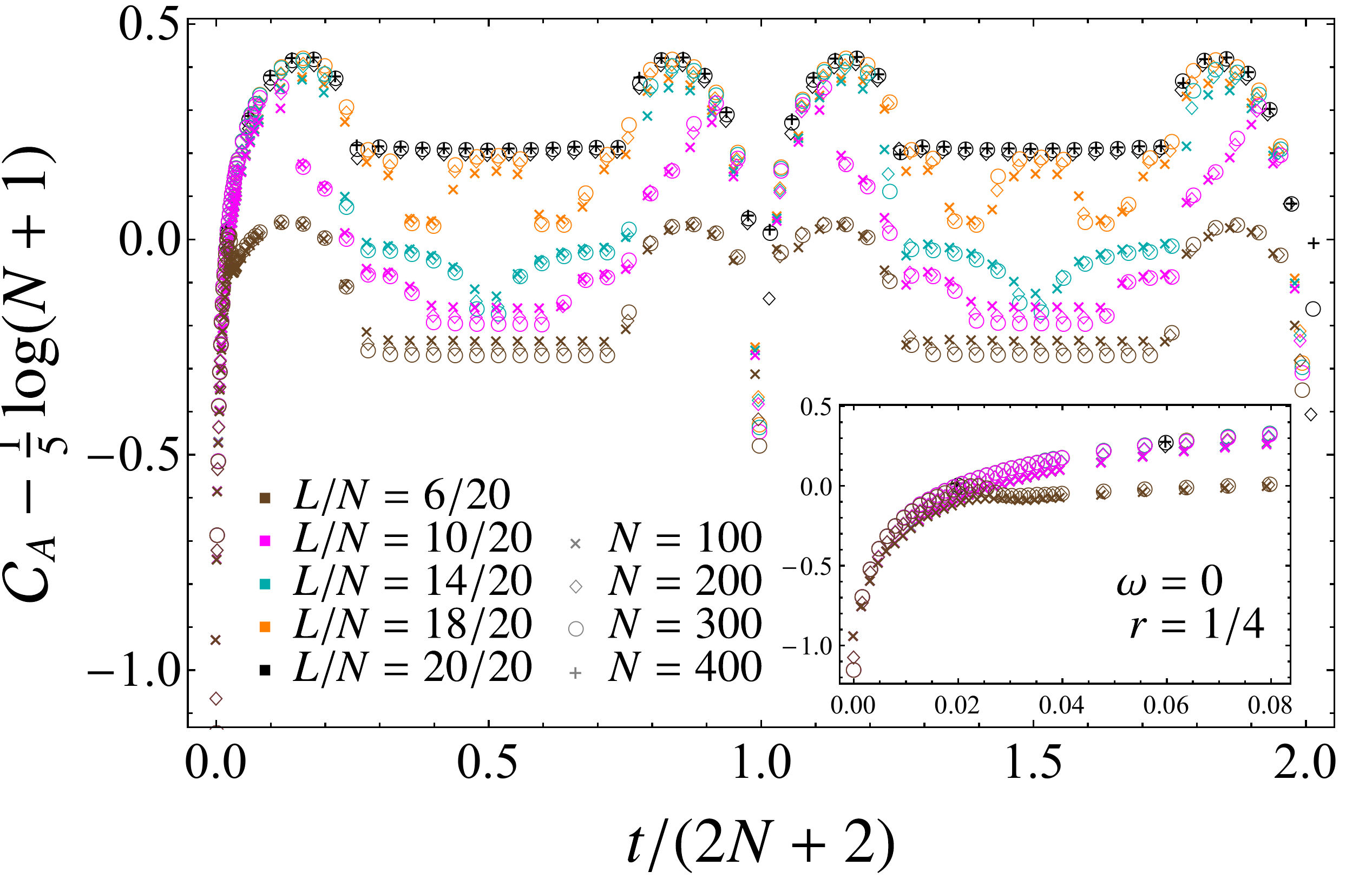}}
\subfigure
{\hspace{-1.25cm}
\includegraphics[width=.57\textwidth]{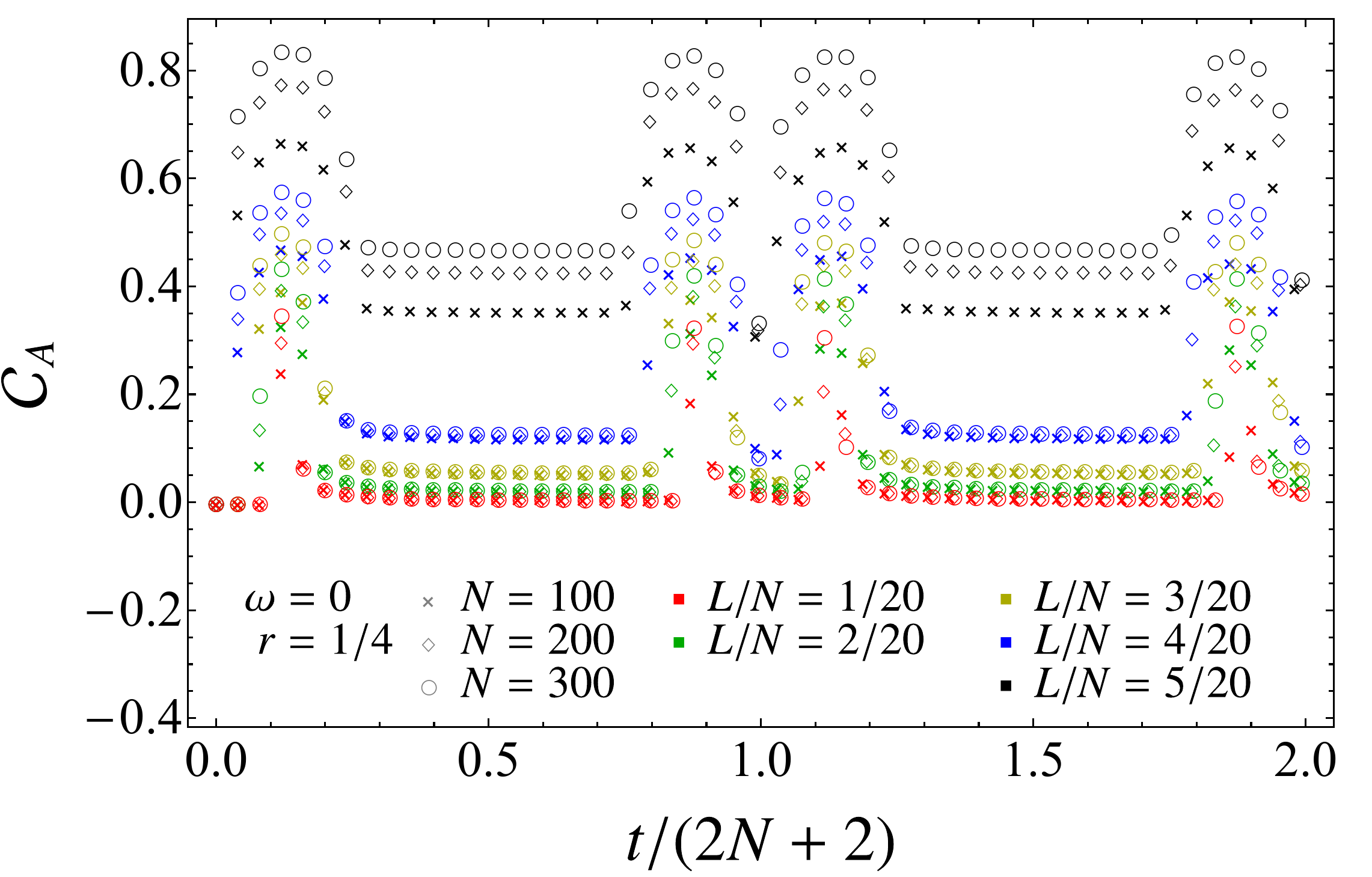}}
\subfigure
{
\hspace{-.0cm}\includegraphics[width=.57\textwidth]{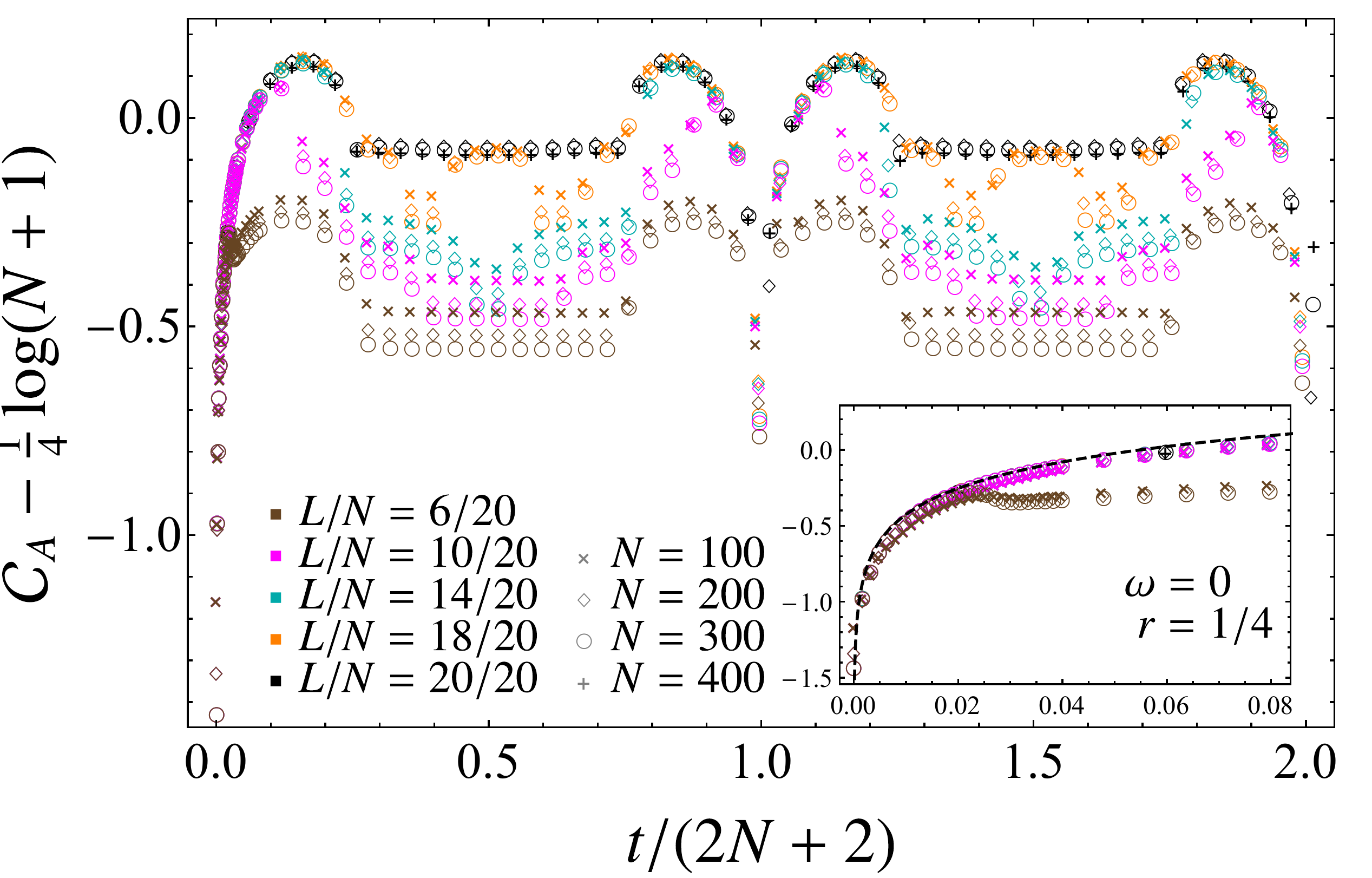}}
\caption{
Temporal evolution of the subsystem complexity $\mathcal{C}_A$ 
in (\ref{c2-complexity-rdm-our-case})
for a block $A$ made by $L$ consecutive sites adjacent to the left boundary
of harmonic chains made by $N$ sites (see Fig.\,\ref{fig:intro-configs}, bottom left panel)
after a local quench with $\omega  = 0$ and $r=1/4$.
The size of the blocks is $L \leqslant N r$ in the left panels and $L>N r$ in the right panels. 
The insets zoom in on the initial growth 
(the dashed curve in the bottom right panel corresponds to (\ref{guess complexity initial})).
The data points corresponding to $L/N=1$ in the right panels
are also reported in the top right panel of Fig.\,\ref{fig:PureStateCritical}.
 }
\vspace{0.4cm}
\label{fig:CompMixedMasslessr1over4}
\end{figure}

\begin{figure}[t!]
\vspace{-.5cm}
\subfigure
{\hspace{-1.25cm}
\includegraphics[width=.57\textwidth]{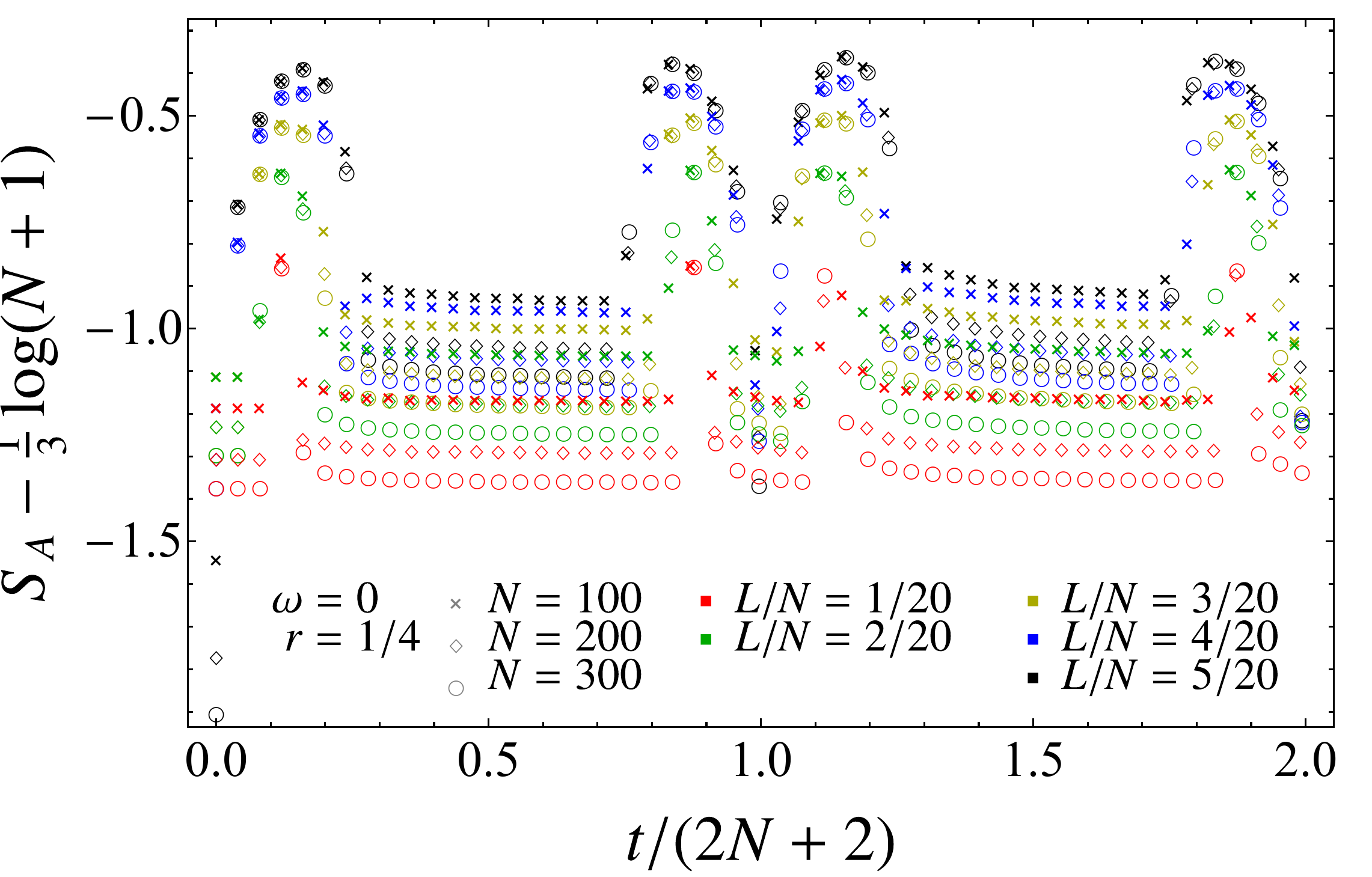}}
\subfigure
{
\hspace{-.0cm}\includegraphics[width=.57\textwidth]{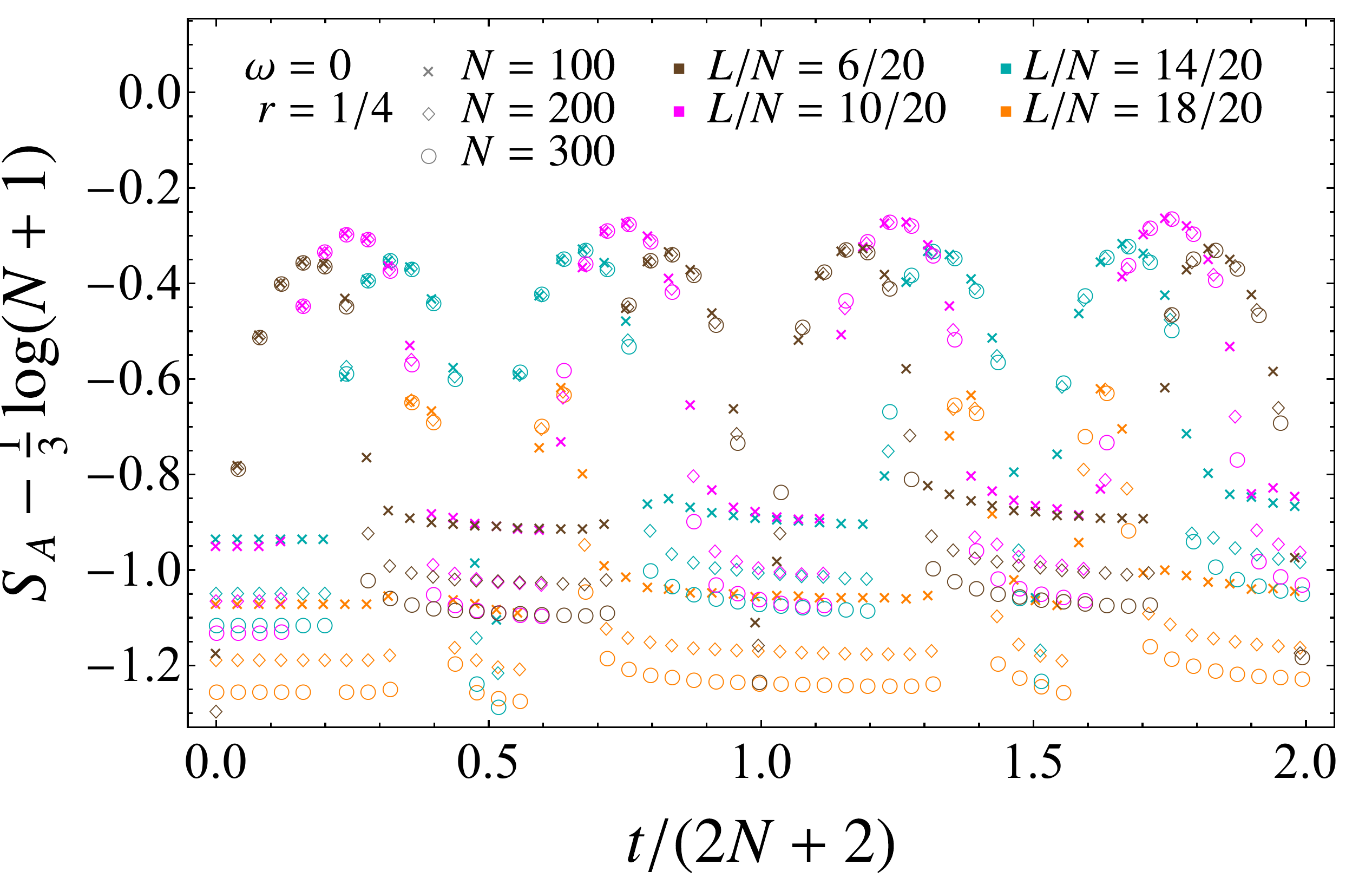}}
\subfigure
{\hspace{-1.25cm}
\includegraphics[width=.57\textwidth]{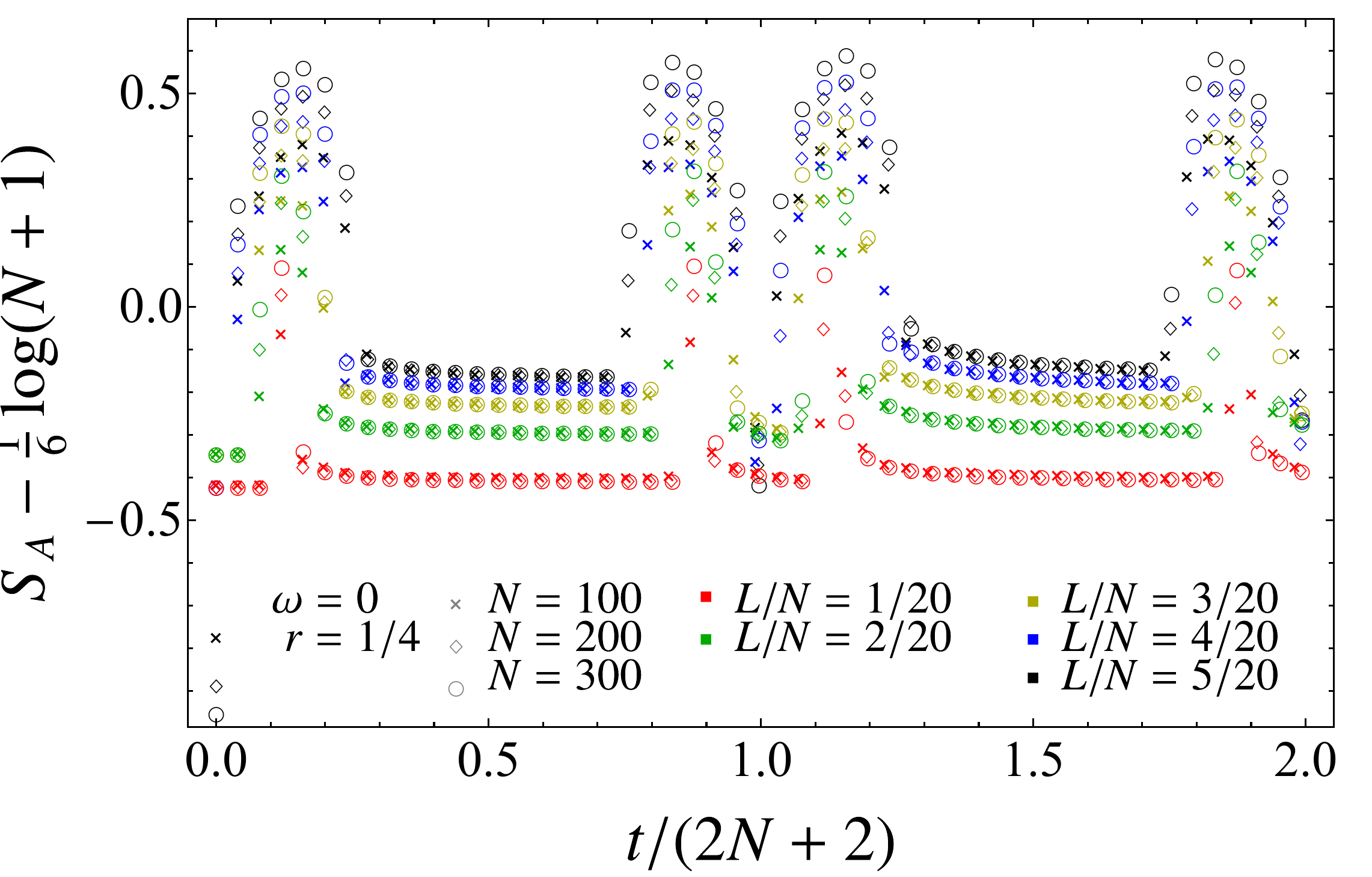}}
\subfigure
{
\hspace{-.0cm}\includegraphics[width=.57\textwidth]{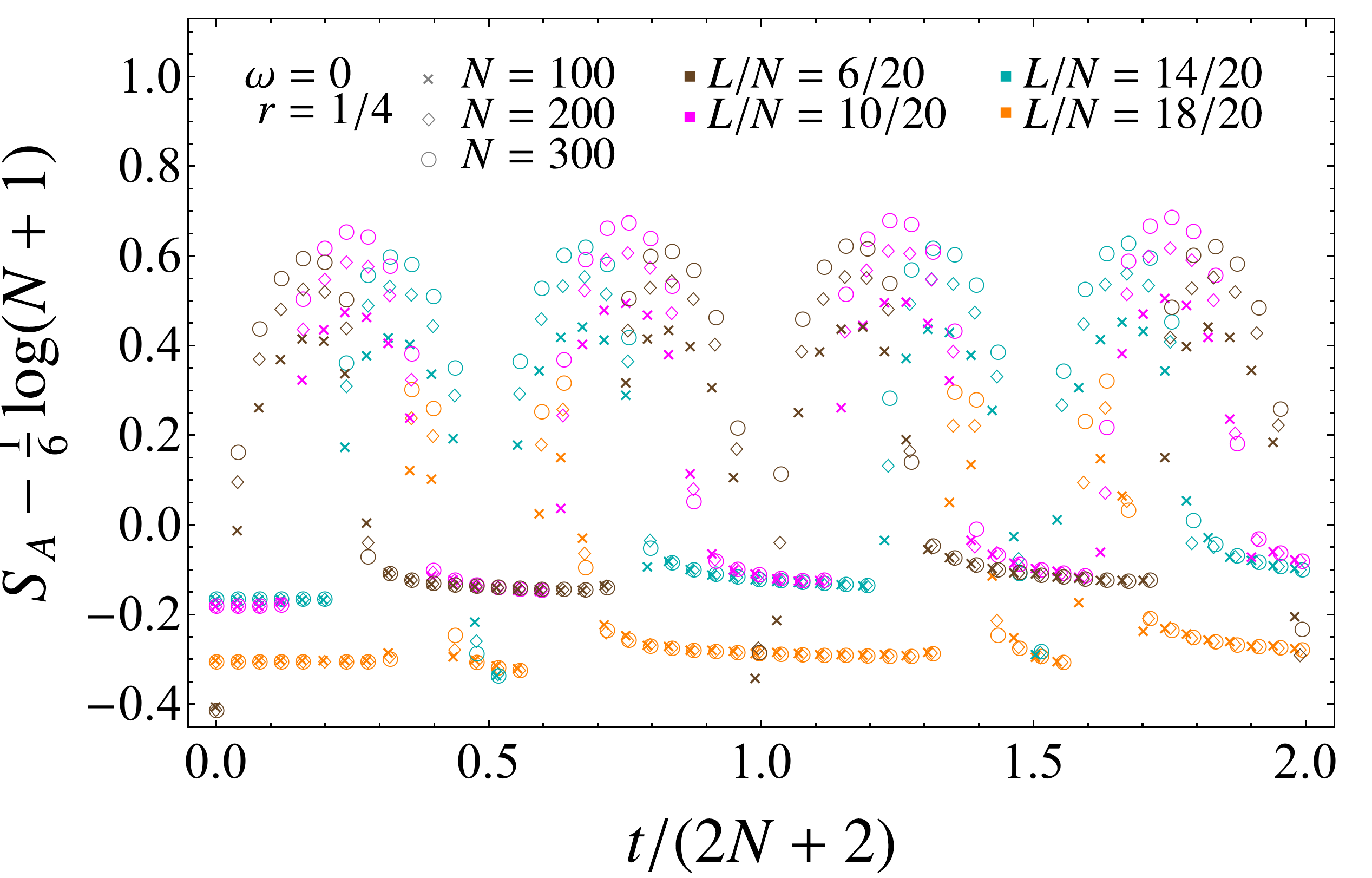}}
\caption{
Temporal evolution of the entanglement entropy $S_A$ 
in the same setup described for Fig.\,\ref{fig:CompMixedMasslessr1over4}.
The size of the blocks is $L \leqslant r N $ in the left panels and $L> r N $ in the right panels. 
}
\vspace{0.4cm}
\label{fig:EntMasslessr1over4}
\end{figure}

\begin{figure}[t!]
\vspace{-.5cm}
\subfigure
{\hspace{-1.25cm}
\includegraphics[width=.57\textwidth]{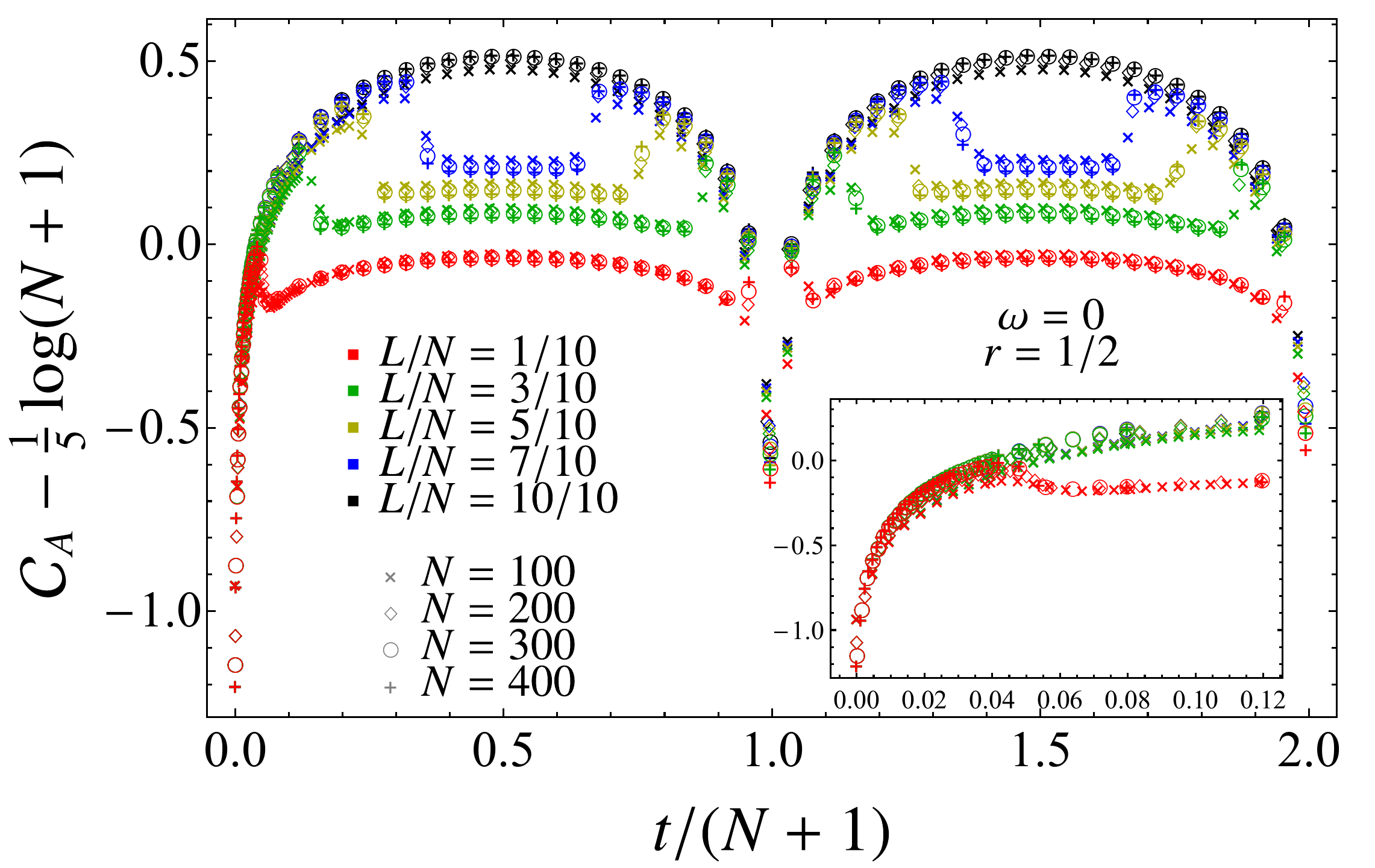}}
\subfigure
{
\hspace{-.0cm}\includegraphics[width=.57\textwidth]{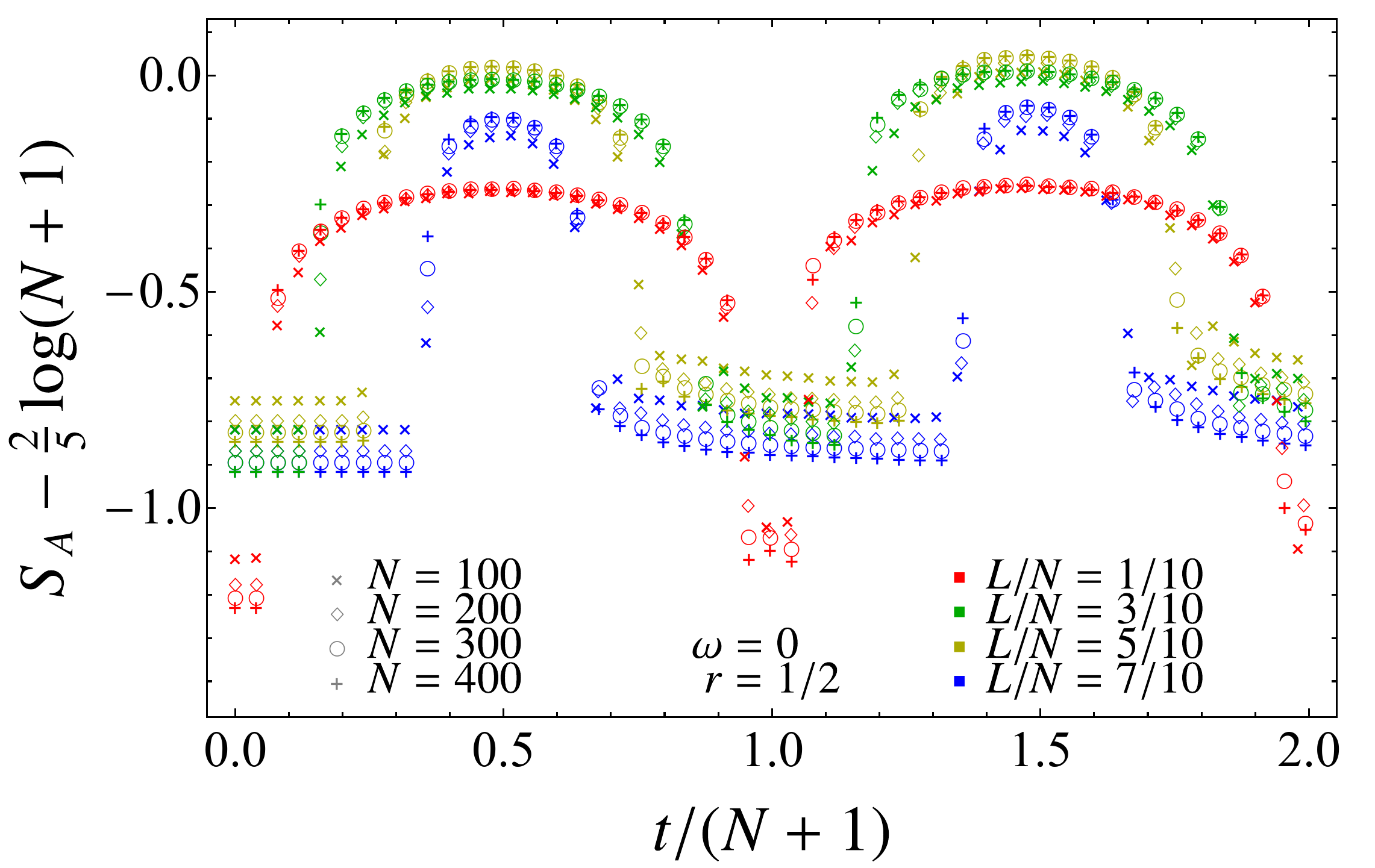}}
\subfigure
{\hspace{-1.25cm}
\includegraphics[width=.57\textwidth]{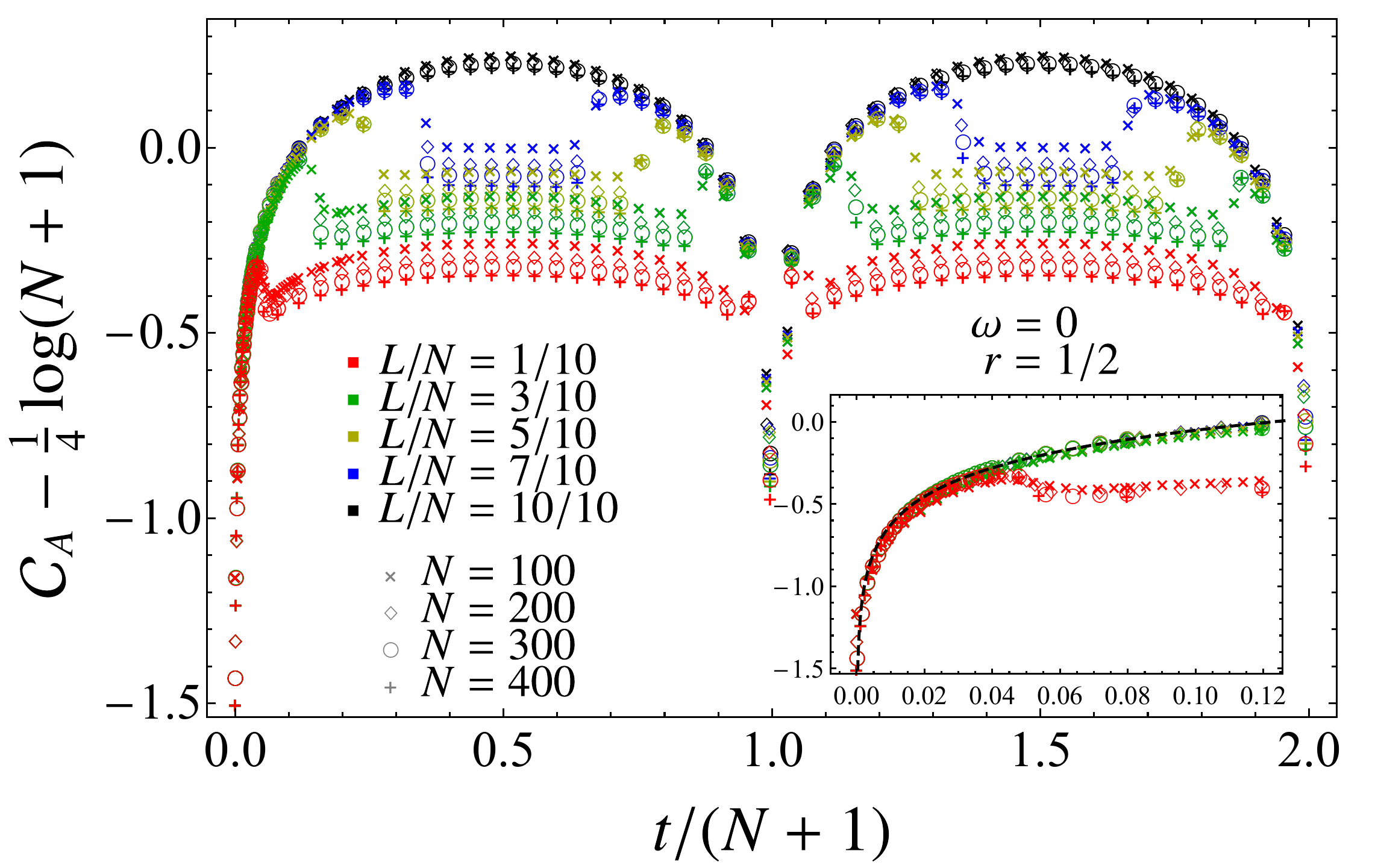}}
\subfigure
{
\hspace{-.0cm}\includegraphics[width=.57\textwidth]{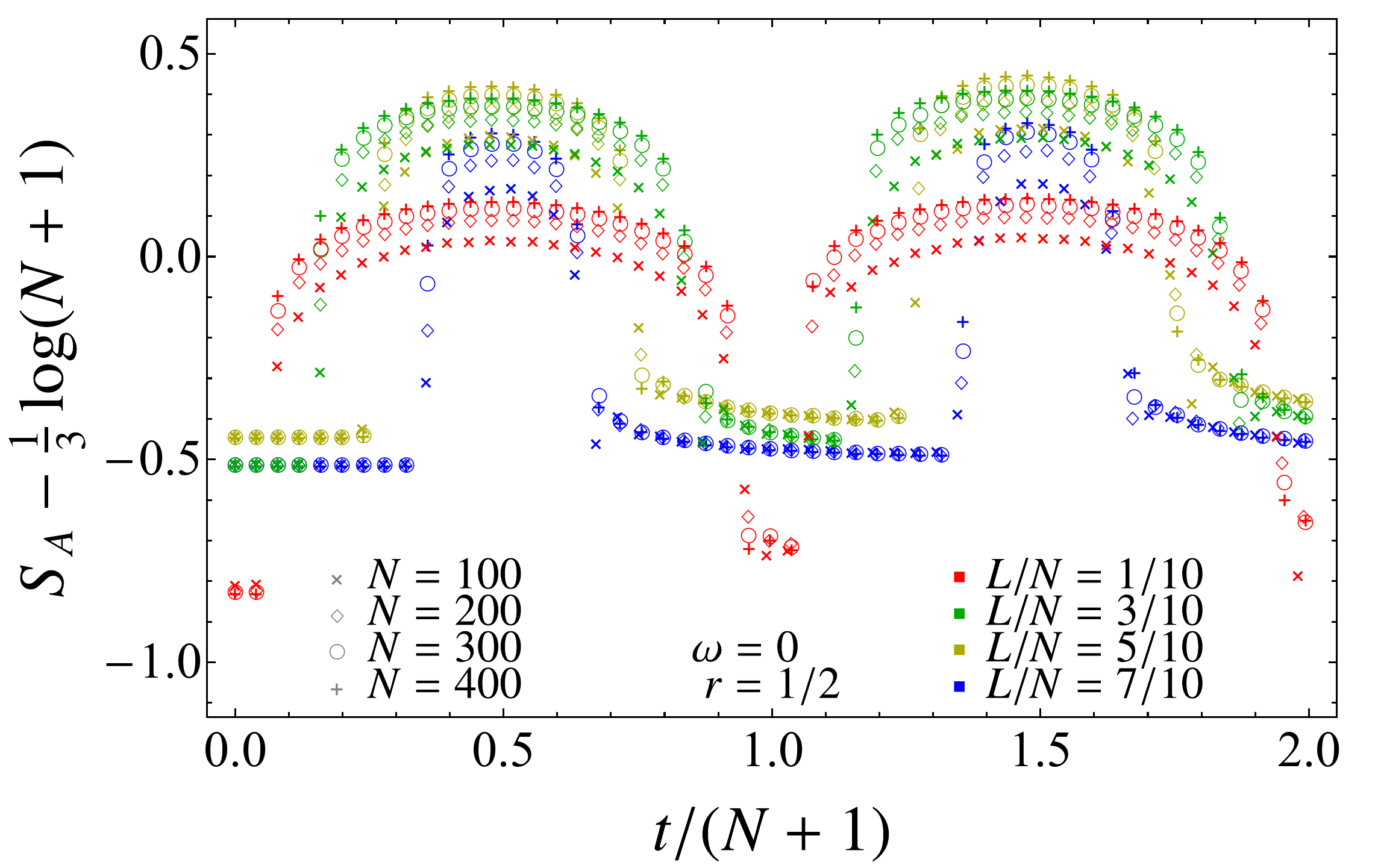}}
\caption{Temporal evolution of the subsystem complexity $\mathcal{C}_A$ (left panels)
and entanglement entropy $S_A$ (right panels) 
for a block $A$ made by $L$ sites 
whose midpoint is given by the joining point at  $r=1/2$
(see Fig.\,\ref{fig:intro-configs}, bottom right panel)
after a local quench with $\omega  = 0$.
The insets zoom in on the initial growth
(the dashed curve in the bottom left panel is given by (\ref{guess complexity initial})).
}
\vspace{0.4cm}
\label{fig:Massless2endpoints}
\end{figure}

\begin{figure}[t!]
\vspace{-.5cm}
\subfigure
{\hspace{-1.25cm}
\includegraphics[width=.57\textwidth]{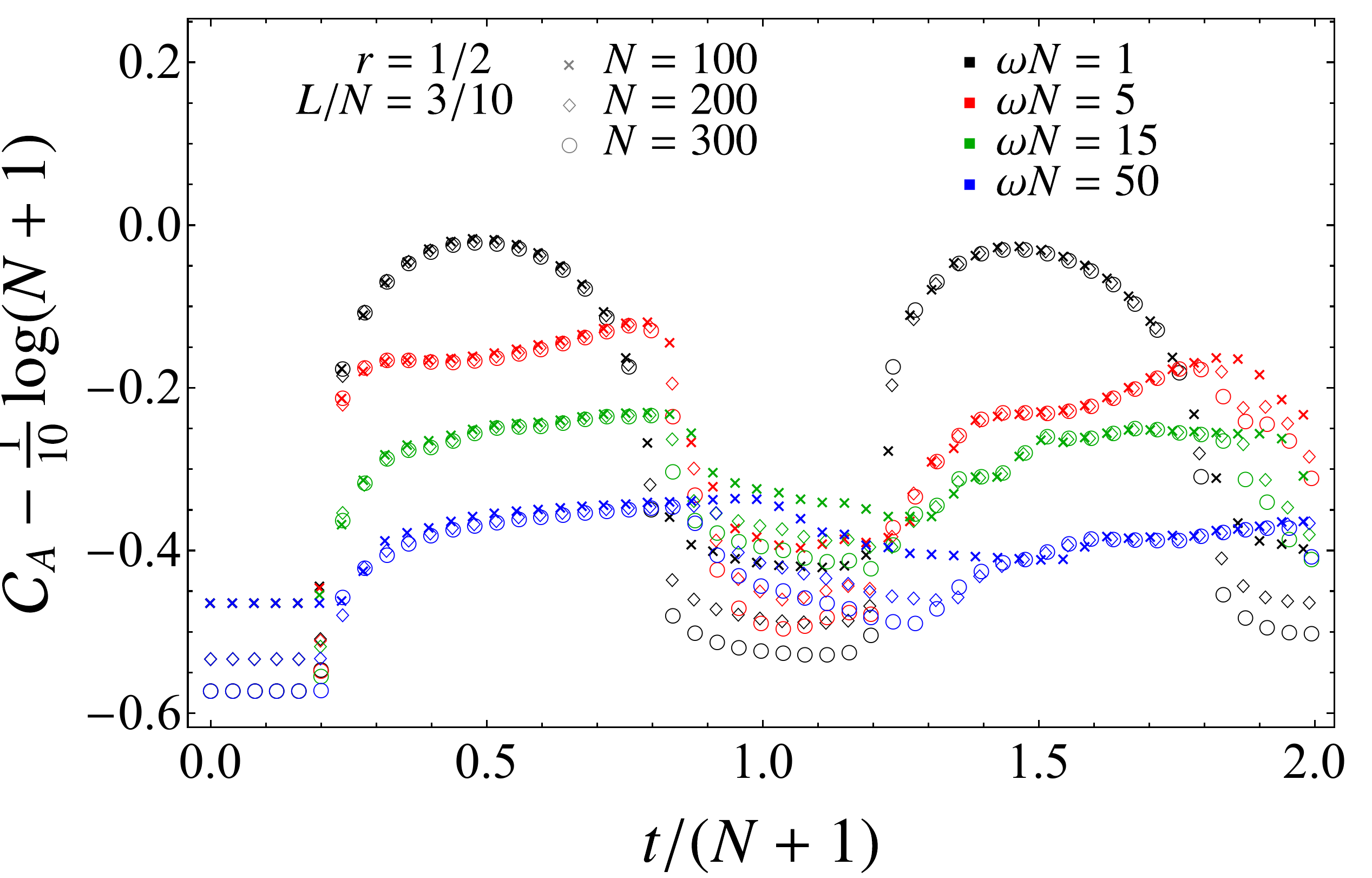}}
\subfigure
{
\hspace{-.0cm}\includegraphics[width=.57\textwidth]{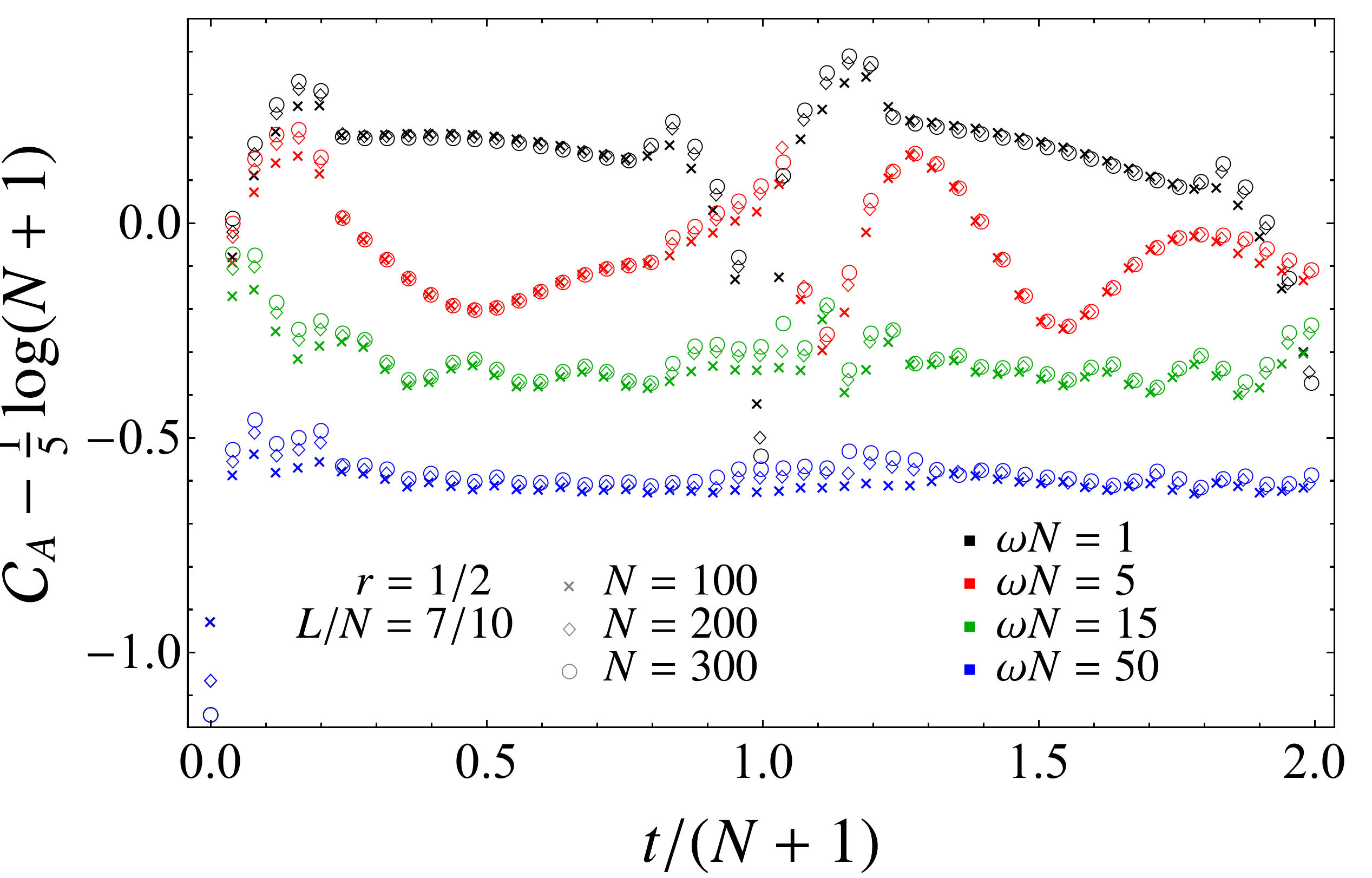}}
\subfigure
{\hspace{-1.25cm}
\includegraphics[width=.57\textwidth]{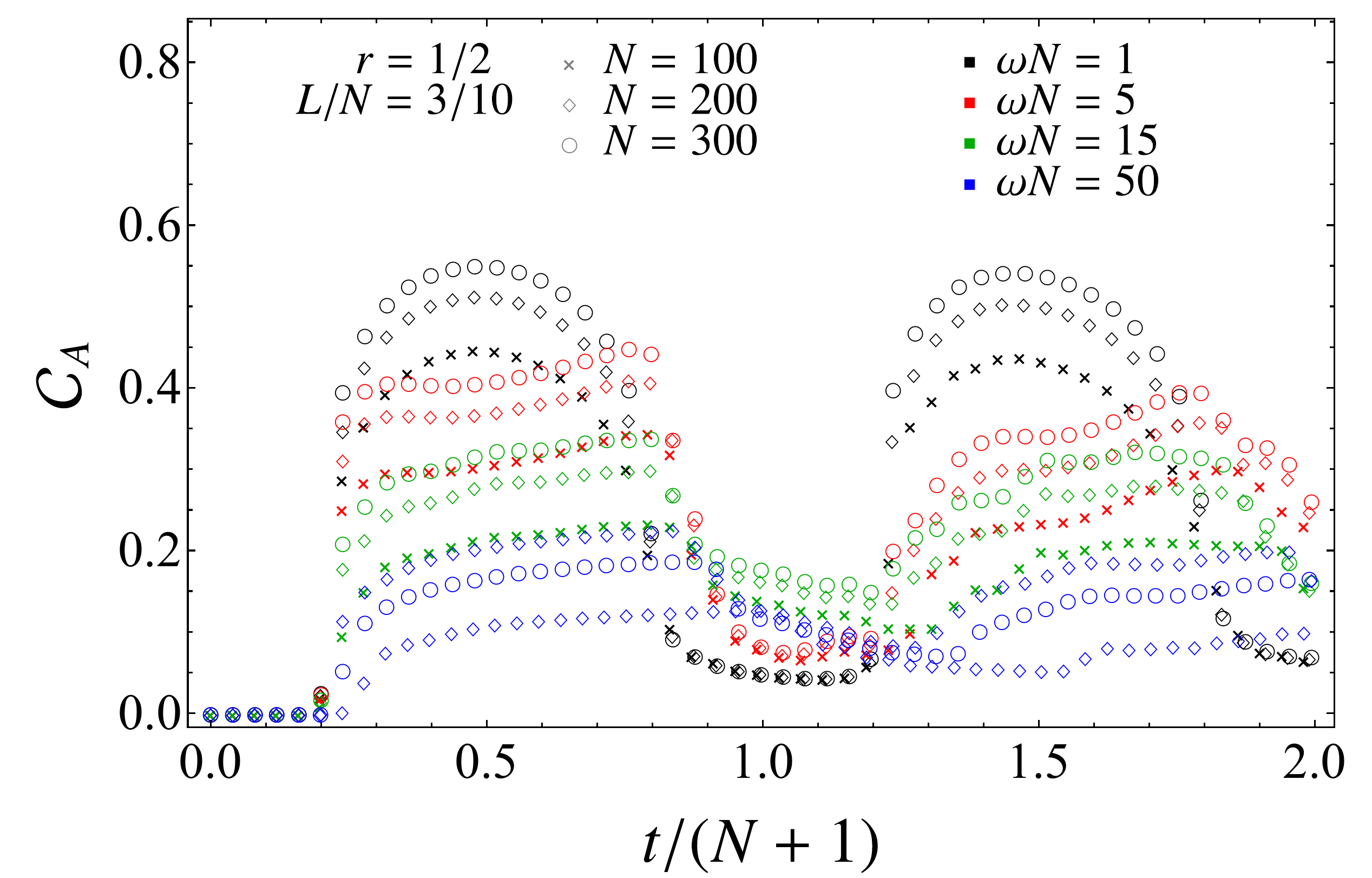}}
\subfigure
{
\hspace{-.0cm}\includegraphics[width=.57\textwidth]{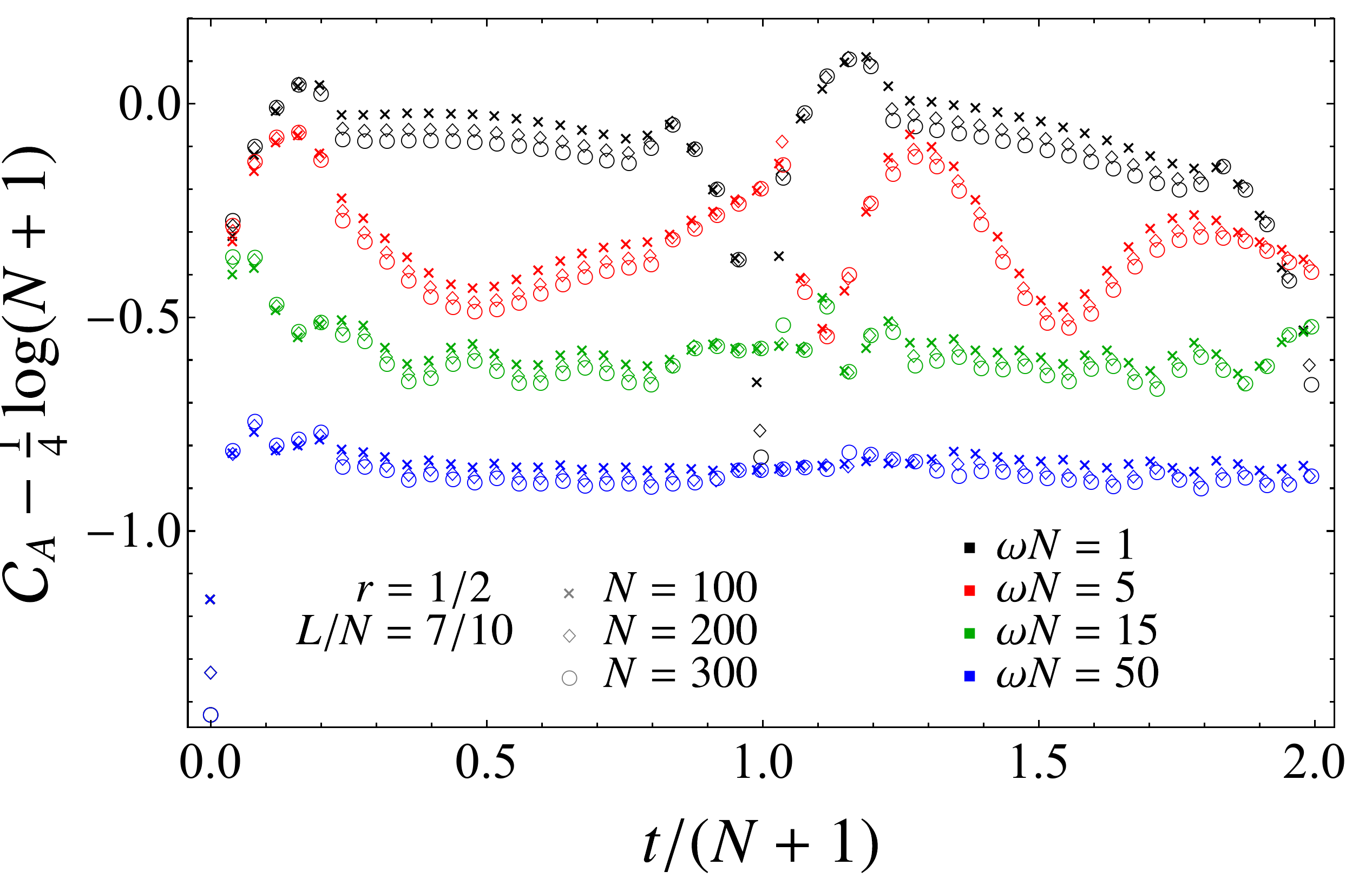}}
\caption{Temporal evolution of the subsystem complexity $\mathcal{C}_A$ 
in (\ref{c2-complexity-rdm-our-case})
for a block $A$ made by $L$ consecutive sites adjacent to the left boundary
of harmonic chains made by $N$ sites 
(see Fig.\,\ref{fig:intro-configs}, bottom left panel)
after a local quench with $\omega  > 0$ and $r=1/2$.
The size of the blocks is fixed to a value $L/N < r$ in the left panels 
and to a value $L/N > r$ in the right panels. 
}
\vspace{0.4cm}
\label{fig:CompMixedMassiver1over2}
\end{figure}

\begin{figure}[t!]
\vspace{-.5cm}
\subfigure
{\hspace{-1.25cm}
\includegraphics[width=.57\textwidth]{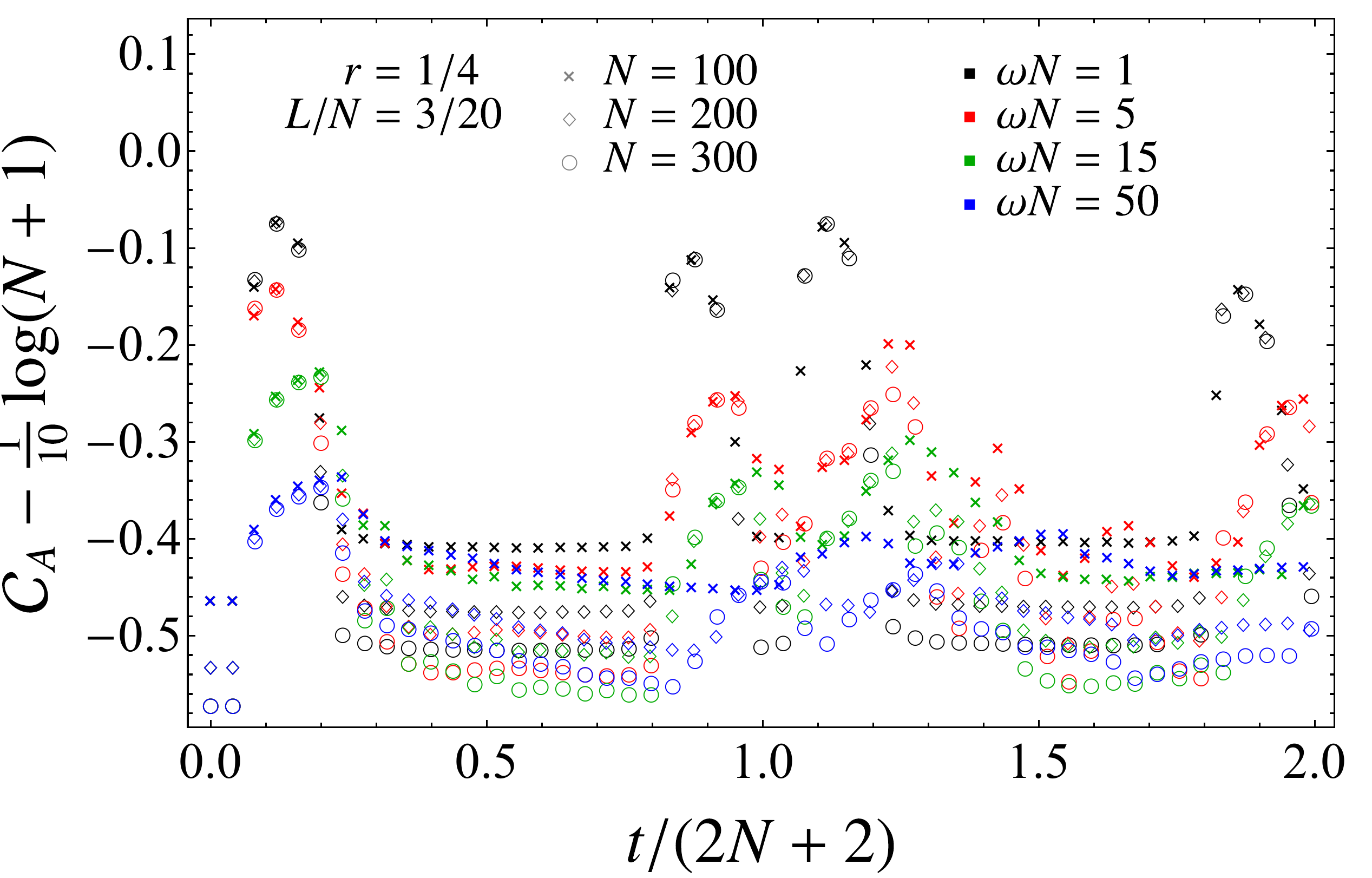}}
\subfigure
{
\hspace{-.0cm}\includegraphics[width=.57\textwidth]{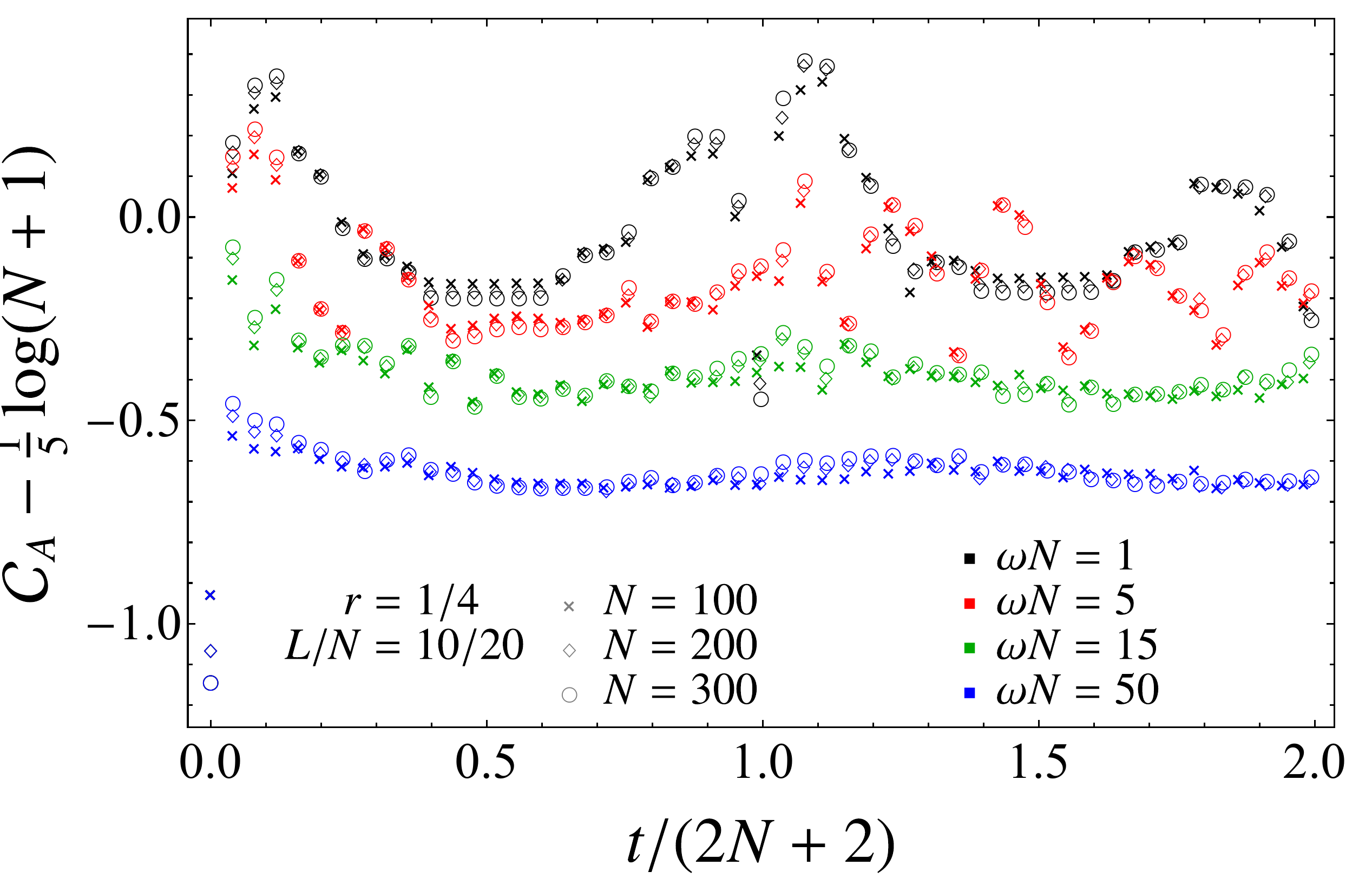}}
\subfigure
{\hspace{-1.25cm}
\includegraphics[width=.57\textwidth]{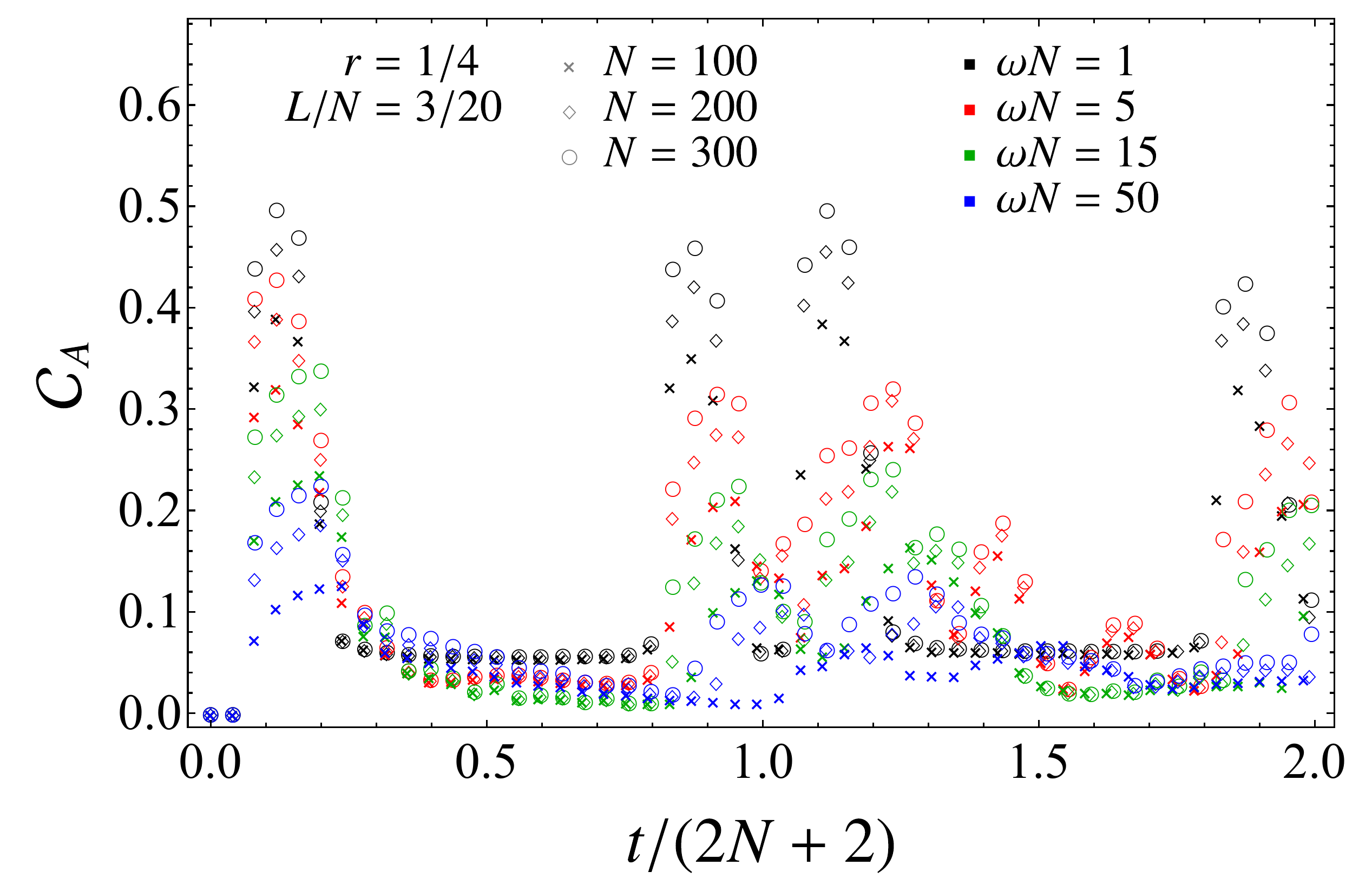}}
\subfigure
{
\hspace{-.0cm}\includegraphics[width=.57\textwidth]{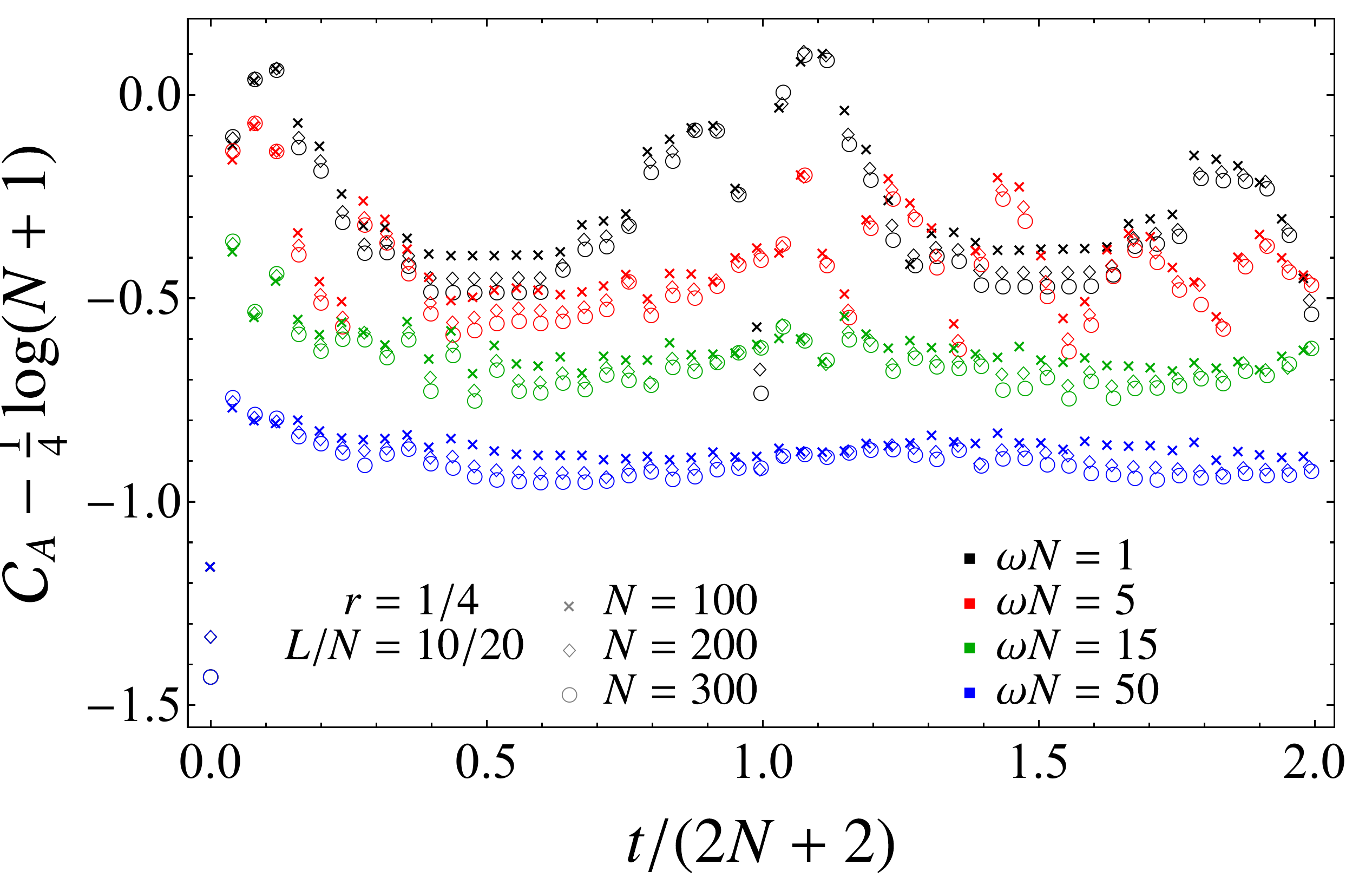}}
\caption{Temporal evolution of the subsystem complexity $\mathcal{C}_A$ 
in (\ref{c2-complexity-rdm-our-case})
for a block $A$ made by $L$ consecutive sites adjacent to the left boundary
of harmonic chains made by $N$ sites 
(see Fig.\,\ref{fig:intro-configs}, bottom left panel)
after a local quench with $\omega  > 0$ and $r=1/4$.
The size of the blocks is fixed to a value $L/N < r$ in the left panels 
and to a value $L/N > r$ in the right panels. 
}
\vspace{0.4cm}
\label{fig:CompMixedMassiver1over4}
\end{figure}

\begin{figure}[t!]
\vspace{-.5cm}
\subfigure
{\hspace{-1.25cm}
\includegraphics[width=.57\textwidth]{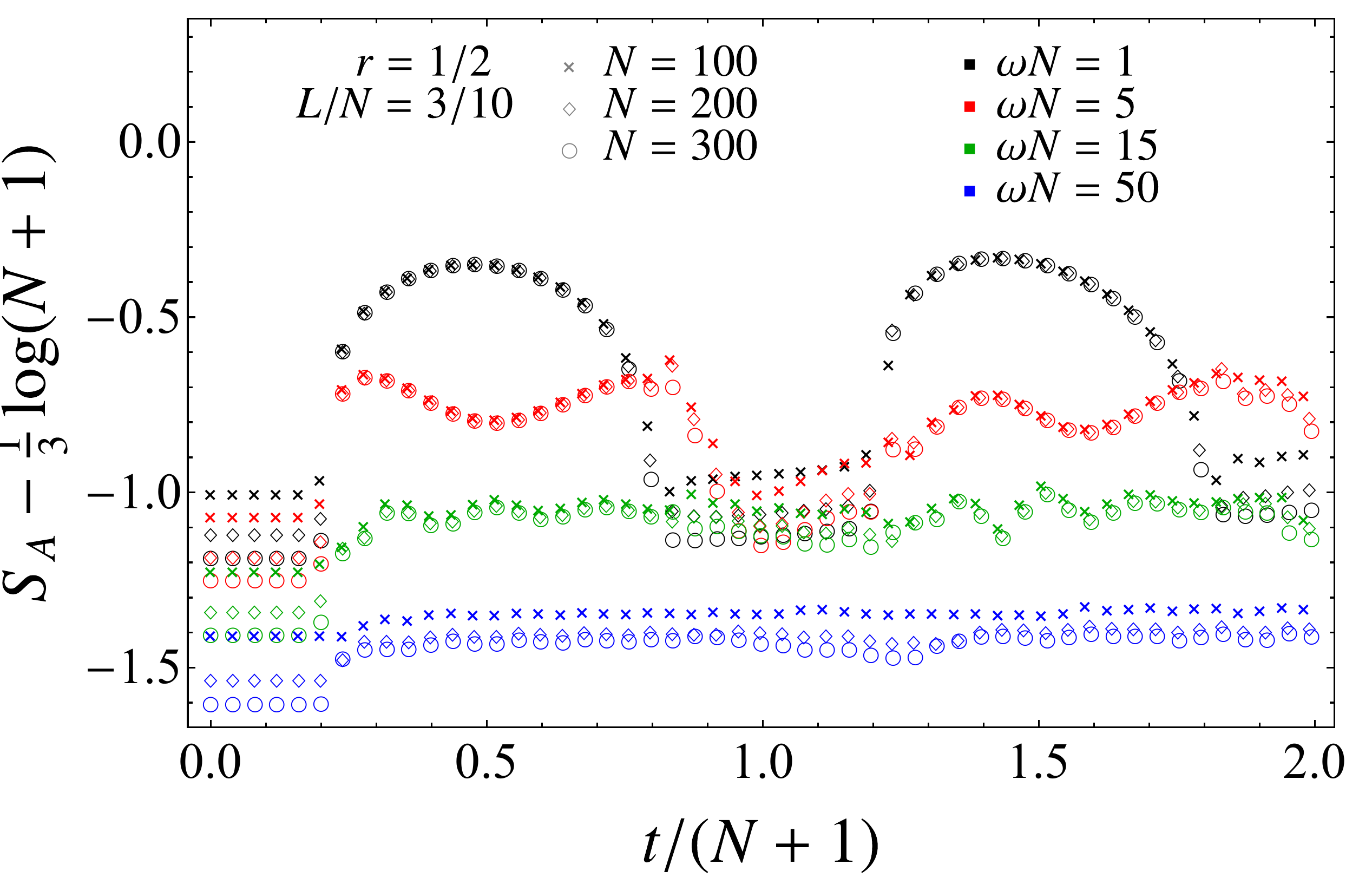}}
\subfigure
{
\hspace{-.0cm}\includegraphics[width=.57\textwidth]{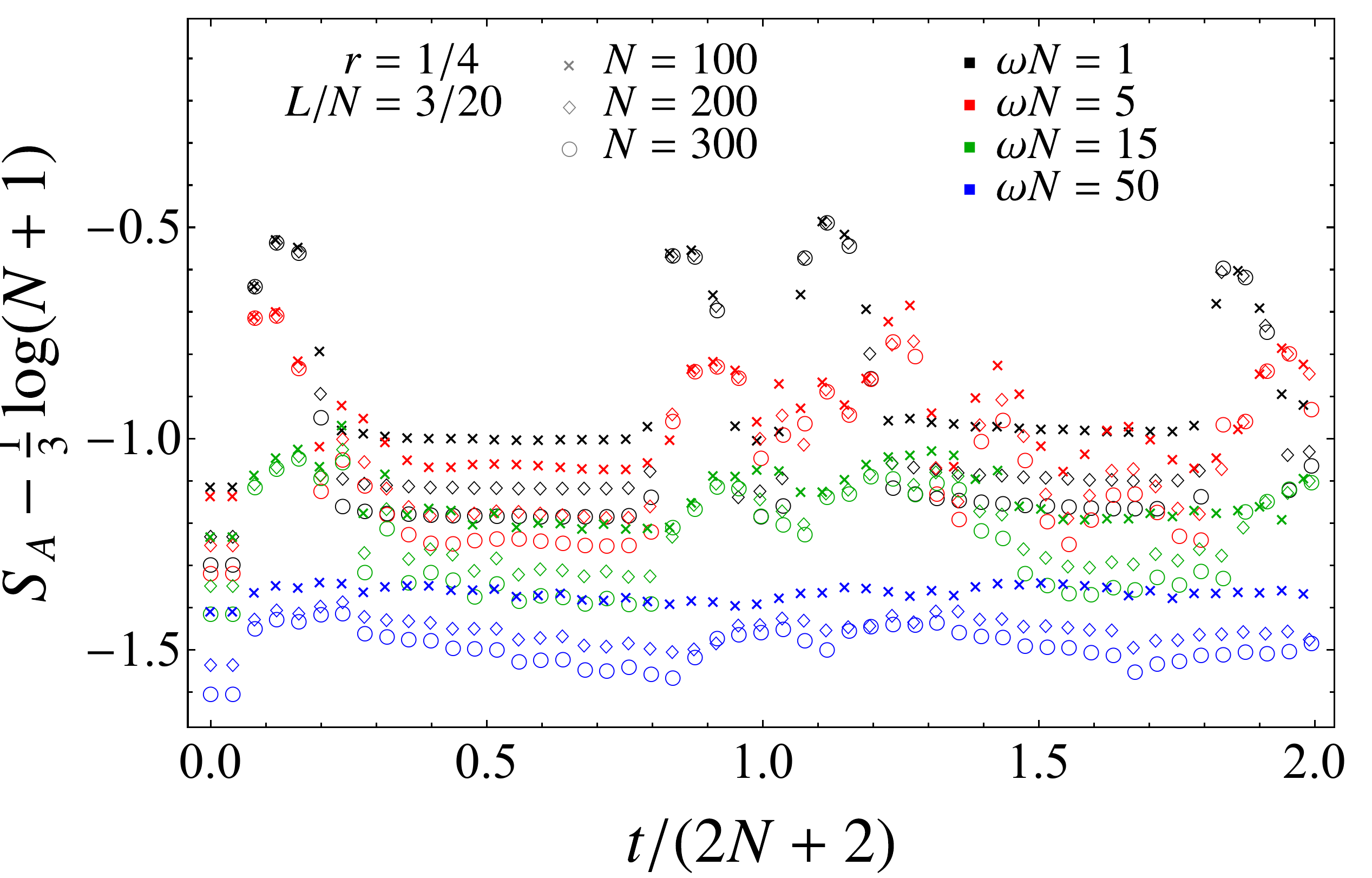}}
\subfigure
{\hspace{-1.25cm}
\includegraphics[width=.57\textwidth]{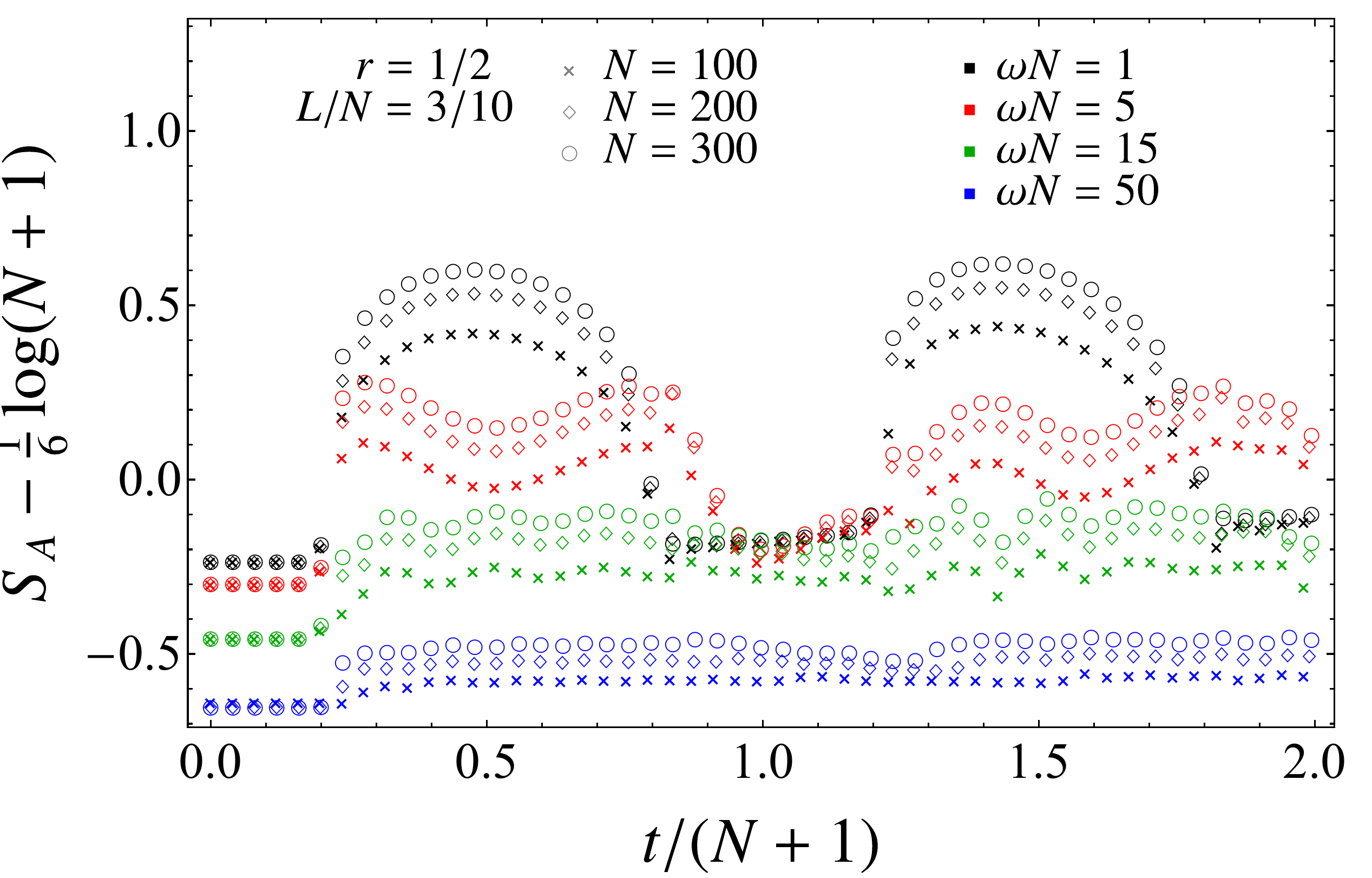}}
\subfigure
{
\hspace{-.0cm}\includegraphics[width=.57\textwidth]{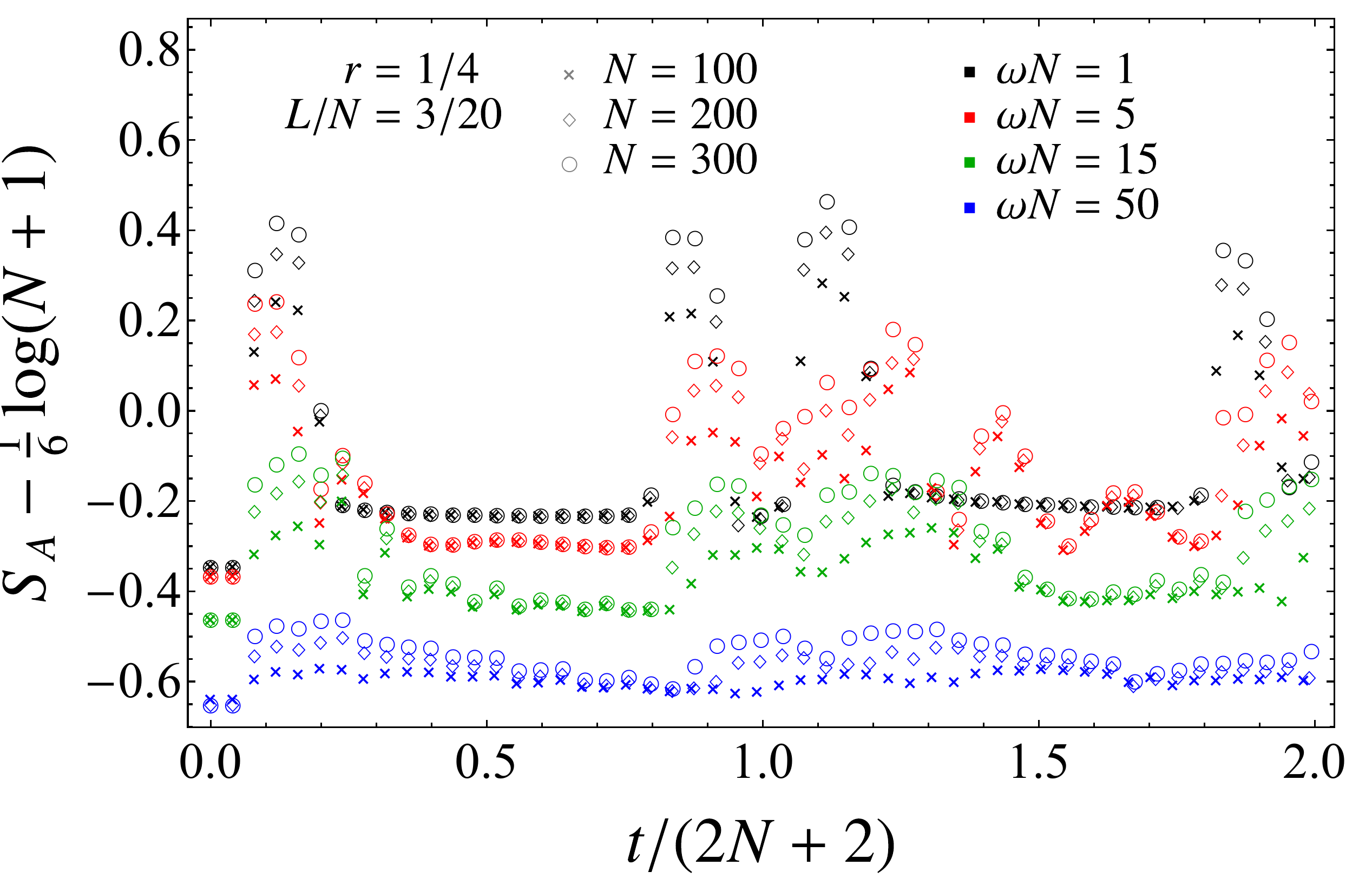}}
\caption{Temporal evolution of the entanglement entropy $S_A$ 
in the setup described either in Fig.\,\ref{fig:CompMixedMassiver1over2} (left panels)
or in Fig.\,\ref{fig:CompMixedMassiver1over4} (right panels).
}
\vspace{0.4cm}
\label{fig:EntrMixedMassive}
\end{figure}

The temporal evolutions in 
Figs.\,\ref{fig:CompMixedMasslessr1over2}, \ref{fig:EntMasslessr1over2}, \ref{fig:CompMixedMasslessr1over4}, 
\ref{fig:EntMasslessr1over4}, \ref{fig:CompMixedMassiver1over2}, \ref{fig:CompMixedMassiver1over4} and \ref{fig:EntrMixedMassive} correspond to blocks $A$ adjacent to the left boundary of the chain
(as shown pictorially in the bottom left panel of Fig.\,\ref{fig:intro-configs})
and for this bipartition the joining point is outside the subsystem whenever $L< r N$.
The temporal evolutions in Fig.\,\ref{fig:Massless2endpoints} correspond to
blocks $A$ whose midpoint coincides with the joining point
(see Fig.\,\ref{fig:intro-configs}, bottom right panel).
While the temporal evolutions in 
Figs.\,\ref{fig:CompMixedMasslessr1over2}, \ref{fig:EntMasslessr1over2}, \ref{fig:CompMixedMasslessr1over4}, \ref{fig:EntMasslessr1over4} and \ref{fig:Massless2endpoints}  
are determined by the critical evolution Hamiltonian,
for the ones in Figs.\,\ref{fig:CompMixedMassiver1over2}, \ref{fig:CompMixedMassiver1over4} 
and \ref{fig:EntrMixedMassive} the evolution Hamiltonian is gapped.

Let us consider first local quenches whose evolution Hamiltonian is critical, i.e. $\omega =0$.
The corresponding temporal evolutions for $\mathcal{C}_A$ and $S_A$ are shown respectively 
in Fig.\,\ref{fig:CompMixedMasslessr1over2} and Fig.\,\ref{fig:EntMasslessr1over2} 
when $r=1/2$
and respectively  in Fig.\,\ref{fig:CompMixedMasslessr1over4} and Fig.\,\ref{fig:EntMasslessr1over4} 
when $r=1/4$.

Both the temporal evolutions of $\mathcal{C}_A$ and $S_A$ exhibit revivals 
because our system has finite volume. 
For a generic values of $r $, 
the cycles correspond to $p<t/(2N+2)<p+1 $, with $p$ non negative integer
(see Fig.\,\ref{fig:CompMixedMasslessr1over4} and Fig.\,\ref{fig:EntMasslessr1over4}),
while only for $r=1/2$ they correspond to $p<t/(N+1)<p+1$
(see Fig.\,\ref{fig:CompMixedMasslessr1over2} and Fig.\,\ref{fig:EntMasslessr1over2})
because of the symmetry provided by the fact that
the joining point coincides with the midpoint of the chain \cite{Cardy:2011zz,Cardy:2014rqa}.

Focussing on the temporal evolution during a single cycle,
as $N$ and $L$ increase with $L/N$ kept fixed, 
two different scalings are observed:
one at the beginning and at the end of the cycle
and another one in its central part.
In these two temporal regimes, 
the curves obtained for different values of $N$ collapse 
when the time independent quantity $\alpha\log(N+1)$
is subtracted, with different values of $\alpha$.

In Fig.\,\ref{fig:CompMixedMasslessr1over2} and Fig.\,\ref{fig:CompMixedMasslessr1over4}
we show some temporal evolutions of $\mathcal{C}_A - \alpha\log(N+1)$
when $r=1/2$ and $r=1/4$ respectively. 
We find that $\alpha$ depends on
(a) whether the joining point is outside ($L<rN$) or inside  ($L> rN$) the subsystem;
(b) the temporal regime within the cycle where the collapse of the data is observed 
(either the central part of the cycle or its extremal parts).

When the entangling point coincides with the joining point, i.e. $L=rN$,
the collapses of the data in the different temporal regimes is observed for values of $\alpha$ 
that are slightly different from the ones adopted in the vertical axes of the panels in
Fig.\,\ref{fig:CompMixedMasslessr1over2} and Fig.\,\ref{fig:CompMixedMasslessr1over4}.
In particular, when $\mathcal{C}_A$ is not constant,
the black curves in the left panels of these figures collapse with $\alpha \simeq 1/7$,
otherwise the data collapse is observed with $\alpha \simeq 1/10$ 
(see Fig.\,\ref{fig:CompMixedMasslessr1over4}, left panels).

The different scalings in the diverse temporal regimes within each cycle pointed out in (b) 
occur also for the temporal evolution of $S_A$ after a local quench
\cite{Eisler_2007,Calabrese:2007mtj,Dubail_2011}. 
Numerical results for the temporal evolution of $S_A$ after the same quench 
and for the same bipartition considered above (see the bottom left panel of Fig.\,\ref{fig:intro-configs})
are reported in Fig.\,\ref{fig:EntMasslessr1over2} and Fig.\,\ref{fig:EntMasslessr1over4}
for $r=1/2$ and $r=1/4$ respectively. 
In the case of $r=1/2$,
these numerical outcomes for $S_A$ are well described 
by the analytic curve discussed in \cite{Dubail_2011},
namely\footnote{See Eq.\,(39) of \cite{Dubail_2011} 
with $c=1$ and $v_\textrm{\tiny F}=1$.}
\be
\label{entropy CFT}
S_A(t)=
\left\{\begin{array}{ll}
\displaystyle
\frac{1}{6}\log(N+1) + a
&
\displaystyle
\hspace{1cm}
\frac{t}{N+1} \in T_0
\\
\rule{0pt}{.9cm}
\displaystyle
\frac{1}{6}\log\Bigg\{
\bigg(\frac{N+1}{\pi}\bigg)^2\,
\bigg[ \bigg(\! \sin\! \bigg(\frac{\pi t}{N+1}\bigg) \! \bigg)^2 \! -  \big( \sin(\pi d) \big)^2 \,\bigg]
\Bigg\}
+\tilde{a}
&
\displaystyle
\hspace{1cm}
\frac{t}{N+1} \in T_1
\end{array}
\right.
\ee
(see the black dashed curves in Fig.\,\ref{fig:EntMasslessr1over2})
within the first cycle (then extended periodically to the subsequent cycles),
where $d\equiv\tfrac{1}{2}-\tfrac{L}{N+1}$ parameterises the distance 
between  the entangling point and the joining point
and
we have introduced  the temporal regimes 
$T_0 \equiv (0,d) \cup (1-d\,, 1) $
and
$T_1 \equiv (d\, , 1-d)$.
The expression (\ref{entropy CFT}) 
holds only when $r=1/2$ and the interval $A$ is adjacent to one of the boundaries of the segment.
The different scalings corresponding to the two different regimes within the cycle,
which lead to subtract 
$\alpha\log(N+1)$ with either $\alpha=1/3$ (top panels) or $\alpha=1/6$ (bottom panels),
agree with (\ref{entropy CFT}).
We remark that, since $r=1/2$, 
the numerical curves in the left panels of Fig.\,\ref{fig:EntMasslessr1over2} 
are identical to the ones in the right panels characterised by the same coloured marker:
this is because the entanglement entropy of a subsystem is equal to the entanglement entropy of its complement
when the entire state is in a pure state
(this is the case for any $t>0$ after the local quench that we are exploring). 

As for the temporal evolution of the subsystem complexity $\mathcal{C}_A$,
when $r=1/2$ and $L<N/2$, hence the joining point is outside the subsystem
(see the left panels of Fig.\,\ref{fig:CompMixedMasslessr1over2}),
we find  that it is qualitatively similar to the temporal evolution of $S_A$.
Combining this observation with the different scalings obtained numerically, 
we are led to consider the following ansatz
\be
\label{complexity guess}
\mathcal{C}_A(t)
=
\left\{\begin{array}{ll}
\displaystyle
0
&
\displaystyle
\hspace{1cm}
\frac{t}{N+1} \in T_0
\\
\rule{0pt}{.9cm}
\displaystyle
\frac{1}{10}\log\Bigg\{
\bigg(\frac{N+1}{\pi}\bigg)^2\,
\bigg[ \bigg(\! \sin\! \bigg(\frac{\pi t}{N+1}\bigg) \! \bigg)^2 \! -  \big( \sin(\pi d) \big)^2 \,\bigg]
\Bigg\}
+\tilde{b}
&
\displaystyle
\hspace{1cm}
\frac{t}{N+1} \in T_1
\end{array}
\right.
\ee
within the first cycle
(the parameter $d$ and the temporal regimes are introduced in (\ref{entropy CFT})),
which is then extended periodically to any value of $t>0$.
In the top left panel of Fig.\,\ref{fig:CompMixedMasslessr1over2},
a remarkable agreement is observed between the numerical data and 
the ansatz (\ref{complexity guess}),
which corresponds to the black dashed curves.

Considering also the right panels of Fig.\,\ref{fig:CompMixedMasslessr1over2}, 
where $r=1/2$ again but $L>N/2$,
we find that the temporal evolutions of $\mathcal{C}_A$ 
for blocks that include the joining point are qualitatively different 
from the ones corresponding to blocks that do not contain the joining point. 
Indeed, in the right panels of Fig.\,\ref{fig:CompMixedMasslessr1over2}, 
focussing e.g. on the first cycle and considering $t/(N+1)<1/2$ 
(the regime $t/(N+1)>1/2$ is obtained straightforwardly through a time reversal), 
we observe three regimes: 
an initial growth until a local maximum, followed by a fast decrease 
and then another growth, milder than the previous one
(it becomes almost flat as $L/N$ increases).

When $r \neq 1/2$, 
the symmetry under a spatial reflection 
with respect to  the midpoint of the chain does not occur
and more regimes are observed within a cycle,
for both $\mathcal{C}_A$ and $S_A$.

The same quantities considered 
in Fig.\,\ref{fig:CompMixedMasslessr1over2} and Fig.\,\ref{fig:EntMasslessr1over2}, where $r=1/2$,
are shown 
in Fig.\,\ref{fig:CompMixedMasslessr1over4} and Fig.\,\ref{fig:EntMasslessr1over4}
for $r=1/4$.
Notice the different periodicity with respect to the case of $r=1/2$,
as already mentioned above. 
The main feature to highlight is the qualitative difference 
between the temporal evolutions of $\mathcal{C}_A$ 
when the joining point lies outside $A$
(see Fig.\,\ref{fig:CompMixedMasslessr1over4}, left panels)
and the ones corresponding to blocks that include the joining point
(see Fig.\,\ref{fig:CompMixedMasslessr1over4}, right panels).
Furthermore, the temporal evolution of $\mathcal{C}_A$ 
when the joining point lies outside $A$
is also qualitatively similar to the one of the corresponding $S_A$ 
(see Fig.\,\ref{fig:EntMasslessr1over4}, left panels).

When $L<rN$, focussing on the temporal evolutions
in the first half of the first cycle (i.e. $0 < \tfrac{t}{2N+2}< 1/2$),
for both $\mathcal{C}_A$ and $S_A$ we observe three regimes
(left panels of Fig.\,\ref{fig:CompMixedMasslessr1over4} and Fig.\,\ref{fig:EntMasslessr1over4}):
first a flat curve, then a growth followed by a decrease 
and finally another regime where the evolution is almost constant.
This means that, when $L<rN$,
for the temporal evolutions within the first cycle we identify five regimes.
The values of $\tfrac{t}{2N+2}$ at which the changes of regime occur are given by
 $rN-L $,  $L+rN $,  $2N-L-rN $ and  $2N-rN+L $,
 whose time ordering depends on the explicit values of $N$, $r$ and $L$.
 In the special case of $r=1/2$,  we have only three regimes within the first cycle
 (first a flat regime, then a growth/decrease regime and finally another flat regime),
 as one can observe from 
 Fig.\,\ref{fig:CompMixedMasslessr1over2} and Fig.\,\ref{fig:EntMasslessr1over2},
 but also from the analytic expressions in (\ref{entropy CFT}) and (\ref{complexity guess}).

When $L>rN$ and therefore the joining point is inside the subsystem, 
by comparing the right panels of Fig.\,\ref{fig:CompMixedMasslessr1over4} 
against the right panels of Fig.\,\ref{fig:EntMasslessr1over4},
it is straightforward to realise that
the temporal evolutions of $\mathcal{C}_A$ and $S_A$ are qualitatively very different.
In particular, 
while the temporal evolutions of $S_A$ in the right panels of Fig.\,\ref{fig:EntMasslessr1over4}
are similar to the ones displayed in the left panels of the same figure
(e.g. the same five regimes mentioned above),
as expected from the fact that $S_A = S_B$ for the spatial bipartition $A \cup B$ of the system in a pure state
(the qualitative difference is only due to the asymmetric position of the joining point),
the temporal evolutions of $\mathcal{C}_A$ in the right panels of Fig.\,\ref{fig:CompMixedMasslessr1over4}
are more complicated than the ones in the left panels of the same figure, 
which correspond to blocks that do not include the joining point. 
For instance, considering the temporal evolution of $\mathcal{C}_A$  immediately after the quench,
a rapid initial growth is observed when $L>rN$ 
(highlighted in the insets in the right panels of Fig.\,\ref{fig:CompMixedMasslessr1over4}),
while it remains stationary when $L<rN$.
We remind that, whenever $L \neq rN$, 
also  the temporal evolution of $S_A$  right after the quench remains stationary
(see Fig.\,\ref{fig:EntMasslessr1over2} and Fig.\,\ref{fig:EntMasslessr1over4}).
As for initial growth of $\mathcal{C}_A$ when $L>rN$, 
an interesting numerical observation that we find it worth remarking is the fact that
the logarithmic curve  (\ref{guess complexity initial}),
which has been first employed in the bottom panels of Fig.\,\ref{fig:PureStateCritical}
to describe the logarithmic growth for the complexity of the entire chain, 
occurs also in the temporal evolution of the subsystem complexity;
indeed it corresponds also to the dashed lines displayed in the 
bottom right panels of Fig.\,\ref{fig:CompMixedMasslessr1over2} and Fig.\,\ref{fig:CompMixedMasslessr1over4}.
Notice that, when the joining point is outside the subsystem
(see the left panels of Fig.\,\ref{fig:CompMixedMasslessr1over2} and Fig.\,\ref{fig:CompMixedMasslessr1over4}),
a logarithmic growth right after the stationary regime is observed,
but in this case the coefficient of the logarithm is different from the one in (\ref{guess complexity initial}),
as one can infer from the second line of (\ref{complexity guess}) when $r=1/2$ and  $t/(N+1)\simeq d$.

Comparing the left and the right panels in Fig.\,\ref{fig:CompMixedMasslessr1over4},
it is straightforward to observe that
the qualitative behaviour of temporal evolutions of $\mathcal{C}_A$ 
is more complicated when the joining point lies inside $A$, as anticipated above.
For instance,  focussing on $t/(2N+2)<1/2$ for the magenta and the cyan curves, 
after the first local maximum two subsequent temporal regimes occur where the curves decrease before becoming constant. 
Furthermore, when $L/N>1-r$  (see the orange curves) and considering again only $t/(2N+2)<1/2$,
after the first local maximum, the two decreases and the flat regime mentioned above,
we observe another growth followed by a regime where $\mathcal{C}_A$ becomes constant again
(at a higher value w.r.t. the previous flat regime).
We remark that, when $L/N \gtrsim r$ (see the brown curves), two local maxima occur in the temporal evolution of 
$\mathcal{C}_A$ for $\tfrac{t}{2N+2} < \tfrac{1}{2}$
(the first one is highlighted in the insets).
A more systematic analysis is needed to determine  the values of $t/(2N+2)$ that
identify the various regimes occurring in these temporal evolutions.

While in Figs.\,\ref{fig:CompMixedMasslessr1over2}, \ref{fig:EntMasslessr1over2},
\ref{fig:CompMixedMasslessr1over4} and \ref{fig:EntMasslessr1over4}
the block $A$ is adjacent to a boundary (see Fig.\,\ref{fig:intro-configs}, bottom left panel)
and therefore only one entangling point occurs,
in Fig.\,\ref{fig:Massless2endpoints}
we consider some temporal evolutions of $\mathcal{C}_A$ and $S_A$
when $r=1/2$ and the joining point coincides with the midpoint of $A$
(see Fig.\,\ref{fig:intro-configs}, bottom right panel),
hence two entangling points separate $A$ from its complement $B$,
which is made by two disjoint intervals adjacent to different boundaries.
By construction,  for this configuration the joining point is always inside the subsystem.
The blocks providing the reduced covariance matrix (\ref{reduced CM})
for this bipartition are obtained by restricting the indices 
of the matrices $Q$, $P$ and $M$  in (\ref{CM block decomposed}) 
to $i,j\in \big\{\frac{N}{2}-\frac{L}{2}+1,\dots,\frac{N}{2}+\frac{L}{2}\big\}$. 

The numerical results shown in Fig.\,\ref{fig:Massless2endpoints}
for some temporal evolutions of $\mathcal{C}_A$ (left panels) and $S_A$ (right panels) 
after local quenches correspond to critical evolution Hamiltonians, i.e. with $\omega=0$.
Also in this numerical analysis 
we subtract $\alpha\log(N+1)$ with the proper value of $\alpha$,
in order to observe collapses of data sets corresponding to the same $L/N$ when $N$ is large enough,
finding that  $\alpha$ depends both on the quantity  (either $\mathcal{C}_A$ or $S_A$) 
and on the temporal regime within the cycle where the data collapses are observed
(either the central regime or the initial and final regimes).
Interestingly, by comparing the left panels of Fig.\,\ref{fig:Massless2endpoints} against 
the right panels of Fig.\,\ref{fig:CompMixedMasslessr1over2},
we observe that,
when the joining point is inside the subsystem $A$,
the temporal evolutions of $\mathcal{C}_A$ are qualitatively very similar, 
despite the fact that the number of entangling points is different in the two figures. 
Moreover,  the values of $\alpha$ employed are the same,
which are therefore independent of the number of the entangling points.
Instead, let us remind that the values of $\alpha$ to employ for $S_A$ depend on the number of the entangling points,
as one realises by comparing
the right panels of Fig.\,\ref{fig:Massless2endpoints} against
the right panels of Fig.\,\ref{fig:EntMasslessr1over2}.

Focussing on bipartitions where the joining point lies inside the subsystem,
by comparing the left and the right panels of Fig.\,\ref{fig:Massless2endpoints},
one notices that, while $S_A$ is constant at the beginning of its evolution, 
$\mathcal{C}_A$ increases immediately. 
This feature has been highlighted also during the comparison 
of the right panels of Fig.\,\ref{fig:CompMixedMasslessr1over2} and Fig.\,\ref{fig:CompMixedMasslessr1over4} 
against the right panels of Fig.\,\ref{fig:EntMasslessr1over2} and Fig.\,\ref{fig:EntMasslessr1over4},
where only one entangling point occurs.

Another interesting difference between the temporal evolutions 
corresponding to the two bottom panels in Fig.\,\ref{fig:intro-configs}
is that the first local minimum occurs at $\tfrac{t}{N+1} \simeq \tfrac{L}{N} - \tfrac{1}{2} $ 
in the right panels of Fig.\,\ref{fig:CompMixedMasslessr1over2} (one entangling point)
and at $\tfrac{t}{N+1} \simeq \tfrac{L}{2N}$
in the left panels of Fig.\,\ref{fig:Massless2endpoints} (two entangling points).

We find it worth mentioning some intriguing similarities between 
the temporal evolution of  $\mathcal{C}_A$ after the local quench discussed above and 
the one after the global quench studied in \cite{DiGiulio:2021oal}.
Let us consider the block made by $L$ consecutive sites in the infinite chain
and compare the temporal evolution of  $\mathcal{C}_A$ 
after the global quench of the mass parameter, as done in \cite{DiGiulio:2021oal},
against the one after the local quench where two half-lines are joined at the midpoint of $A$.
The latter temporal evolution can be inferred by taking 
e.g. the red curves in the left panels of Fig.\,\ref{fig:Massless2endpoints} 
for $\tfrac{t}{N+1} < \tfrac{1}{2}$,
while the former one corresponds e.g. to the black data points in the top panel of Fig.\,14 of \cite{DiGiulio:2021oal}.
These temporal evolutions are qualitatively very similar.
However, important differences occur when these curves are studied quantitatively. 
For instance, while at the beginning a logarithmic growth is observed
in the case of the local quench, as remarked above, 
a power law behaviour occurs in the case of the global quench \cite{DiGiulio:2021oal}.
Notice that, by performing the same comparison 
for the temporal evolutions of the corresponding $S_A$, 
qualitatively different behaviours are observed
(see the red data points in the right panels of Fig.\,\ref{fig:Massless2endpoints}
against the black data points in the bottom panel of Fig.\,14 in \cite{DiGiulio:2021oal}).
It would be interesting to explore further these comparisons 
by considering different kind of quenches 
and performing a quantitative analysis.

In the final part of this discussion
we consider temporal evolutions of $\mathcal{C}_A$ and $S_A$
after local quenches characterised by gapped evolution Hamiltonians,
for some fixed values of $\omega N > 0$.

Since for $\omega =0 $ the qualitative behaviour of the temporal evolution of $\mathcal{C}_A$
depends on whether the joining point is located inside or outside the block $A$,
let us explore these two cases also when $\omega >0$.
Considering the bipartition shown in the bottom left panel of Fig.\,\ref{fig:intro-configs},
where $r=1/2$, 
in Fig.\,\ref{fig:CompMixedMassiver1over2} 
we display some temporal evolutions of $\mathcal{C}_A$ 
for two fixed values of $L/N$ 
such that the joining point is either outside (left panels) or inside (right panels) the block $A$. 
In Fig.\,\ref{fig:CompMixedMassiver1over4}
the same analysis is performed in the case of $r=1/4$.
These numerical results show that the temporal evolution of $\mathcal{C}_A$
depends on whether the joining point is inside or outside the subsystem.
The temporal evolutions of $S_A$ for these quenches 
are reported in  Fig.\,\ref{fig:EntrMixedMassive} and,
since $S_A = S_B$ for any $t>0$ (the entire chain is $A\cup B$),
whether the joining point is inside or outside the block does
not influence the qualitative temporal evolution of $S_A$,
as already remarked above
(once the eventual asymmetric position of the joining point is taken into account).

The approximate periodicity highlighted in the evolutions corresponding to $\omega=0$
is not observed in general when $\omega >0$.
For small values of $\omega N$ an approximate periodicity can be identified
for a temporal regime whose duration decreases as $\omega N$ increases. 

When the block $A$ contains the joining point, $\mathcal{C}_A$  has a non-trivial initial growth, 
while the evolution of the corresponding $S_A$ is constant at the beginning. 
This is the same feature highlighted for the critical evolution through the comparison 
of Fig.\,\ref{fig:CompMixedMasslessr1over2} against Fig.\,\ref{fig:EntMasslessr1over2}, 
of Fig.\,\ref{fig:CompMixedMasslessr1over4} against  Fig.\,\ref{fig:EntMasslessr1over4} 
and of the left panels of Fig.\,\ref{fig:Massless2endpoints} against the right panels of the same figure.

Approximate collapses of the data points corresponding to large values of $N$ while $L/N$ is kept fixed
are observed when the constant $\alpha\log(N+1)$ is subtracted,
with the same values of $\alpha$ employed  
in Fig.\,\ref{fig:CompMixedMasslessr1over2} and Fig.\,\ref{fig:CompMixedMasslessr1over4} for the subsystem complexity 
and in Fig.\,\ref{fig:EntMasslessr1over2} and Fig.\,\ref{fig:EntMasslessr1over4} for the entanglement entropy.
Because of the absence of clear revivals, 
it is more difficult to identify different temporal regimes
as done in the case of critical evolution Hamiltonians. 
These difficulties arise also in the analysis of the temporal evolutions of $S_A$ when $\omega >0$
displayed in Fig.\,\ref{fig:EntrMixedMassive}.
\\

Let us conclude our discussion by mentioning some results about the 
temporal evolutions of the subsystem complexity 
obtained within the gauge/gravity correspondence
\cite{Chen:2018mcc,Auzzi:2019mah,Ling:2019ien,Zhou:2019xzc,Ageev:2018nye}.

In the Vaidya gravitational spacetimes,
the temporal evolution of the holographic subsystem complexity 
has been studied through the prescription based on the volume 
of a particular spacetime slice \cite{Chen:2018mcc,Auzzi:2019mah},
finding curves that qualitatively agree with the 
temporal evolution of the subsystem complexity after a global quench
of the mass parameter in the harmonic chains discussed in \cite{DiGiulio:2021oal} . 

It would be interesting to perform a comparison between 
the qualitative behaviour of the temporal evolutions of the subsystem complexity 
discussed in this manuscript and
the one of the temporal evolutions of the holographic subsystem complexity 
in the spacetime describing the gravitational dual of the joining local quench
\cite{Ugajin:2013xxa,Shimaji:2018czt}.

\subsection{Single site in the chain made by two sites}
\label{subsec:single-site-comp}

In the following we discuss the temporal evolution of $\mathcal{C}_A$ 
after the local quench that we are exploring 
for the chain made by two sites, described in Sec.\,\ref{subsec:twositespure},
and the subsystem made by a single site.

 From (\ref{gamma0 2 sites}) and  (\ref{VEV block dec})
we have that the covariance matrices of the reference and the target states are respectively
\be
\label{gamma T gamma R 2 sites}
\gamma_{\textrm{\tiny R}}
=
\Gamma_{0}
\;\;\qquad\;\;
\gamma_{\textrm{\tiny T}}
=
V \, \mathcal{E}\, V \,
\Gamma_{0} \,
V \, \mathcal{E}^{\textrm{t}} \, V\,.
\ee
The blocks occurring in (\ref{VEV block dec}) are $2\times 2$ matrices 
that can be written as
\be
\label{VFV 2times2}
\widetilde{V}_2 \, \mathcal{F} \, \widetilde{V}_2
\,=\,
\frac{1}{2}
\bigg( \begin{array}{cc}
\mathcal{F}_1+\mathcal{F}_2\;\;  & \mathcal{F}_1-\mathcal{F}_2
\\
\mathcal{F}_1-\mathcal{F}_2  \;\;  & \mathcal{F}_1+\mathcal{F}_2
\end{array} \bigg)
\;\;\;\qquad\;\;\;
\mathcal{F}_j \in \big\{ \mathcal{A}_j , \mathcal{B}_j, \mathcal{D}_j  \big\}
\;\qquad\;
j \in \{1,2\}
\ee 
where the expressions for $\mathcal{A}_{j}$, $\mathcal{B}_{j}$ and $\mathcal{D}_{j}$ 
are obtained by specifying  (\ref{blocks diagonal matrices}) to $N=2$
and $\widetilde{V}_2$ has been defined in (\ref{Vtilde few sites}).

Considering the subsystem $A$ made by the first oscillator of the chain, 
for the 
$2\times 2$ reduced covariance matrices
of the reference and of the target states we find respectively
\be
\gamma_{\textrm{\tiny R},A}=
\frac{1}{2}\,\textrm{diag}
\Big( \big[ m\Omega^{(1)}_1\big]^{-1} , \,m\Omega^{(1)}_1\Big)
\ee
and
\be
\gamma_{\textrm{\tiny T},A}
\,=\,
\frac{1}{2 m \Omega^{(1)}_1}\,
\left(
\begin{array}{cc}
\mathcal{D}_+^2+\mathcal{D}_-^2
+
\big(m\Omega^{(1)}_1\big)^2
\big( \mathcal{A}_+^2+\mathcal{A}_-^2\big)
  & 
\Pi_{\mathcal{B}}+\big(m\Omega^{(1)}_1\big)^2 \,\Pi_{\mathcal{A}}
\\
\rule{0pt}{.6cm}
\Pi_{\mathcal{B}}+\big(m\Omega^{(1)}_1\big)^2\, \Pi_{\mathcal{A}}
  & 
  \mathcal{B}_+^2+\mathcal{B}_-^2
  +\big(m\Omega^{(1)}_1\big)^2
  \big[\mathcal{D}_+^2+\mathcal{D}_-^2\big]
\end{array}
\,\right)
\ee
where $\Omega^{(1)}_1$ has been defined in  (\ref{gamma0 2 sites})
and we have introduced
$\mathcal{A}_\pm \equiv \mathcal{A}_1 \pm \mathcal{A}_2$\,, 
$\mathcal{B}_\pm \equiv \mathcal{B}_1 \pm \mathcal{B}_2$
and 
$\mathcal{D}_\pm \equiv \mathcal{D}_1 \pm \mathcal{D}_2$\,, 
which allow to construct 
\be
\label{PiA PiB def}
\Pi_{\mathcal{A}}\equiv \mathcal{A}_-\mathcal{D}_-+\mathcal{A}_+\mathcal{D}_+
\;\;\qquad\;\;
\Pi_{\mathcal{B}}\equiv \mathcal{B}_-\mathcal{D}_-+\mathcal{B}_+\mathcal{D}_+\,.
\ee
The quantities $\mathcal{A}_{j}$, $\mathcal{B}_{j}$ and $\mathcal{D}_{j}$, with $j \in \{1,2\}$,
depend on $t$ and on the parameters of the local quench $m$, $\kappa$ and $\omega$
as reported in (\ref{blocks diagonal matrices}).
The eigenvalues of the matrix 
$\gamma_{\textrm{\tiny T},A} \,\gamma_{\textrm{\tiny R},A}^{-1} $ 
can be written in terms of these quantities
as follows
\bea
\label{gTR reduced relativeCM}
g_{\textrm{\tiny TR},\pm}
&=& 
\mathcal{D}_+^2+\mathcal{D}_-^2
+\frac{1}{2\big(m\Omega^{(1)}_1\big)^2} \;
\Bigg\{\,
\mathcal{B}_+^2+\mathcal{B}_-^2
+
\big(m\Omega^{(1)}_1\big)^4
\big(\mathcal{A}_+^2+\mathcal{A}_-^2\big)
\\
\rule{0pt}{.8cm}
&&
\pm\,
\sqrt{
\Big[
\mathcal{B}_+^2+\mathcal{B}_-^2
-
\big(m\Omega^{(1)}_1\big)^4
\big(\mathcal{A}_+^2+\mathcal{A}_-^2\big)
\Big]^2
+
\big(2 m\Omega^{(1)}_1\big)^2
\Big[\Pi_{\mathcal{B}} +\big(m\Omega^{(1)}_1\big)^2\, \Pi_{\mathcal{A}} \Big]^2
}\,
\; \Bigg\}\,.
\nonumber
\eea
By employing these results into (\ref{c2-complexity-rdm-our-case}), 
we obtain the subsystem complexity
\be
\label{ComplexityMixed 2 sites}
\mathcal{C}_A
\,=\,
\frac{1}{2\sqrt{2}} \, \sqrt{\big[\log g_{\textrm{\tiny TR},+}\big]^2+\big[\log g_{\textrm{\tiny TR},-}\big]^2}
\ee
whose explicit expression in terms of $m$, $\kappa$, $\omega$ and $t$ is quite cumbersome; 
hence we have not reported it here.

\begin{figure}[t!]
\vspace{-.7cm}
\subfigure
{\hspace{-1.25cm}
\includegraphics[width=.57\textwidth]{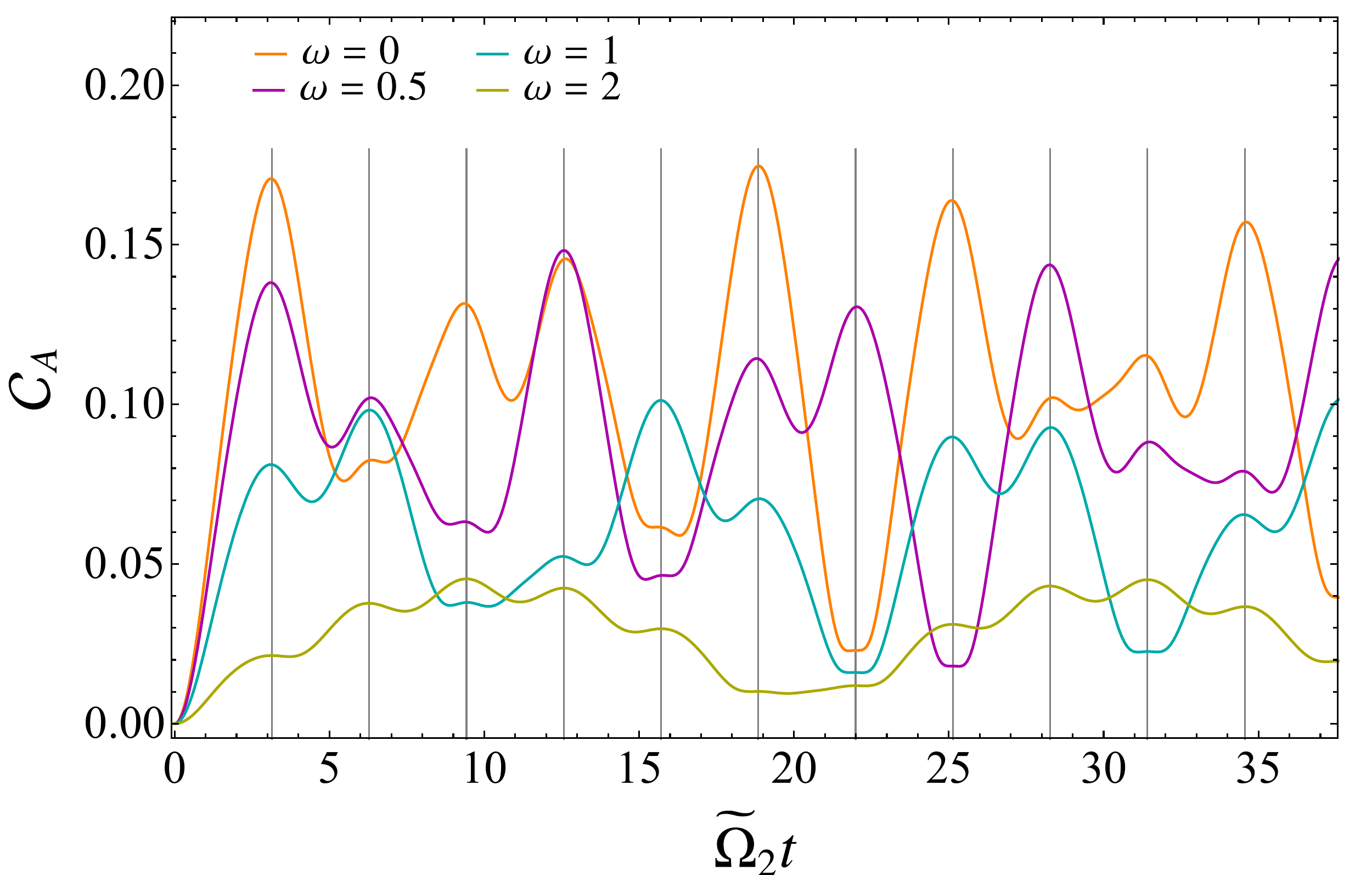}}
\subfigure
{
\hspace{0.cm}\includegraphics[width=.57\textwidth]{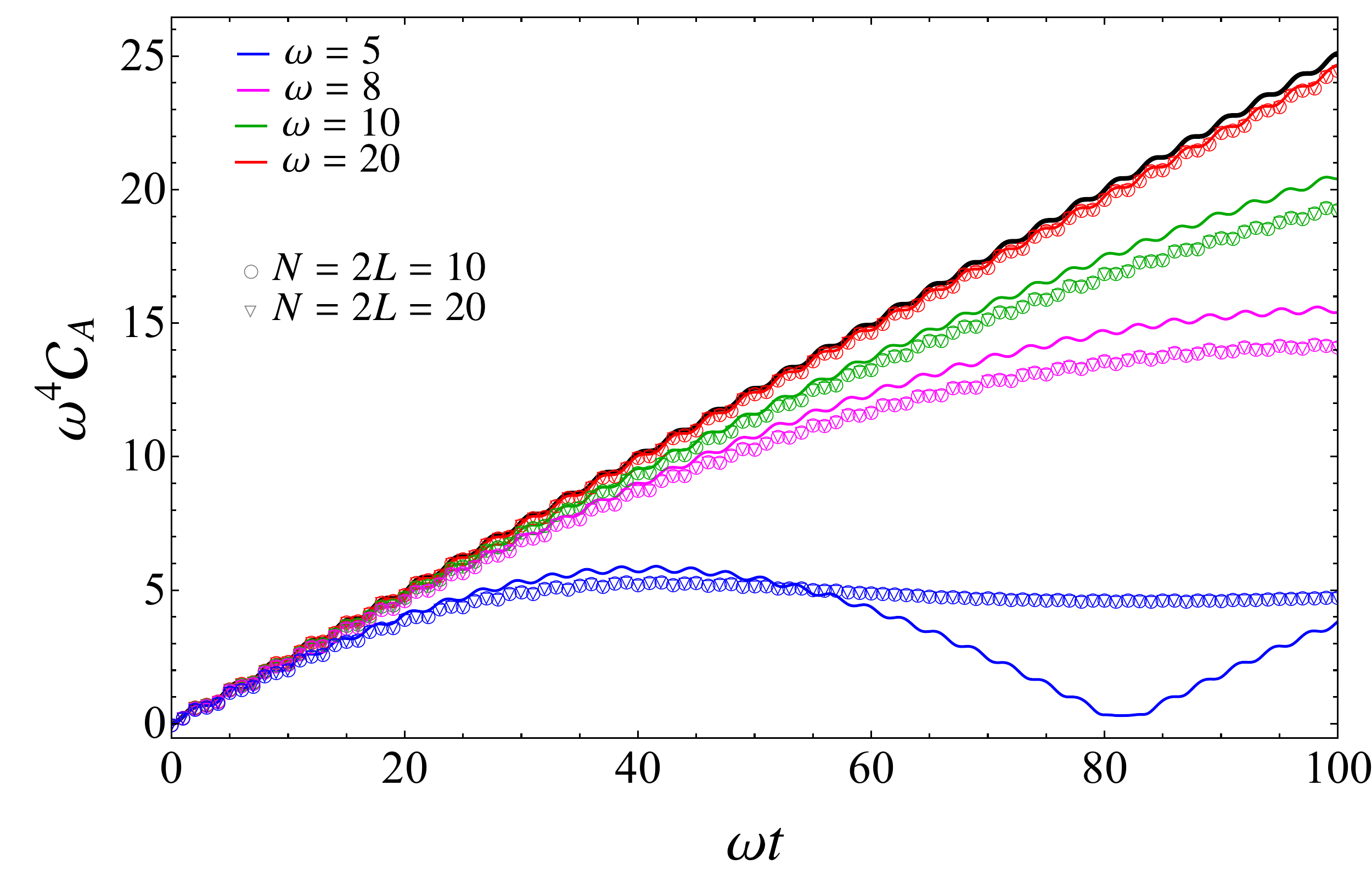}}
\\
\subfigure
{
\hspace{3.25cm}\includegraphics[width=.57\textwidth]{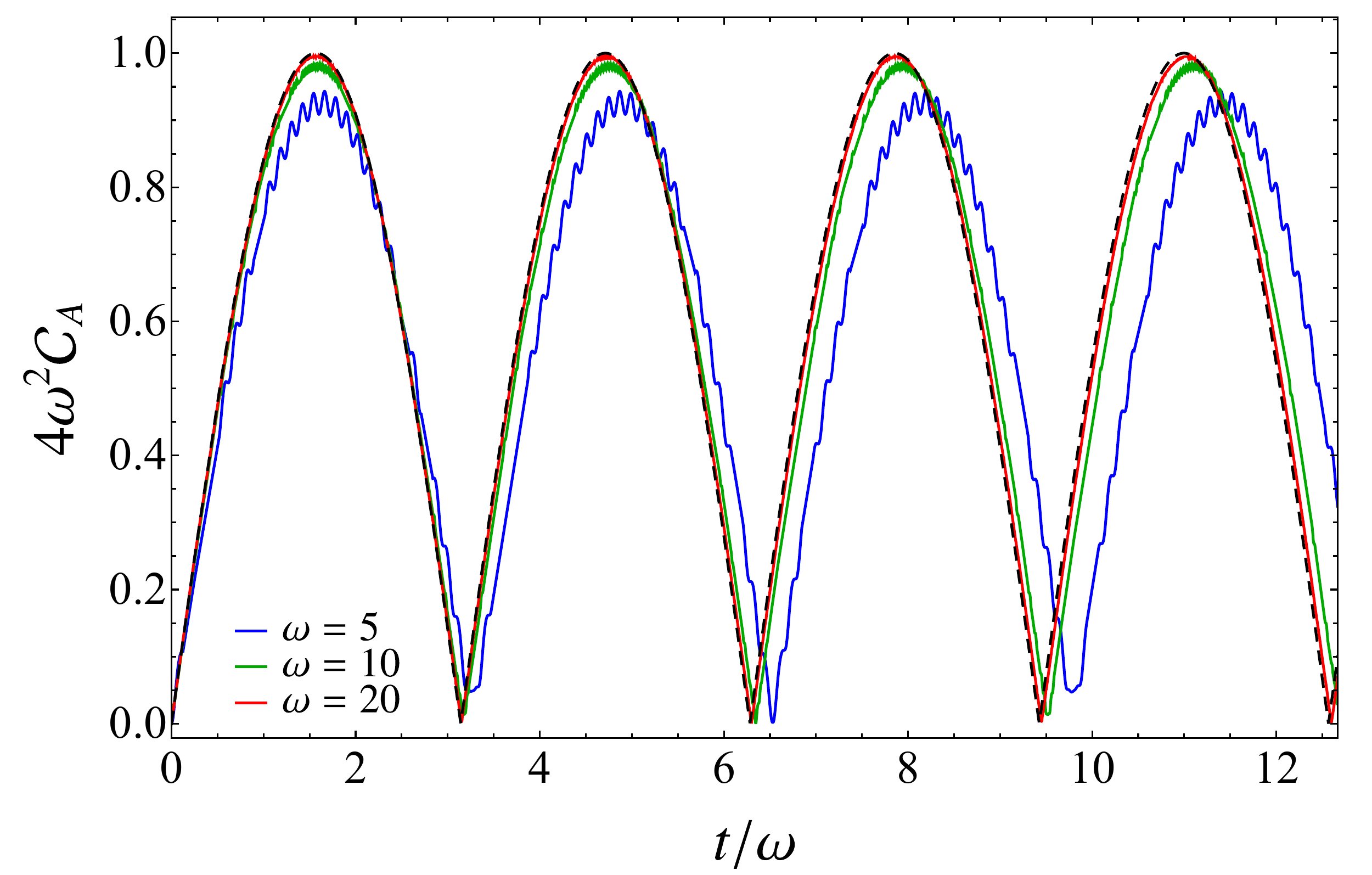}}
\caption{Temporal evolution of the subsystem complexity $\mathcal{C}_A$ 
after the local quench  w.r.t. the initial state at $t=0$
when $A$ contains a single site and the chain is made by two sites.  
In the top left panel (\ref{ComplexityMixed 2 sites}) is shown for small values of  $\omega$.
In the top right and in the bottom panel 
we compare (\ref{ComplexityMixed 2 sites}) for larger values of $\omega$ 
against (\ref{ComplexityMixed 2 sites largem}) (black solid line) 
and (\ref{ComplexityMixed 2 sites verylargem}) (black dashed line)
respectively.
In the top right panel, numerical data for 
the subsystem complexity of half chains with $N>2$ are also reported.
In all the panels $\kappa=1$ and $m=1$.
}
\vspace{0.4cm}
\label{fig:MixedState2sites}
\end{figure}

In the top left panel of Fig.\,\ref{fig:MixedState2sites}, 
the subsystem complexity (\ref{ComplexityMixed 2 sites}) is shown for various $\omega \leqslant 2$ 
and $\kappa=1$ and $m=1$. 
The local maxima of the curves corresponding to different $\omega$ 
occur at the same values of $\widetilde{\Omega}_2 t$ given by multiple integers of $\pi$ 
(see the vertical lines),
where $\widetilde{\Omega}_k$ with $k\in \{1,2\}$ is defined below (\ref{eigenvalues deltaTR 2 sites}).
The same feature is observed also for the local minima, 
if $\widetilde{\Omega}_1 t$ is employed as the independent variable 
on the horizontal axis instead of $\widetilde{\Omega}_2 t$.

In the asymptotic regime given by $\tilde{\omega} \to\infty$,
where $\tilde{\omega}$ has been introduced in (\ref{eigenvalues deltaTR 2 sites largemass}), 
while $\omega t$ is kept fixed and finite
(introduced in Sec.\,\ref{subsec:twositespure}),
the expansion of  (\ref{gTR reduced relativeCM}) reads
\be
\label{gTR reduced relativeCM largem}
g_{\textrm{\tiny TR},\pm}
\,=\,
1
+\frac{1}{2\tilde{\omega}^4}\, 
\bigg(
\big[ \sin(\omega t) \big]^2 \pm 
\sqrt{\big[ \sin(\omega t) \big]^2 -\omega t \, \sin(2\omega t)+\omega^2 t^2}
\,\bigg)
+
O\big( 1/\tilde{\omega}^{6} \big)\,.
\ee
By employing this result, for the expansion of  (\ref{ComplexityMixed 2 sites}) one finds
\be
\label{ComplexityMixed 2 sites largem}
\mathcal{C}_A
\,=\,
\frac{1}{4\tilde{\omega}^4}
\sqrt{\big[ \sin(\omega t) \big]^4 + \big[ \sin(\omega t) \big]^2
+\omega t \, \big[\omega t- \sin(2\omega t)\big]}
+
O\big( 1/\tilde{\omega}^{6} \big)\,.
\ee
In the top right panel of Fig.\,\ref{fig:MixedState2sites}, 
this expression corresponds to the black solid line,
while the other curves have been drawn through
the exact formula (\ref{ComplexityMixed 2 sites}).
In the same panel, we have also reported $\mathcal{C}_A$ of half chains with $N=2L>2$ (coloured symbols).
We find remarkable their agreement with (\ref{ComplexityMixed 2 sites}) for early times,
which improves as $\omega$ increases.

Comparing Fig.\,\ref{fig:MixedState2sites} and Fig.\,\ref{fig:PureState2sites},
we notice that,
while for the complexity of the entire chain made by two sites only a main oscillatory behaviour is observed,
for the subsystem complexity we can identify two kinds of oscillations:
one has a larger amplitude and period $ \pi\tilde{\omega}^2/\omega $
and another one is characterised by a  smaller amplitude and period $\pi/\omega$.
When $\tilde{\omega} \gg 1$, the amplitude of the latter oscillation becomes negligible 
and we find that the temporal evolution of the single site subsystem complexity is 
nicely described by the following ansatz
\be
\label{ComplexityMixed 2 sites verylargem}
\mathcal{C}_A\,=\,
\frac{1}{4\tilde{\omega}^2}\,\big|\sin\big[(\omega t)/\tilde{\omega}^2\big]\big|
\ee
which is compared against the exact result (\ref{ComplexityMixed 2 sites}) in the bottom panel of Fig.\,\ref{fig:MixedState2sites}.

\section{Conclusions}
\label{sec:conclusions}

We studied the temporal evolutions of the circuit complexity and of the subsystem complexity 
after a local quench
by considering harmonic chains in a segment with Dirichlet boundary conditions
and the local quench where
two finite chains made by $rN$ and $(1-r)N$ sites (with $0<r<1$) are joined at $t=0$
\cite{Eisler_2014}.
The subsystem complexity $\mathcal{C}_A$ in (\ref{c2-complexity-rdm})
has been evaluated 
by employing the complexity of mixed bosonic Gaussian states based on the Fisher information geometry
\cite{DiGiulio:2020hlz},
which provides also the optimal circuit (\ref{optimal circuit rdm}).
For the sake of simplicity, we considered only the case where the subsystem is a block of $L$ consecutive sites
and we mainly studied the complexity of circuits whose reference state is the initial state at $t=0$.

The covariance matrices of the reference state and of the target state at time $t>0$ along the temporal evolution
have been introduced in Sec.\,\ref{subsec:setup}.
Then, in Sec.\,\ref{sec:PureStateComp} they have been employed to evaluate 
some temporal evolutions of the circuit complexity for the entire harmonic chain.

For any value $\omega \geqslant 0$ of the mass occurring in the evolution Hamiltonian,
we found that the initial growth of the complexity immediately after the quench is linear
(see (\ref{initial growth})) with a slope given by (\ref{slope smallt final}),
which can be evaluated numerically, as done in the bottom panel of Fig.\,\ref{fig:PureStateMassiveLargetimes}.
When the evolution Hamiltonian is critical (i.e. $\omega =0$),
after the above mentioned initial growth
we observe a logarithmic growth independent of $r$
(see (\ref{guess complexity initial}) and the bottom panels of Fig.\,\ref{fig:PureStateCritical}).
We expect to observe this feature also when the system is infinite.  
In our numerical analysis we have considered only finite systems.
The temporal evolutions of the complexity for finite systems and $\omega=0$ display revivals, 
independently of $r$.
Three temporal regimes are observed within the first half of the temporal interval containing a single revival:
a growth followed by a decrease and finally a regime where the complexity does not evolve
(see the top right panel of Fig.\,\ref{fig:PureStateCritical} and Fig.\,\ref{fig:PureStateCriticalLargeTimes}).
In the case of $r=1/2$, the latter regime does not occur; hence this choice halves the duration of a revival
(see the top left panel of Fig.\,\ref{fig:PureStateCritical}).
When $\omega >0$, the temporal evolutions of the complexity are more complicated; 
indeed, for instance, 
an approximate periodicity is not observed (see Fig.\,\ref{fig:PureStateMassiveLargetimes}).
When $\omega N$ is large, the complexity rapidly changes through small variations about a constant value 
that is independent of $r$.
Importantly, we have identified different temporal regimes 
where different scaling behaviours are observed as $N$ increases.
It would be interesting to explain these scalings through quantum field theory methods.

In Sec.\,\ref{sec-subsystem-comp} we have explored
the temporal evolutions of $\mathcal{C}_A$ 
for the bipartitions shown in the bottom panels of Fig.\,\ref{fig:intro-configs},
where either one entangling point or two entangling points occur.
One of our main results is given by the numerical evidences 
that the qualitative behaviour of the temporal evolutions of $\mathcal{C}_A$ 
depends on whether the block $A$ contains the joining point. 
 In the case of the spatial bipartition shown in the bottom left panel of Fig.\,\ref{fig:intro-configs},
 where one entangling point occurs, 
 this qualitative difference is evident once the left panels are compared against the corresponding right panels
both in Fig.\,\ref{fig:CompMixedMasslessr1over2} and  Fig.\,\ref{fig:CompMixedMasslessr1over4}
 (where $r=1/2$ and $r=1/4$ respectively) when $\omega=0$
 and both
 in Fig.\,\ref{fig:CompMixedMassiver1over2} and  Fig.\,\ref{fig:CompMixedMassiver1over4}
 (where, again, $r=1/2$ and $r=1/4$ respectively) when $\omega>0$.
When the evolution Hamiltonian is critical and the joining point is inside the block, 
during the initial regime of the temporal evolution of $\mathcal{C}_A$ we observe
the same logarithmic growth (\ref{guess complexity initial})
occurring in the temporal evolution of 
the complexity of the entire chain 
(compare the insets of Fig.\,\ref{fig:CompMixedMasslessr1over2} and Fig.\,\ref{fig:CompMixedMasslessr1over4} 
against the bottom panels of Fig.\,\ref{fig:PureStateCritical}).
Furthermore, in the case of $r=1/2$ and when the joining point lies outside the block, 
we find that the analytic expression (\ref{complexity guess}) for the temporal evolution of $\mathcal{C}_A$ 
nicely reproduces the behaviour of the numerical data.

It is very instructive to compare a temporal evolution of $\mathcal{C}_A$ 
(in this manuscript we have considered only circuits where the reference state is the initial state)
against the corresponding temporal evolution of $S_A$,
obtained for the same bipartition and the same quench protocol.
The temporal evolutions of $S_A$ for the bipartition 
shown in the bottom left panel of Fig.\,\ref{fig:intro-configs} have been reported 
in Fig.\,\ref{fig:EntMasslessr1over2} and Fig.\,\ref{fig:EntMasslessr1over4} 
(where $r=1/2$ and $r=1/4$ respectively) for $\omega =0$
and in Fig.\,\ref{fig:EntrMixedMassive} for $\omega >0$.
We remark that, whenever the block $A$ does not contain the joining point,
the temporal evolutions of $\mathcal{C}_A$ and of $S_A$ are qualitatively similar 
(for instance, both these quantities do not evolve immediately after the quench whenever $L \neq rN$).
Instead,  they are qualitatively very different when the joining point is inside the subsystem;
indeed, for instance, when $L \neq rN$ we find that $\mathcal{C}_A$ rapidly grows immediately after the quench,
while $S_A$ remains constant for a while.

In this manuscript we have considered both the spatial bipartitions shown in the bottom panels of Fig.\,\ref{fig:intro-configs},
in order to investigate the influence of the number of entangling points on the temporal evolution of $\mathcal{C}_A$.
By comparing the right panels of Fig.\,\ref{fig:CompMixedMasslessr1over2} 
against the left panels of Fig.\,\ref{fig:Massless2endpoints},
where $r=1/2$ and $\omega=0$,
we observed that, when the joining point is located inside the block,
both the qualitative behaviour of the temporal evolution of $\mathcal{C}_A$
and the values of the scaling parameter $\alpha$
are not influenced by the number of entangling points.
Instead, the values of the scaling parameter $\alpha$ for $S_A$
do depend on the number of entangling points
(see the right panels of Fig.\,\ref{fig:EntMasslessr1over2} and Fig.\,\ref{fig:Massless2endpoints}).

We find it worth remarking that the logarithmic growth of $\mathcal{C}_A$
highlighted in the inset of the bottom right panels 
of Fig.\,\ref{fig:CompMixedMasslessr1over2} and of Fig.\,\ref{fig:CompMixedMasslessr1over4}
for one entangling point 
and of the bottom left panel of Fig.\,\ref{fig:Massless2endpoints} for two entangling points
is described by the same curve (\ref{guess complexity initial}), including the additive constant,
which has been found for the temporal evolution of the complexity of the entire chain
(see the bottom panels of Fig.\,\ref{fig:PureStateCritical}).

We have also explored 
the temporal evolutions of the complexity for the entire chain 
and of the subsystem complexity after the local quench 
in the minimal setup where the chain is made by two sites,
and therefore the subsystem $A$ contains only one site
(see Sec.\,\ref{subsec:twositespure} and Sec.\,\ref{subsec:single-site-comp}).
In this simple setup, 
we have obtained the analytic expressions given 
by  (\ref{eigenvalues deltaTR 2 sites}) and (\ref{ComplexityPure 2 sites})
(see also Fig.\,\ref{fig:PureState2sites}) for the complexity of the chain
and by  (\ref{gTR reduced relativeCM}) and (\ref{ComplexityMixed 2 sites})
(see also Fig.\,\ref{fig:MixedState2sites}) for the subsystem complexity.
While these analyses are useful to get some insights about some regimes of the parameters
(e.g. small $t$ and large $\tilde{\omega}$), they do not capture many important features observed for large values of $N$.

As for the local quench considered in this manuscript, 
it would be interesting to explore more systematically the temporal evolutions of the subsystem complexity 
when $\omega >0$
or for asymmetric bipartitions involving two or even more entangling points,
to obtain analytic results in the thermodynamic limit,
to find bounds that still describe some essential features of the temporal evolution of the subsystem complexity 
and also to study the thermalisation of the subsystem complexity,
as done in \cite{DiGiulio:2021oal} for a global quench of the mass parameter.

The Gaussian states can be employed 
to investigate the temporal evolution of the subsystem complexity 
after some local quenches in higher dimensions and also in free fermionic lattice models.
It could be instructive to explore these temporal evolutions
by employing the entanglement spectrum or the entanglement Hamiltonians 
\cite{Peschel03,Peschel_2009, Casini:2009sr,Casini:2011kv,Lauchli:2013jga, Cardy:2016fqc,Arias:2016nip, Eisler:2017cqi,
Tonni:2017jom, Alba:2017bgn, Eisler:2019rnr,DiGiulio:2019cxv, DiGiulio:2019lpb, Surace:2019mft,Eisler:2020lyn},
as done in \cite{DiGiulio:2020hlz} at equilibrium.
It would be insightful to explore also the dependence of the temporal evolution 
of $\mathcal{C}_A$ on the reference state, 
by adopting a state different from the initial one 
as the reference state (e.g. the unentangled product state).

In this manuscript we have compared the temporal evolutions 
of the subsystem complexity against the ones 
of the corresponding entanglement entropy,
but it could be interesting to perform analogue comparisons against
the temporal evolutions of other entanglement quantifiers like
the entanglement negativity \cite{VidalWerner-neg,Plenio:2005cwa,Calabrese:2012ew,Calabrese:2012nk,Calabrese:2014yza,Coser:2014gsa,Eisler_2014,Eisler_2015_2dneg,DeNobili:2016nmj},
the entanglement contours \cite{Chen_2014,Coser:2017dtb} 
and the relative entropies \cite{Blanco:2013joa}.

We remark that investigating the temporal evolutions of the subsystem complexity after various quantum quenches
through lattice methods and quantum field theory techniques  in interacting models is a very challenging task
that deserves future studies.
Holography can provide important benchmarks.
Interesting analyses have been performed \cite{Caputa:2017urj,Caputa:2017yrh,
Chapman:2017rqy, Bhattacharyya:2018wym,Caputa:2018kdj,Chapman:2018bqj,Camargo:2019isp,Ge:2019mjt,
Erdmenger:2020sup,Flory:2020eot,Flory:2020dja,Chagnet:2021uvi,Bernamonti:2019zyy,Bernamonti:2020bcf}
and it would be interesting to employ these methods to explore also the out-of-equilibrium dynamics
of the circuit complexity.

\vskip 20pt 
\centerline{\bf Acknowledgments} 
\vskip 10pt

We are grateful to Federico Galli and Francesco Gentile for insightful discussions. 
ET's work has been conducted within the framework of the 
Trieste Institute for Theoretical Quantum Technologies.


\vskip 30pt 
\appendix

\section{Global quenches}
\label{app:Newglobalquench}

The approach developed in \cite{Eisler_2014} and discussed in Sec.\,\ref{subsec:setup}
allows to study the temporal evolution of the covariance matrix after various quenches.
While in the main text of this manuscript we mainly consider the temporal evolution of the
complexity after a local quench where two 
harmonic chains are joined at $t=0$,
in this appendix we explore the evolution of the same quantity after
two global quenches. 
In Appendix\,\ref{subapp:massquench} we study the quench of the frequency parameter.
In Appendix\,\ref{subapp:kappaquench}, 
considering the unentangled product state as the initial state (and also as reference state),
at $t=0$ we perform a quench of the spring constant and of the frequency.

\subsection{Mass quench}
\label{subapp:massquench}

Consider the ground state of the Hamiltonian defined by (\ref{HC ham-1d}) where 
$\omega$ is replaced by $\omega_0$.
Given this pure state as the initial state for the evolution, 
at $t=0$ the sudden change $\omega_0 \to \omega$ is performed;
hence the evolution Hamiltonian becomes (\ref{HC ham-1d}).
The temporal evolutions after this global quench 
of the complexity \cite{Alves:2018qfv,Camargo:2018eof,Ali:2018fcz}
and of the subsystem complexity \cite{DiGiulio:2021oal}
have been investigated.

The initial state of this global quench 
is obtained by setting the parameters introduced in Sec.\,\ref{subsec:setup} to
$(N_\textrm{\tiny l},N_\textrm{\tiny r}) =(N,0)$,
$\kappa_\textrm{\tiny l}=\kappa_\textrm{\tiny r}=\kappa$, $m_\textrm{\tiny l}=m_\textrm{\tiny r}=m$ and $\omega_\textrm{\tiny l}=\omega_\textrm{\tiny r}=\omega_0$;
hence its covariance matrix (\ref{initial CM general}) becomes
\be
\label{initial CM global} 
\gamma_0=V^{\textrm{t}} \,\Gamma_0 \,V 
\;\;\qquad\;\;
\Gamma_0=\mathcal{Q}_0\oplus \mathcal{P}_0
\ee
with $V$ being the orthogonal matrix defined in (\ref{mat E def}) and
\be
\label{Q0 P0 mat diag global}
\mathcal{Q}_0
= 
\frac{1}{2} \;\textrm{diag}
\Big( 
\big( m \Omega_{0,1}\big)^{-1},\dots , \big( m \Omega_{0,N}\big)^{-1}
\Big)
\;\;\qquad\;\;
\mathcal{P}_0
=
\frac{1}{2} \;\textrm{diag}
\Big( 
m \Omega_{0,1}\,,\dots ,  m \Omega_{0,N}
\Big)
\ee
where $\Omega_{0,k}$ is  (\ref{dispersion relation evolution}) 
with $\omega$
replaced by $\omega_0$.

From (\ref{initial CM global}), (\ref{CM-local-evolved general}) and (\ref{mat E def}), 
for the temporal evolution of the covariance matrix in (\ref{initial CM global}) we find
\be
\label{evolved CM global} 
\gamma(t) 
= 
E(t) \,\gamma_{\textrm{0}} \, E(t)^{\textrm t}
=V^{\textrm{t}} \, \mathcal{E}(t)\, \Gamma_0\, \mathcal{E}(t)^{\textrm t} \,V\,.
\ee
A characteristic feature of this quench is the occurrence of the same matrix $V$ both in $E(t)$ and in $\gamma_0$,
which leads to the crucial simplification highlighted in (\ref{evolved CM global}).
This feature is not verified when both $N_{\textrm{\tiny l}}$ and $N_{\textrm{\tiny r}}$ are non vanishing, as discussed in Sec.\,\ref{subsec:PureStateCompTheory}.
From (\ref{initial CM global}) and (\ref{diagmat E}), we obtain
\be
\mathcal{E}(t) \, \Gamma_0 \, \mathcal{E}(t)^{\textrm t}
\,=\,
\bigg(\begin{array}{cc}
\mathcal{Q}(t)  & \mathcal{M}(t)
\\
 \mathcal{M}(t)  &  \mathcal{P}(t)
\end{array}\bigg)
\ee
where the block matrices $\mathcal{Q}(t)$, $\mathcal{P}(t)$ and $\mathcal{M}(t)$
in the r.h.s. are diagonal matrices whose diagonal elements are respectively
\be
\label{QPM mat diag global}
\begin{array}{l}
\displaystyle
\mathcal{Q}_{k,k}(t)=
\frac{1}{ 2m\Omega_k}
\left( \,\frac{\Omega_k}{\Omega_{0,k}} \, [ \cos(\Omega_k t) ]^2
+ \frac{\Omega_{0,k}}{\Omega_k} \, [\sin(\Omega_k t)]^2 \right)
\\
\displaystyle
\rule{0pt}{.9cm}
 \mathcal{P}_{k,k}(t)=
 \frac{m\Omega_k}{2}
\left( \,\frac{\Omega_k}{\Omega_{0,k}}  \, [\sin(\Omega_k t)]^2
+ \frac{\Omega_{0,k}}{\Omega_k} \, [\cos(\Omega_k t)]^2 \right)
 \\
 \displaystyle
 \rule{0pt}{.9cm}
  \mathcal{M}_{k,k}(t)=
\frac{1}{2}\left(  \frac{\Omega_{0,k}}{\Omega_k} -\,\frac{\Omega_k}{\Omega_{0,k}} \right)
\sin(\Omega_k t) \cos(\Omega_k t) 
\end{array}
\ee
and $\Omega_k$ is defined in (\ref{dispersion relation evolution}).
The expressions (\ref{QPM mat diag global}) have been first obtained in \cite{Calabrese:2007rg} and recently employed in 
\cite{DiGiulio:2021oal} to study the subsystem complexity after the global quench of the mass.
Thus, also the global quench of the mass can be described
through the formalism of \cite{Eisler_2014}.

\subsection{Quench of the spring constant}
\label{subapp:kappaquench}

In the following we consider the temporal evolution of the complexity 
when the initial state of the global quench
is given by an unentangled product state.
In terms of the parameters introduced in Sec.\,\ref{subsec:setup}, 
this quench corresponds to 
$(N_\textrm{\tiny l},N_\textrm{\tiny r}) =(N,0)$, 
$\kappa_\textrm{\tiny l}=\kappa_\textrm{\tiny r}=0$, $m_\textrm{\tiny l}=m_\textrm{\tiny r}=m$ 
and $\omega_\textrm{\tiny l}=\omega_\textrm{\tiny r}=\mu$. 
At $t=0$ all the spring constants of the chain are suddenly switched on
and the evolution Hamiltonian becomes (\ref{HC ham-1d}).
For the sake of generality, we consider $\mu\neq\omega$;
hence we suddenly change both the spring constant to $\kappa >0$ and the frequency from $\mu$ to $\omega$.
Setting $\mu=\omega$ provides the global quench where only the spring constant is changed.

Since $(N_\textrm{\tiny l},N_\textrm{\tiny r}) =(N,0)$,
we have $V=V_0$ (see (\ref{Vtilde0 mat}) and (\ref{mat E def})).
By specialising (\ref{T0 mat diag}), (\ref{initial CM general}) and (\ref{Q0-P0 mat diag}) to this case, 
one finds the following covariance matrix for the initial state
\be
\label{Q0 P0 mat diag quench kappa}
\gamma_0\,=\,V^{\textrm{t}} \,\Gamma_0 \,V
\;\;\qquad\;\;
\Gamma_0= 
\bigg(\frac{1}{2m\mu}\;\boldsymbol{1}\bigg)\oplus\bigg(\frac{m\mu}{2 }\;\boldsymbol{1}\bigg) \,.
\ee
Since $\widetilde{V}_N$ in (\ref{mat E def}) is orthogonal, from (\ref{mat E def}) and (\ref{Q0 P0 mat diag quench kappa}) we obtain that $\gamma_0=\Gamma_0$.

By employing (\ref{CM-local-evolved general}) and (\ref{mat E def}) (where $\widetilde{V}_N$ is orthogonal),
we find that the temporal evolution of (\ref{Q0 P0 mat diag quench kappa}) reads
\be
\label{evolved CM quench kappa}
\gamma(t) \,=\, V^{\textrm{t}} \,\mathcal{E}(t) \,\Gamma_0 \, \mathcal{E}(t)^{\textrm t} \,V\,.
\ee
From (\ref{Q0 P0 mat diag quench kappa}) and the block decomposition in (\ref{diagmat E}), we get
\be
\mathcal{E}(t) \, \Gamma_0\, \mathcal{E}(t)^{\textrm t}=
\bigg( 
\begin{array}{cc}
\widetilde{\mathcal{Q}}(t) \, & \widetilde{\mathcal{M}}(t)
\\
 \widetilde{\mathcal{M}}(t) \, &  \widetilde{\mathcal{P}}(t)
\end{array}
\bigg)
\ee
where $\widetilde{\mathcal{Q}}(t) $, $\widetilde{\mathcal{P}}(t) $ and $\widetilde{\mathcal{M}}(t) $ 
are diagonal matrices whose diagonal elements are given respectively 
by (\ref{QPM mat diag global}) with $\Omega_{0,k}$ replaced by $\mu$.

\begin{figure}[t!]
\subfigure
{
\hspace{-1.25cm}\includegraphics[width=.57\textwidth]{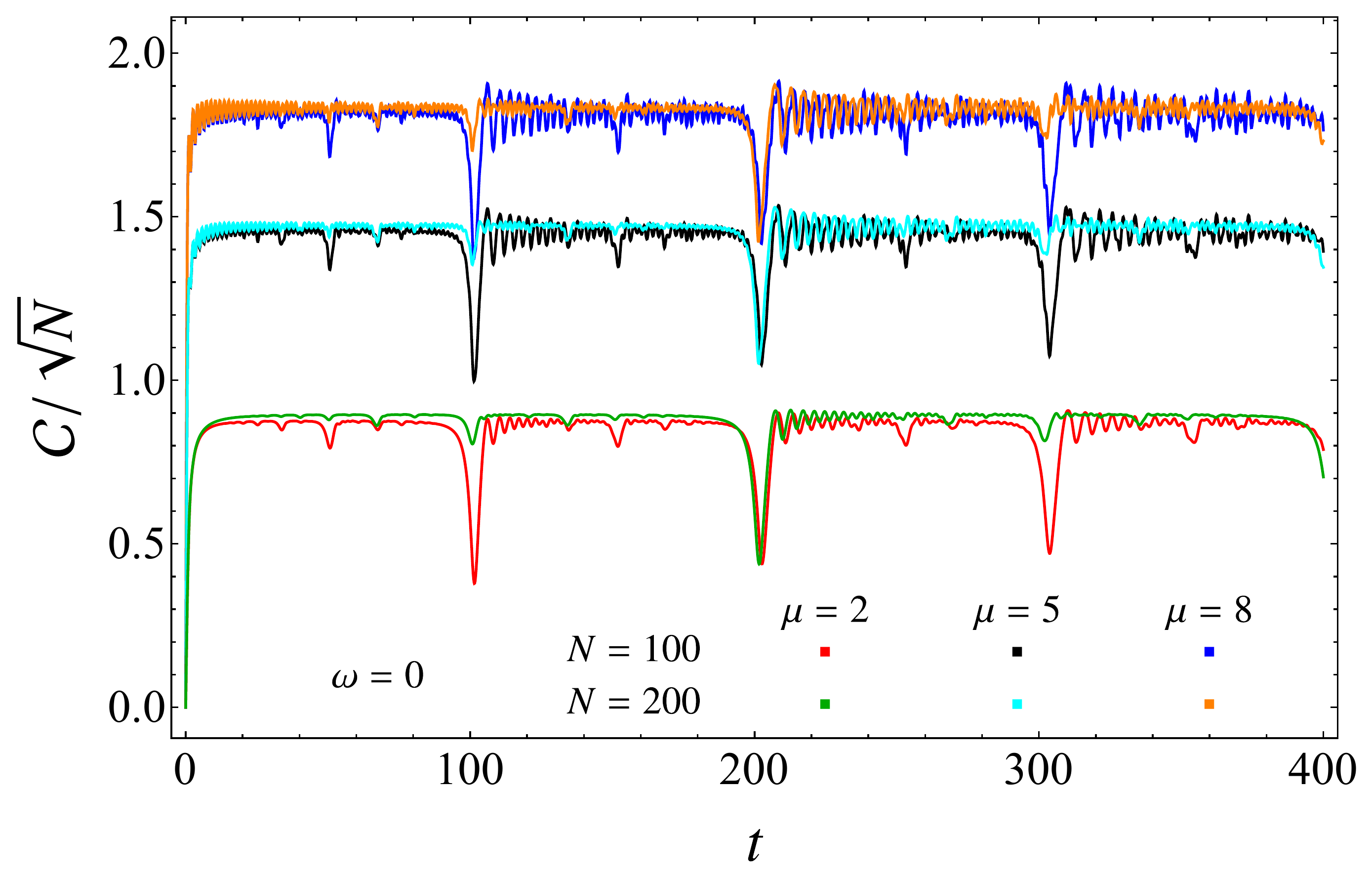}}
\subfigure
{
\hspace{-0.05cm}\includegraphics[width=.57\textwidth]{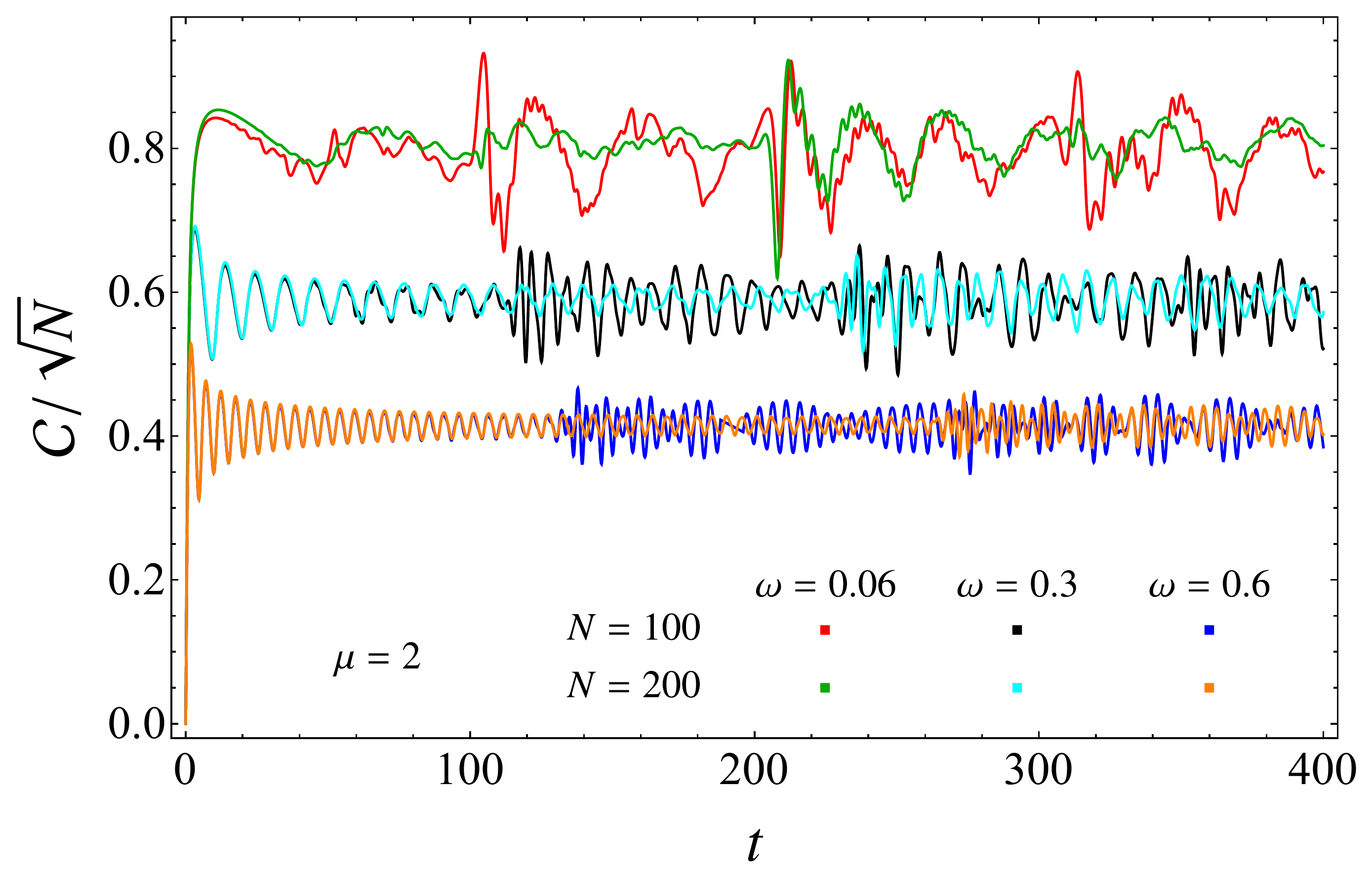}}
\subfigure
{
\hspace{-1.25cm}\includegraphics[width=.57\textwidth]{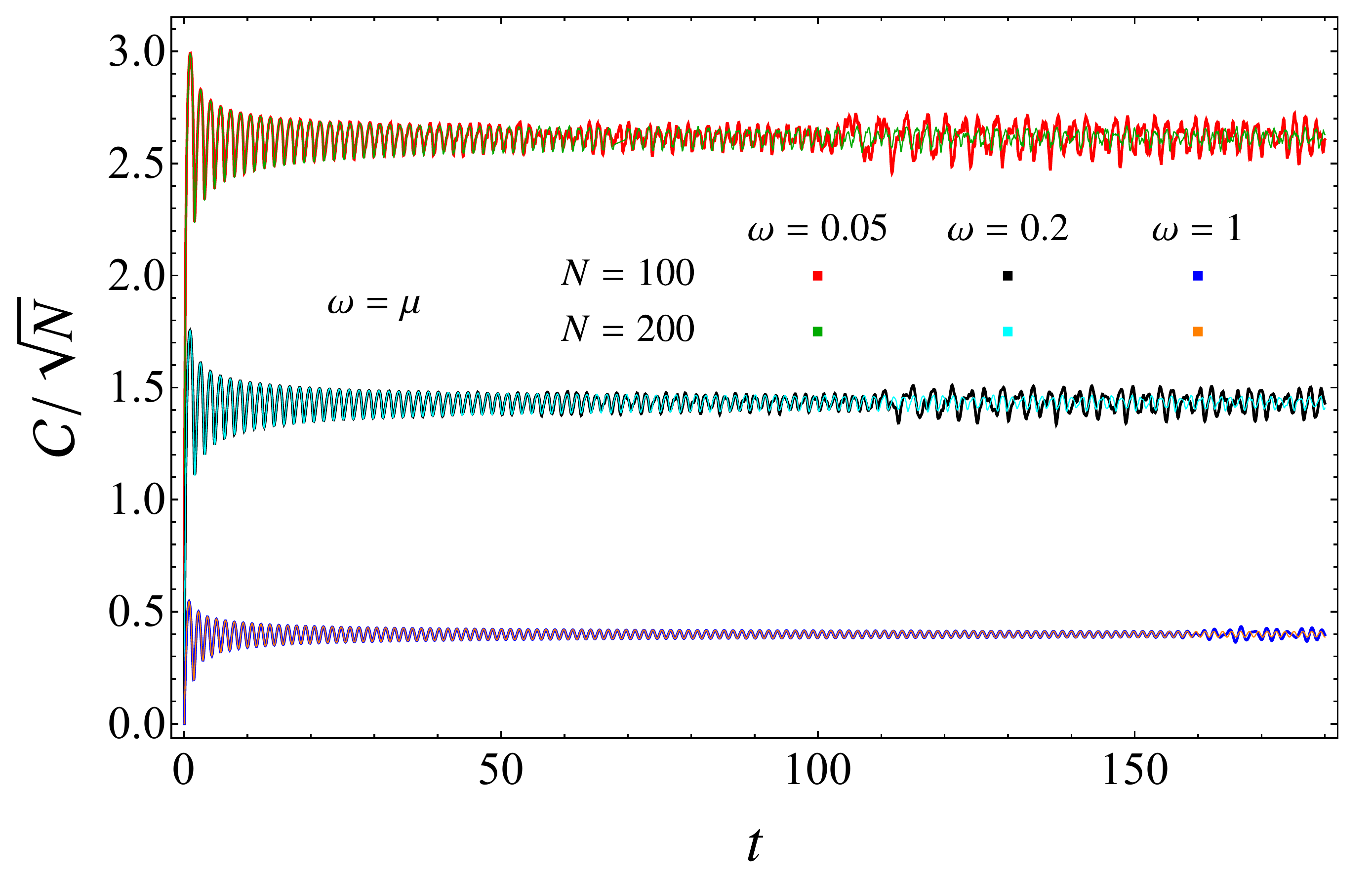}}
\subfigure
{
\hspace{-0.05cm}\includegraphics[width=.57\textwidth]{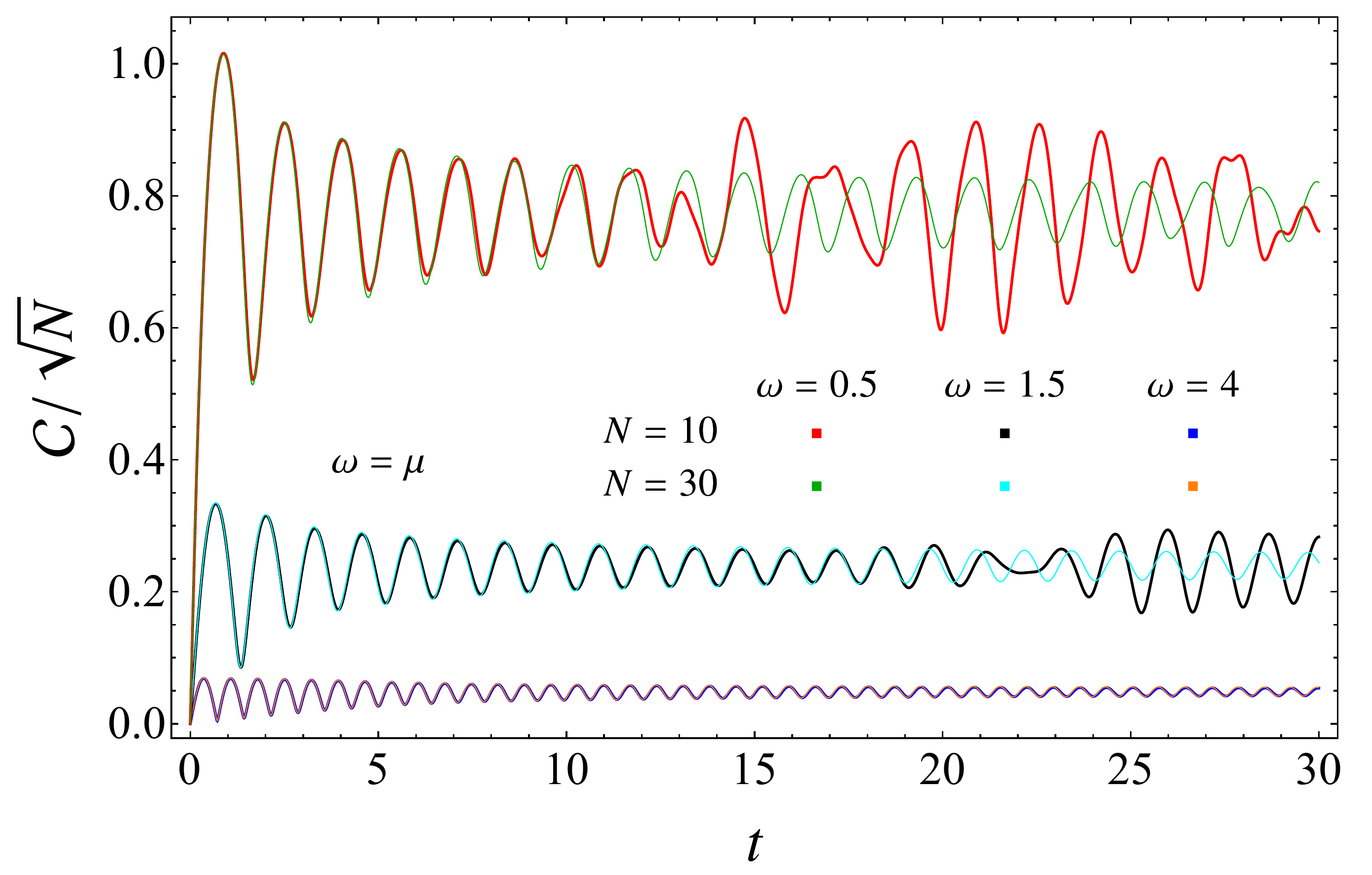}}
\caption{
Temporal evolution of the complexity (\ref{c2 quench kappa}) for the entire chain
after the global quench of the frequency and of the spring constant. 
The expression for $C_{\textrm{\tiny TR},k}$ reported 
in (\ref{c2 quench kappa C_TR}) and in (\ref{CTR quench kappa only})
has been used respectively in the top and in the bottom panels. 
}
\vspace{0.4cm}
\label{fig:PureStateQuenchKappa}
\end{figure}

We consider the temporal evolution of the complexity
where the reference and the target states are the states 
at the values of time $t_{\textrm{\tiny R}}$ and $t_{\textrm{\tiny T}}$ respectively 
along a given quench, 
namely $\gamma_{\textrm{\tiny R}}=\gamma(t_{\textrm{\tiny R}})$ 
and $\gamma_{\textrm{\tiny T}}=\gamma(t_{\textrm{\tiny T}})$,
with $\gamma(t)$ given in (\ref{evolved CM quench kappa}).

Following the analysis reported in  \cite{DiGiulio:2021oal} 
(see also \cite{Camargo:2018eof,Ali:2018fcz}),
we compute the eigenvalues of $\gamma_{\textrm{\tiny T}}\, \gamma_{\textrm{\tiny R}}^{-1}$ 
which provide the complexity (\ref{c2 complexity}), finding
\be
\label{c2 quench kappa}
\mathcal{C}\,=\,
\frac{1}{2}\; \sqrt{\sum_{k=1}^N \big[ \textrm{arccosh}(C_{\textrm{\tiny TR},k})\big]^2}
\ee
with
\be
\label{c2 quench kappa C_TR}
C_{\textrm{\tiny TR},k}
=
1+\frac{1}{2}\bigg(\frac{\Omega_k^2-\mu^2}{\Omega_k \,\mu}\,
\sin\!\big[\Omega_k(t_{\textrm{\tiny T}}-t_{\textrm{\tiny R}})\big]
\bigg)^2
\ee
where
$\Omega_{k}$ is given in (\ref{dispersion relation evolution}).
Notice that, when $\mu\neq \omega$, 
we can set $\omega=0$ and obtain a finite result for the complexity in (\ref{c2 quench kappa}). 
Since $C_{\textrm{\tiny TR},k}$ in (\ref{c2 quench kappa C_TR}) is an oscillating function of $|t_{\textrm{\tiny T}}-t_{\textrm{\tiny R}}|$ for any $k$, 
the complexity $\mathcal{C}$ is finite, also for large values of $|t_{\textrm{\tiny T}}-t_{\textrm{\tiny R}}|$.

In Fig.\,\ref{fig:PureStateQuenchKappa} we show some
temporal evolutions of the complexity (\ref{c2 quench kappa})
when $t_{\textrm{\tiny R}} =0$ and $t_{\textrm{\tiny T}}=t$.
In the top panels we keep $\mu\neq\omega$,
setting  either $\omega=0$ (left panel) and $\omega> 0$ (right panel).
These temporal evolutions are qualitatively similar to the ones observed for the global quench discussed in 
Appendix\,\ref{subapp:massquench},
as expected from the fact that the role of $\Omega_{0,k}$ is played by $\mu$ in this case.

In the case where only the quench of the spring constant is performed, i.e. when $\mu=\omega$
(see the bottom panels of Fig.\,\ref{fig:PureStateQuenchKappa}),
by using the explicit expression of $\Omega_{k}$ in (\ref{dispersion relation evolution}),
one obtains 
\be
\label{CTR quench kappa only}
C_{\textrm{\tiny TR},k}
\,=\,
1+\frac{1}{2}\Bigg(\,
\frac{4\,\kappa}{m \omega \, \Omega_k}\,
\big[ \sin\!\big(\pi k /[2(N+1)]\big) \big]^2\,
\sin\!\big[\Omega_k(t_{\textrm{\tiny T}}-t_{\textrm{\tiny R}})\big]\Bigg)^2
\ee
which provides the complexity through (\ref{c2 quench kappa}).
In this case the critical evolution cannot be explored
because (\ref{CTR quench kappa only}) diverges for any $k$ as $\omega\to 0$.

For a given $\omega >0$,
the coefficient of $ ( \sin[\Omega_k(t_{\textrm{\tiny T}}-t_{\textrm{\tiny R}})])^2$ 
in the r.h.s. of (\ref{CTR quench kappa only}) 
becomes negligible when $\tfrac{k}{N+1}\ll 1$,
while it reaches its maximum when $\tfrac{k}{N+1}\simeq 1$.
Thus, the main contributions to 
the complexity given by (\ref{c2 quench kappa}) and (\ref{CTR quench kappa only}) 
come from the modes such that $\tfrac{k}{N+1}\simeq 1$. 
For these modes $| \sin[\tfrac{\pi k}{2(N+1)}] |\simeq 1$; hence 
$\sin[\Omega_k(t_{\textrm{\tiny T}}-t_{\textrm{\tiny R}})]
\simeq\sin[\sqrt{\omega^2+4}\,(t_{\textrm{\tiny T}}-t_{\textrm{\tiny R}})]$.
This leads us to identify two regimes in $\omega$:
when $\omega^2\ll 4$, the complexity oscillates with frequency approximately equal to $ 2$, 
independently of $\omega$; 
instead, when $\omega^2\gtrsim 4$, 
the frequency of the oscillations is $ \sqrt{\omega^2+4}$.
This behaviour can be observed
in the bottom panels of Fig.\,\ref{fig:PureStateQuenchKappa}, 
where some temporal evolutions of the complexity given by 
(\ref{c2 quench kappa}) and (\ref{CTR quench kappa only}) 
are shown, in the case of $t_{\textrm{\tiny R}}=0$ and $t_{\textrm{\tiny T}}=t$,
i.e. when the reference state is the (initial) unentangled product state.
When $\omega\in \{ 0.05,0.2,0.5\}$, in the initial part of the evolution the curves 
for $\mathcal{C}/\sqrt{N}$ collapse displaying the same oscillatory behaviour
with frequency independent of $\omega$ and approximately equal to 2.
Instead, when $\omega\in \{ 1.5,4\}$ and therefore $\omega^2 \sim  4$,
the frequency of the oscillations in the initial part of the curves 
depends also on $\omega$ and it is given by
$\sqrt{\omega^2+4}$.


\section{Euler decomposition for a class of symplectic matrices}
\label{app:EulerDecomposition}

The Euler decomposition is a powerful tool to evaluate the circuit complexity for pure states.
Indeed, (\ref{c2 complexity}) can be written also in terms of the squeezing parameters 
of the symplectic matrix  $W_{\textrm{\tiny TR}}$ (see (\ref{complexity-squeezing})),
as discussed in Sec.\,\ref{subsec:PureStateCompTheory} 
\cite{Chapman:2018hou,Camargo:2018eof,DiGiulio:2021oal}.
In this appendix we derive the analytical expressions for the matrices  in 
the Euler decomposition (\ref{EulerDec}) for a specific class of symplectic matrices. 
This provides the Euler decomposition of the matrix $\mathcal{E}$ defined in (\ref{diagmat E}), 
which is exploited to get (\ref{WTR v2}).

Consider a $2N\times 2N$ matrix $M$ partitioned into the 
four $N\times N$ blocks $\mathcal{S}$, $\mathcal{U}$, $\mathcal{Y}$ and $\mathcal{Z}$ 
which are diagonal matrices
(we denote respectively by $s_k$, $u_k$, $y_k$, $z_k$ 
their $k$-th element on the diagonal)
whose elements can be reorganised into a block diagonal matrix having non vanishing 
$2\times 2$ blocks $M_k$ along the diagonal, where $k=1,\dots N$, namely
\be
\label{matrix block diags}
M=
\bigg( 
\begin{array}{cc}
\mathcal{S} \,\, & \mathcal{U}
\\
 \mathcal{Y} \,\, &  \mathcal{Z}
\end{array} 
\bigg)
\;\;\qquad\;\;
M_k=
\bigg( \begin{array}{cc}
s_k \! &\,\, u_k
\\
 y_k \! &\,\,  z_k
\end{array} \bigg)\,.
\ee

It is straightforward to check that $M$ is symplectic if and only if $M_k$ is symplectic for any $k=1,\dots,N$.
A $2\times 2$ symplectic matrix has three independent parameters;
hence let us consider the following parametrisation\footnote{This parametrisation does not include the $2\times 2$ symplectic matrices having three non vanishing elements. 
However, the analysis reported in this appendix can be easily adapted to this class of symplectic matrices.}
\be
\label{submat sympl}
M_k
=
\bigg(\begin{array}{cc}
\alpha_k \cos\theta_k  & \;\; \beta_k \sin\theta_k
\\
 -\,\beta_k^{-1}\sin\theta_k  & \;\;  \alpha_k^{-1}\cos\theta_k
\end{array} \bigg)
\ee
where $\alpha_k$ and $\beta_k$ are non vanishing real numbers. 
We also assume that $(\alpha_k,\beta_k) \neq (1,1)$ because in this case $M_k$ is  orthogonal,
therefore its Euler decomposition is trivial. 

For the Euler decomposition  of (\ref{submat sympl}) we find 
\be
\label{EulerDec 2by2-dec}
M_k=
\,L_k \, \mathcal{X}_k \, R_k
\ee
where  (see (\ref{EulerDec})) 
\be
\label{EulerDec 2by2}
L_k \equiv
\bigg(
\begin{array}{cc}
\cos\theta_k^{(\textrm{\tiny L})} \! & \;\;  \sin\theta_k^{(\textrm{\tiny L})}
\\
 -\sin\theta_k^{(\textrm{\tiny L})} \! & \;\;  \cos\theta_k^{(\textrm{\tiny L})}
\end{array}
\bigg)
\qquad
\mathcal{X}_k \equiv
\bigg(
\begin{array}{cc}
e^{\Lambda_k} \! & \;\;  0
\\
 0\! & \;\;  e^{-\Lambda_k}
\end{array}
\bigg)
\qquad
R_k \equiv
\bigg(\begin{array}{cc}
\cos\theta_k^{(\textrm{\tiny R})} \! & \;\;  \sin\theta_k^{(\textrm{\tiny R})}
\\
 -\sin\theta_k^{(\textrm{\tiny R})} \! & \;\;  \cos\theta_k^{(\textrm{\tiny R})}
\end{array}
\bigg)
\ee
with the non-vanishing elements of $\mathcal{X}_k$ given by 
\be
\label{squeezing 2by2}
e^{\pm\Lambda_k}=g_k\pm\sqrt{g_k^2-1}
\;\;\qquad\;\;
g_k
=
\frac{\big(\alpha_k^2+\beta_k^2\big) \big(\alpha_k^2\beta_k^2+1\big)
+\big(\alpha_k^2-\beta_k^2\big) \big(\alpha_k^2\beta_k^2-1\big)
\cos(2\theta_k)}{4 \,\alpha_k^2\, \beta_k^2}\,.
\ee
The matrix $L_k$ is the symplectic and orthogonal matrix whose columns are the eigenvectors of $M_k M_k^\textrm{t}$, 
while $R_k$ is the symplectic and orthogonal matrix whose rows are the eigenvectors of $M_k^\textrm{t} M_k$. 
Evaluating the eigenvectors of $M_k^\textrm{t} M_k $ and $M_k M_k^\textrm{t} $ leads to
\bea
\label{Lk components}
& &
\cos\theta_k^{(\textrm{\tiny L})}=\pm\frac{v_k^{(\textrm{\tiny L})}}{\sqrt{1+\big(v_k^{(\textrm{\tiny L})}\big)^2}}
\qquad
\sin\theta_k^{(\textrm{\tiny L})}=\mp\frac{1}{\sqrt{1+\big(v_k^{(\textrm{\tiny L})}\big)^2}}
\\
\rule{0pt}{.8cm}
\label{Rk components}
& &
\cos\theta_k^{(\textrm{\tiny R})}=\frac{v_k^{(\textrm{\tiny R})}}{\sqrt{1+\big(v_k^{(\textrm{\tiny R})}\big)^2}}
\qquad\;\;\,
\sin\theta_k^{(\textrm{\tiny R})}=\frac{1}{\sqrt{1+\big(v_k^{(\textrm{\tiny R})}\big)^2}}
\eea
where
\bea
v_k^{(\textrm{\tiny L})}
&=&
\frac{(\beta_k^2-\alpha_k^2) \,\cos(2\theta_k)}{2\,\alpha_k \beta_k \big(e^{\Lambda_k}-\alpha_k^2\cos^2\theta_k-\beta_k^{2}\sin^2\theta_k\big)}
\\
\rule{0pt}{.8cm}
v_k^{(\textrm{\tiny R})}
&=&
\frac{(\alpha_k^2\beta_k^2-1) \,\cos(2\theta_k)}{2\,\alpha_k \beta_k \big(e^{\Lambda_k}-\alpha_k^2\cos^2\theta_k-\beta_k^{-2}\sin^2\theta_k\big)}\,.
\eea
The sign in (\ref{Lk components}) has to be fixed case by case,
checking that the correct matrix $M_k$ is obtained through the decomposition
(\ref{EulerDec 2by2-dec}),
once (\ref{EulerDec 2by2}) with (\ref{Lk components}) and (\ref{Rk components}) is employed. 

Finally, by using 
(\ref{submat sympl}), (\ref{EulerDec 2by2-dec}) and (\ref{EulerDec 2by2}),
we can write the Euler decomposition of $M$
in terms of $\Lambda_k$, $\theta_k^{(\textrm{\tiny L})}$, $\theta_k^{(\textrm{\tiny R})}$ 
given respectively in (\ref{squeezing 2by2}), (\ref{Lk components}) and (\ref{Rk components}) 
as follows
\be
\label{EulerDec M}
M=\bigg(\begin{array}{cc}
C_\textrm{\tiny L} \! & \;\;  S_\textrm{\tiny L}
\\
 -S_\textrm{\tiny L} \! & \;\;  C_\textrm{\tiny L}
\end{array}
\bigg)\,
\bigg(
\begin{array}{cc}
e^{\Lambda} \! & \;\;  0
\\
 0\! & \;\;  e^{-\Lambda}
\end{array}
\bigg)\,
\bigg(
\begin{array}{cc}
C_\textrm{\tiny R} \! & \;\;  S_\textrm{\tiny R}
\\
 -S_\textrm{\tiny R} \! & \;\;  C_\textrm{\tiny R}
\end{array}
\bigg)
\ee
where
$e^{\Lambda}=\textrm{diag}\big(e^{\Lambda_1},\dots,e^{\Lambda_N}\big)$ and
\bea
& &
C_\textrm{\tiny L}=\diag\!\big(\cos\theta_1^{(\textrm{\tiny L})},\dots,\cos\theta_N^{(\textrm{\tiny L})}\big)
\,\,\qquad\,\,\,
S_\textrm{\tiny L}=\diag\!\big(\sin\theta_1^{(\textrm{\tiny L})},\dots,\sin\theta_N^{(\textrm{\tiny L})}\big)
\\
\rule{0pt}{.5cm}
& &
C_\textrm{\tiny R}=\diag\!\big(\cos\theta_1^{(\textrm{\tiny R})},\dots,\cos\theta_N^{(\textrm{\tiny R})}\big)
\,\,\qquad\,\,
S_\textrm{\tiny R}=\diag\!\big(\sin\theta_1^{(\textrm{\tiny R})},\dots,\sin\theta_N^{(\textrm{\tiny R})}\big)\,.
\eea

\section{Derivation of the initial growth}
\label{app:Initialgrowth}

In this appendix we report the derivation of (\ref{initial growth}) and (\ref{slope smallt final}), 
which provide the initial linear growth of the complexity (\ref{c2 complexity}) after a local quench.

As $t \to 0$, from (\ref{DeltaTR smallt}) we find the following expansion 
\bea
\label{logDeltaTR small t}
\big[\log\big(\gamma_{\textrm{\tiny T}}\,\gamma_{\textrm{\tiny R}}^{-1}\big)\big]^2
&=&
t^2
\Big[ \,E_{(1)}+ \gamma_0^{\textrm{\tiny (l,r)}}E_{(1)}^{\,\textrm{t}}\big(\gamma_0^{\textrm{\tiny (l,r)}}\big)^{-1}
\,\Big]^2
+O\big(t^3\big)
\\
\rule{0pt}{.6cm}
&& \hspace{-2.cm}
=\;
t^2
\Big[\,E_{(1)}^2
+
 \gamma_0^{\textrm{\tiny (l,r)}}\big(E_{(1)}^{\,\textrm{t}}\big)^2\big(\gamma_0^{((\textrm{l},\textrm{r})}\big)^{-1}
+
 \Big\{
 E_{(1)}\, , \,\gamma_0^{\textrm{\tiny (l,r)}}E_{(1)}^{\,\textrm{t}}\big(\gamma_0^{\textrm{\tiny (l,r)}}\big)^{-1}
 \Big\}
\, \Big]
 +O\big(t^3\big)
 \nonumber
\eea
where $\big\{ A , B \big\}$ is the anticommutator of two  matrices.
From this expansion, the complexity (\ref{c2 complexity}) becomes (\ref{initial growth}) with 
\be
\label{slope smallt}
c_1
\,=\,
\frac{1}{2\sqrt{2}} \; \sqrt{\textrm{Tr}\Big[\,
E_{(1)}^2
+
 \gamma_0^{\textrm{\tiny (l,r)}}\big(E_{(1)}^{\,\textrm{t}}\big)^2\big(\gamma_0^{\textrm{\tiny (l,r)}}\big)^{-1}
+
 \Big\{E_{(1)}\, , \,\gamma_0^{\textrm{\tiny (l,r)}}E_{(1)}^{\,\textrm{t}}\big(\gamma_0^{\textrm{\tiny (l,r)}}\big)^{-1}\Big\}
\, \Big]}
\ee
where the matrix within the argument of the trace 
has non-negative eigenvalues; hence $c_1 \geqslant 0$.

By exploiting the cyclicity of the trace, one finds
\be
\label{trace E1square}
\textrm{Tr}\Big[\gamma_0^{\textrm{\tiny (l,r)}}\big(E_{(1)}^{\,\textrm{t}}\big)^2\big(\gamma_0^{\textrm{\tiny (l,r)}}\big)^{-1}\Big]
\,=\,
\textrm{Tr}\, E_{(1)}^2
\,=\,
-\,2\sum_{k=1}^N \Omega_k^2
\ee
where the last step has been obtained by using that
\be
E_{(1)}^2
=
\bigg( 
\begin{array}{cc}
\tfrac{1}{m}\,\widetilde{V}_N^{\textrm{t}}\,\mathcal{N}\,\widetilde{V}_N  \,& \boldsymbol{0}
\\
\boldsymbol{0}  \,&  \tfrac{1}{m}\,\widetilde{V}_N^{\textrm{t}}\,\mathcal{N}\,\widetilde{V}_N
\end{array}\bigg)
\ee
whose trace can be easily computed
by using that $\widetilde{V}_N$ is orthogonal
and the matrix $\mathcal{N}$ defined in (\ref{Ediag smalltimes}) is diagonal.

From (\ref{trace E1square}) we have that the first two terms within the square root in (\ref{slope smallt}) are negative; 
thus, in order to have $c_1 \geqslant 0$,
the term containing the anticommutator under the square root must be positive. 
For this term we find
\be 
\label{trace anticomm}
\textrm{Tr}
\Big[ \Big\{
E_{(1)}\, , \,
\gamma_0^{\textrm{\tiny (l,r)}}E_{(1)}^{\,\textrm{t}}\big(\gamma_0^{\textrm{\tiny (l,r)}}\big)^{-1}
\Big\}
\Big]
\,=\,
2\,
\textrm{Tr}
\Big[ 
E_{(1)}\, \gamma_0^{\textrm{\tiny (l,r)}}E_{(1)}^{\,\textrm{t}}\big(\gamma_0^{\textrm{\tiny (l,r)}}\big)^{-1}
\,\Big]\,.
\ee
From (\ref{Emat smalltimes}), (\ref{initial CM general}) and (\ref{Q0-P0 mat diag}), we get
\be
E_{(1)}\, \gamma_0^{\textrm{\tiny (l,r)}}E_{(1)}^{\,\textrm{t}}\big(\gamma_0^{\textrm{\tiny (l,r)}}\big)^{-1}
\,=\,
\bigg( 
\begin{array}{cc}
\tfrac{1}{m^2}\,\widetilde{V}_0^{\textrm{t}}\, \mathcal{P}_0\,\mathcal{Q}_0^{-1}\,\widetilde{V}_0   \,
& 
\boldsymbol{0}
\\
\boldsymbol{0}  \,
&  
\widetilde{V}_N^{\textrm{t}}\,\mathcal{N}\,\widetilde{V}_N\,\widetilde{V}_0^{\textrm{t}}\,\mathcal{Q}_0\,\widetilde{V}_0\,\widetilde{V}_N^{\textrm{t}}\,\mathcal{N}\,\widetilde{V}_N\,\widetilde{V}_0^{\textrm{t}}\,\mathcal{P}_0^{-1}\,\widetilde{V}_0
\end{array}\bigg)
\ee
where $\widetilde{V}_0$ has been defined in (\ref{Vtilde0 mat}).
Then, by exploiting (\ref{Q0-P0 mat diag}) and (\ref{T0 mat diag}), we obtain
\bea
\label{trace anticomm v2}
\textrm{Tr}
\Big[\, E_{(1)}\, \gamma_0^{\textrm{\tiny (l,r)}}E_{(1)}^{\,\textrm{t}}\big(\gamma_0^{\textrm{\tiny (l,r)}}\big)^{-1}
\,\Big]
&=&
\sum_{k=1}^{N_{\textrm{\tiny l}}}\big[\Omega^{\textrm{\tiny (l)}}_k\big]^2
+
\sum_{k=1}^{N_{\textrm{\tiny r}}}\big[\Omega^{\textrm{\tiny (r)}}_k\big]^2
\\
\rule{0pt}{.6cm}
& &
+\;
\textrm{Tr}\Big[\,
\widetilde{V}_N^{\textrm{t}}\,\mathcal{N}\,\widetilde{V}_N\,\widetilde{V}_0^{\textrm{t}}\,\mathcal{Q}_0\,\widetilde{V}_0\,\widetilde{V}_N^{\textrm{t}}\,\mathcal{N}\,\widetilde{V}_N\,\widetilde{V}_0^{\textrm{t}}\,\mathcal{P}_0^{-1}\,\widetilde{V}_0
\,\Big]
\nonumber
\eea
where $\Omega^{\textrm{\tiny (l)}}_k $ and $\Omega^{\textrm{\tiny (r)}}_k$ 
are given by (\ref{dispersion relation general}) 
and $m_{\textrm{\tiny l}}=m_{\textrm{\tiny r}}=m$
for the local quench that we are considering.
Since we are not able to simplify $\widetilde{V}_N\,\widetilde{V}_0^{\textrm{t}}$,
we cannot write an analytic expression for the last term in the r.h.s. of (\ref{trace anticomm v2}). 
Finally, by using (\ref{trace E1square}), (\ref{trace anticomm}) and (\ref{trace anticomm v2}), 
we obtain that the slope $c_1$  in (\ref{slope smallt}) becomes the expression
(\ref{slope smallt final}) in the main text.


\bibliographystyle{nb}

\bibliography{refsMSC}

\end{document}